\documentclass{aa}  

\usepackage{graphicx}
\usepackage{epsfig}
\usepackage{epstopdf}

\usepackage{xcolor}

\usepackage{multirow} 

\usepackage{array} 

\usepackage{txfonts}
\begin{document}

   \title{Gemini-GRACES high-quality spectra of \textit{Kepler} evolved stars with transiting planets\thanks{Based on observations obtained at the Gemini Observatory, which is operated by the Association of Universities for Research in Astronomy, Inc., under a cooperative agreement with the NSF on behalf of the Gemini partnership: the National Science Foundation (United States), National Research Council (Canada), CONICYT (Chile), Ministerio de Ciencia, Tecnolog\'{i}a e Innovaci\'{o}n Productiva (Argentina), Minist\'{e}rio da Ci\^{e}ncia, Tecnologia e Inova\c{c}\~{a}o (Brazil), and Korea Astronomy and Space Science Institute (Republic of Korea).} \fnmsep \thanks{The reduced spectra (FITS files) are only available at the CDS via anonymous ftp to cdsarc.u-strasbg.fr (130.79.128.5) or via
http://cdsweb.u-strasbg.fr/cgi-bin/qcat?J/A+A/.}}

  \subtitle{I. Detailed characterization of multi-planet systems Kepler-278 and Kepler-391}



 \author{E. Jofr\'{e}\inst{1,2,10}, J. M. Almenara\inst{3}, R. Petrucci\inst{1,2,10}, R. F. D\'iaz\inst{4,5}, Y. G\'omez Maqueo Chew \inst{2}, E. Martioli\inst{6}, I. Ramírez\inst{7}, L. Garc\'ia\inst{1}, C. Saffe\inst{8,9,10}, E. F. Canul\inst{2}, A. Buccino\inst{4,5}, M. G\'{o}mez\inst{1,10}, and E. Moreno Hilario\inst{2}  
           }

\institute{Universidad Nacional de C\'ordoba - Observatorio Astron\'{o}mico de C\'{o}rdoba, Laprida 854, X5000BGR, C\'ordoba, Argentina \\
    \email{emiliano@astro.unam.mx}
         \and
   Instituto de Astronomía, Universidad Nacional Autónoma de México, Ciudad Universitaria, CDMX, C.P. 04510, México
\and
             Observatoire de Gen\`eve, D\'epartement d'Astronomie, Universit\'e de Gen\`eve, Chemin des Maillettes 51, 1290 Versoix, Switzerland
             \and
Universidad de Buenos Aires, Facultad de Ciencias Exactas y Naturales, Buenos Aires, Argentina.         
                  \and         
CONICET - Universidad de Buenos Aires, Instituto de Astronom\'ia y F\'isica del Espacio (IAFE), Buenos Aires, Argentina.
\and                  
              Laboratorio Nacional de Astrofísica, rua Estados Unidos 154, Itajubá, MG, Brazil
             \and
             Tacoma Community College, 6501 South 19th Street, Tacoma, WA 98466, United States
             \and
            Instituto de Ciencias Astronómicas, de la Tierra y del Espacio, C.C 467, 5400, San Juan, Argentina
             \and
             Universidad Nacional de San Juan, Facultad de Ciencias Exactas, Físicas y Naturales, San Juan, Argentina
                          \and
            Consejo Nacional de Investigaciones Cient\'{i}ficas y T\'{e}cnicas (CONICET), Godoy Cruz 2290, CABA, CPC 1425FQB, Argentina 
                         }

   \date{Received August 2, 2019; accepted December 17, 2019}

\titlerunning{Stellar and planetary characterization of the multiplanet systems Kepler-278 and Kepler-391}
\authorrunning{Jofr\'e et al.}

 
 \abstract
  {}
  {Kepler-278 and Kepler-391 are two of the three evolved stars known to date on the red giant branch (RGB) to host multiple short-period transiting planets. Moreover, the planets orbiting Kepler-278 and Kepler-391 are among the smallest discovered around RGB stars. Here we present a detailed stellar and planetary characterization of these remarkable systems. }
  {Based on high-quality spectra from Gemini-GRACES for Kepler-278 and Kepler-391, we obtained refined stellar parameters and precise chemical abundances for 25 elements. Nine of these elements and the carbon isotopic ratios, $^{12}\mathrm{C}/^{13}\mathrm{C}$, had not previously been measured. Also, combining our new stellar parameters with a photodynamical analysis of the \textit{Kepler} light curves, we determined accurate planetary properties of both systems.}
 {Our revised stellar parameters agree reasonably well with most of the previous results, although we find that Kepler-278 is $\sim$15\% less massive than previously reported. The abundances of C, N, O, Na, Mg, Al, Si, S, Ca, Sc, Ti, V, Cr, Mn, Co, Ni, Cu, Zn, Sr, Y, Zr, Ba, and Ce, in both stars, are consistent with those of nearby evolved thin disk stars. Kepler-391 presents a relatively high abundance of lithium (A(Li)$_{NLTE}$ = 1.29 $\pm$ 0.09 dex), which is likely a remnant from the main-sequence phase. The precise spectroscopic parameters of Kepler-278 and Kepler-391, along with their high $^{12}\mathrm{C}/^{13}\mathrm{C}$ ratios, show that both stars are just starting their ascent on the RGB.  The planets Kepler-278b, Kepler-278c, and Kepler-391c are warm sub-Neptunes, whilst Kepler-391b is a hot sub-Neptune that falls in the hot super-Earth desert and, therefore, it might be undergoing photoevaporation of its outer envelope. The high-precision obtained in the transit times allowed us not only to confirm Kepler-278c's TTV signal, but also to find evidence of a previously undetected TTV signal for the inner planet Kepler-278b. From the presence of gravitational interaction between these bodies we constrain, for the first time, the mass of Kepler-278b ($M_{\mathrm{p}}$ = 56	$\substack{+37\\-13}$ $M_{\mathrm{\oplus}}$) and Kepler-278c ($M_{\mathrm{p}}$ = 35	$\substack{+9.9\\ -21} $ $M_{\mathrm{\oplus}}$). The mass limits, coupled with our precise determinations of the planetary radii, suggest that their bulk compositions are consistent with a significant amount of water content and the presence of H$_{2}$ gaseous envelopes. Finally, our photodynamical analysis also shows that the orbits of both planets around Kepler-278 are highly eccentric ($e \sim$ 0.7) and, surprisingly, coplanar. Further observations (e.g., precise radial velocities) of this system are needed to confirm the eccentricity values presented here.}
  {}


   \maketitle
%

\section{Introduction}

To date, radial velocity (RV) surveys for planets around stars that evolved off the main-sequence, such as subgiants and giants, have resulted in the discovery of more than 150 planets \citep[e.g.,][]{Johnson2007, Niedzielski2009, Dollinger2009, Sato2010}. These detections have been crucial for extending the studies of planet-star connections to stars more massive than the Sun. The analysis of precise spectroscopic metallicities of the subgiant hosts has revealed that these stars follow the same gas-giant planet-metallicity correlation found for dwarf stars \citep[e.g.,][]{Fischer2005, Ghezzi2010b, Jofre2010, Maldonado2013, Jofre2015a}, but the presence of planets around  giant stars does not seem to be sensitive to the metallicity of their hosts \citep[e.g.,][]{Ghezzi2010b, Maldonado2013, Mortier2013, Jofre2015a}; see, however, the discussion in \citet{Reffert2015}. Moreover, precise determinations of stellar masses in evolved stars together with the results from the surveys around FGKM dwarfs show that giant planet occurrence also increases with stellar mass \citep[e.g.,][]{Lovis2007b, Johnson2010, Ghezzi2018}.

Regarding the properties of planets around evolved stars, one of the most important trends revealed by the RV searches is a paucity of close-in planets. In particular, there is a lack of planets orbiting closer than $\sim$0.5 AU (P $\lesssim$ 100 days) to giant or subgiant stars \citep[e.g.,][]{Johnson2007, Niedzielski2009, Sato2008, Sato2010}. Several scenarios have been proposed to explain the observed distribution. The first idea suggests that planets are destroyed as they spiral into their host stars as a result of tidal interactions \citep{Villaver2009, Kunitomo2011}. In a second scenario, planet formation and evolution mechanisms around stars more massive than the Sun, including the shorter lifetime of the inner protoplanetary disks, promote the lower frequency of gas giant planets at short orbital distances \citep[e.g.,][]{Johnson2007, Burkert2007, Currie2009, Kretke2009}. Another possibility is that short period RV stellar oscillations may mask the detection of short period planets that still might reside very close to their stars \citep{Pasquini2008}.

Detections of planetary transits around evolved stars are extremely challenging because their large radii cause not only shallow transit depths\footnote{Transit depth scales inversely with the square of the stellar radius.} but also long transit durations. Recently, high-precision photometry obtained with the \textit{Kepler} and \textit{K2} missions have allowed the discovery of a handful ($\sim$ 20) transiting planets around stars with $\log g <$ 3.7 dex\footnote{Around 15 planets are orbiting RGB stars.}. Interestingly, in striking contrast to the radial velocity results, most of these are close-in planets with semi-major axis between $\sim$0.06 AU and 0.3  AU. The detection and detailed characterization of these planetary systems are crucial for constraining theories of planet-star interaction \citep{Lillo-box2014, Quinn2015, VanEylen2016, Chontos2019}, models of planet inflation \citep[e.g.,][]{Lopez2016, VanEylen2016, Grunblatt2016, Grunblatt2018}, and scenarios of planet formation in intermediate and high-mass stars \citep[e.g.,][]{Burkert2007}.

Accurate atmospheric stellar parameters ($T_{\mathrm{eff}}$, $\log g$, and [Fe/H]), derived from both high resolution and high signal-to-noise ratio (S/N) spectra \citep[e.g.,][]{Sousa2011, Bedell2014}, can be combined with stellar models \citep[e.g.,][]{Demarque2004} to derive precise stellar masses, radii, and ages. However, given that most of the \textit{Kepler} planet-candidate hosts (KOIs) are too faint (V $\gtrsim 12$) to obtain high-quality spectra for all of them, their atmospheric parameters were derived first based on broadband photometric calibrations \citep{Brown2011, Pinsonneault2012}.  It has been shown that these initial parameters present limited accuracies \citep{Molenda2011, Bruntt2011, Bruntt2012, Thygesen2012, Huber2014} causing uncertainties of $\approx$ 42\% in the determination of the stellar mass and $\approx$ 16\% in radius \citep{Johnson2017}, and ultimately affecting the derived planetary properties significantly. 

\citet{Johnson2017} showed that a significant improvement in the precision of the stellar radius ($\sim$ 11\%) and stellar mass ($\sim$ 4\%) can be obtained when using high-resolution HIRES spectra to derive the fundamental parameters of a large sample of planet-candidate \textit{Kepler} hosts \citep{Petigura2017}. Most of these stars, however, have spectra with S/N $\sim$ 40-70 \citep{Martinez2019}, which may not be suitable for a detailed and precise chemical analysis that could reveal not only observable signatures of planet accretion \citep[e.g.,][]{Adamow2012, Carlberg2012, Adamow2014, Adamow2015, Aguilera2016, Melendez2017} but also to provide better constraints on their evolutionary status \citep[e.g.,][]{Gilroy1991, Carlberg2012}.

In this context, we present a detailed stellar and planetary characterization of the exceptional planetary systems Kepler-278 and Kepler-391. These are two of the three short-period multi-transiting planet systems, known to date, that transit evolved stars in the red giant branch (RGB)\footnote{Kepler-432 is a red giant star that also hosts two planets, but the outer one does not transit the star.}. Furthermore, the planet sizes are among the smallest discovered around RGB stars. Both Kepler-278 (KOI-1221, KIC 3640905; V = 11.8) and Kepler-391 (KOI-2541, KIC 12306058; V = 13.2) were  observed by \textit{Kepler} from 2009 until the end of its primary mission in 2013. Kepler-278 was identified by \citet{Borucki2011} as hosting multiple Neptune-size planet candidates with periods of 30.1 and 51.1 days. \citet{Batalha2013} revealed a single periodic transit signal with a period of 7.4 days and depth consistent with a sub-Neptune size planet candidate around Kepler-391. \citet{Rowe2014}, later reported an additional sub-Neptune size companion around Kepler-391 with a period of 20.5 days. All planets around both stars were statistically validated \citep{Rowe2014, Lissauer2014, Morton2016}. In addition, \citet{VanEylen2015} first reported that Kepler-278c exhibits sinusoidal transit timing variations (TTV).

Our characterization includes the first determination of refined stellar parameters and precise photospheric chemical abundances of 25 elements based on high-quality Gemini-GRACES spectra. We additionally performed a photodynamical modeling of the \textit{Kepler} light curves  that, in combination  with our new stellar parameters, provides improved planetary properties. In particular, for the system Kepler-278 we were able to derive the eccentricity of both planets and, thanks to the presence of dynamical interactions in the system, we constrained, for the first time, the masses of the two planets.

In Section \ref{observations}, we summarize the observations and data reduction. We present the determination of stellar parameters and the detailed chemical analysis in Section \ref{stellar-parameters}. We describe our photodynamical model and present our refined planetary parameters in Section \ref{planetary-parameters}. We discuss the resulting stellar and planetary properties of Kepler-278 and Kepler-391 in the context of other systems in Section \ref{discussion}. Finally, in Section \ref{conclusions}, we summarize our findings and conclusions.

\begin{figure*}[th!]
   \centering
   \includegraphics[width=0.85 \textwidth]{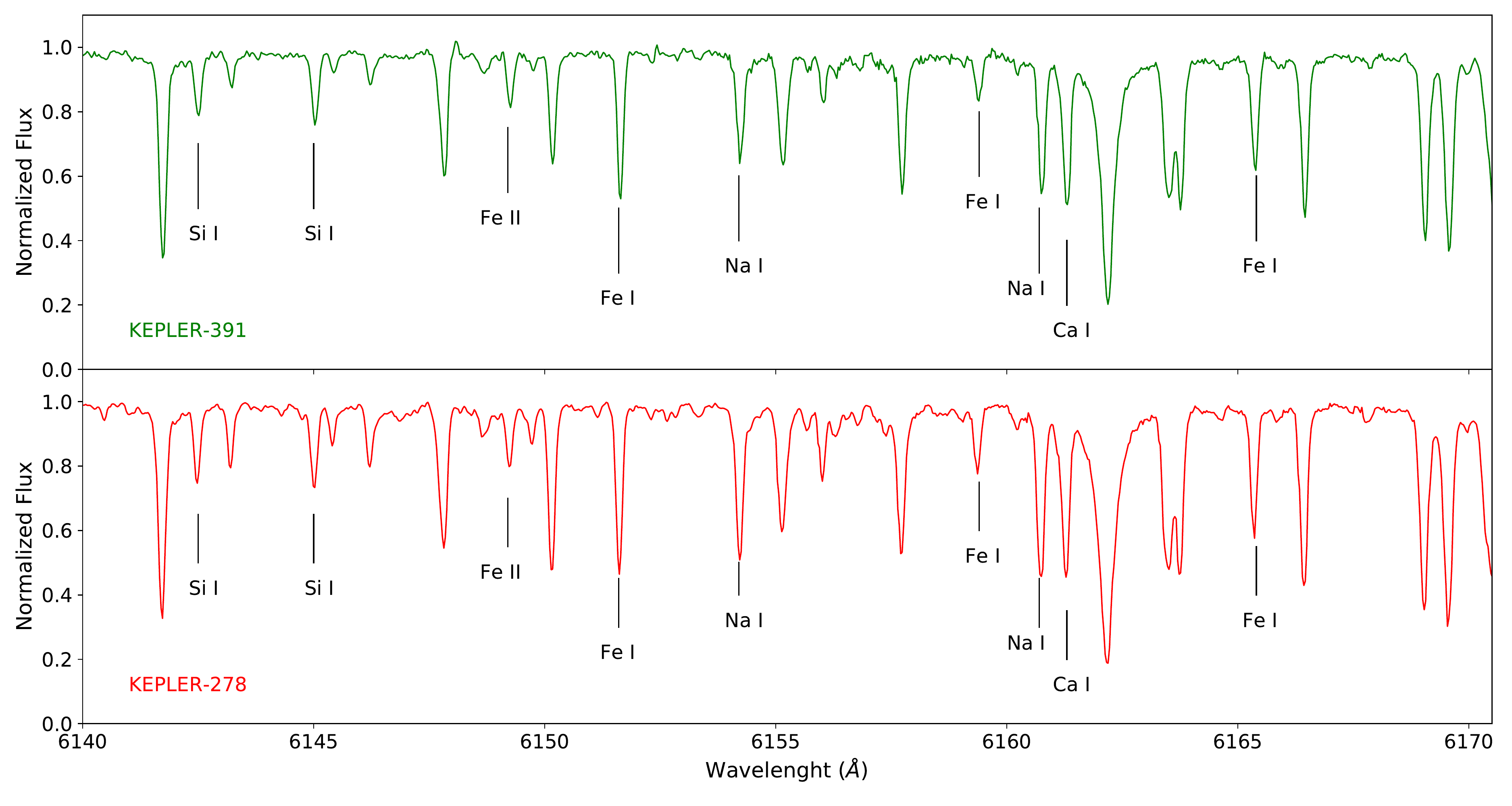}

   \caption{Narrow range of the high-quality spectra of Kepler-391 (top) and Kepler-278 (bottom)  obtained with GRACES at Gemini North Observatory. Some of the lines measured in this region to derive chemical abundances or T$_{\mathrm{eff}}$ via the line depth ratios technique are labeled.}
              \label{spectra}%
    \end{figure*}


\section{Observations}
\label{observations}
\subsection{GRACES spectra}
\label{reduction}
We observed Kepler-278 and Kepler-391 on 2016 September 10 UT with the Gemini Remote Access to CFHT ESPaDOnS Spectrograph \citep[GRACES;][]{Chene2014} at the 8.1-m Gemini North telescope. Observations were carried out in the queue mode (GN-2016B-Q-11, PI: E. Jofr\'e) in the one-fiber mode (object only), which achieves a resolving power of R$\sim$ 67,500 between 400 and 1,050 nm. We obtained consecutive exposures of 2 $\times$ 534 s for Kepler-278 and 2 $\times$ 1124 s for Kepler-391. 

These observations along with a series of calibrations including 6 $\times$ ThAr arc lamp, 10 $\times$ bias, and 7 $\times$ flat-field exposures were used as input in the OPERA (Open source Pipeline for ESPaDOnS Reduction and Analysis) code\footnote{OPERA is available at \url{http://wiki.lna.br/wiki/espectro}} software \citep{Martioli2012} to obtain the reduced data. The reduction includes optimal extraction of the orders using a tilted oversampled aperture. The aperture tilt angle is calibrated on an oversampled instrument profile, which is measured from the ThAr spectrum. The reduction also comprises the wavelength calibration where the set of calibration lines are automatically detected by computing the cross-correlation between the ThAr spectrum and the measured instrument profile. The lines are identified using the \cite{Lovis2007} ThAr atlas. The wavelength solution is obtained with an average accuracy of $\sim$60 m/s. Also, OPERA performs the normalization of the spectra as follows. First, it divides the raw flux by the normalized blaze function obtained from flat-field exposures. Then it bins the spectrum by calculating the median of every $\sim$100 points. For each bin, it calculates a local robust linear fit using two neighbor bins on each side, where the linear function is evaluated at the mean wavelength of the central bin providing an estimate to the continuum flux. Thus, the continuum is evaluated at each bin. To obtain the continuum value for each spectral element, it performs a cubic spline interpolation. Finally, it divides the flux by the continuum to obtain the normalized flux. 

The two individual exposures of each target were co-added to obtain the final spectra with signal-to-noise ratio per resolution element of S/N $\sim$ 450 and S/N $\sim$ 300 around 6700 \AA, for Kepler-278 and Kepler-391, respectively. We derived absolute radial velocities (RVs) by cross-correlating our program stars with standard stars using the IRAF\footnote{IRAF is distributed by the National Optical Astronomy Observatories, which are operated by the Association of Universities for Research in Astronomy, Inc., under cooperative agreement with the National Science Foundation.} task \texttt{fxcor}, obtaining VR = $-$46.98 $\pm$ 0.04 km s$^{-1}$ for Kepler-278 and VR = 21.59 $\pm$ 0.05 km s$^{-1}$ for Kepler-391. These values are in excellent agreement with the absolute RVs from \textit{Gaia} Data Release 2  \citep[DR2][]{Gaia2018} of $-$46.73 $\pm$ 0.83 km s$^{-1}$ and 21.55 $\pm$ 2.18 km s$^{-1}$ for Kepler-278 and Kepler-391, respectively. Finally, the combined spectra were corrected for the radial velocity shifts with the  IRAF task \texttt{dopcor}. A small portion of the spectra of both stars is shown in Fig. \ref{spectra}.

\subsection{Kepler light curve}

We retrieved the data release 25 of the \textit{Kepler} light curves \citep{Twicken2016} from the Mikulski Archive for Space Telescopes (MAST)\footnote{\url{http://archive.stsci.edu}}. Kepler-278 was observed from quarter Q1 to Q8 in long-cadence data (about one point every 29.4 min), and from Q9 to Q17 in short-cadence data (about one point per minute). Kepler-391 was observed from quarter Q0 to Q17 only in long-cadence data. We used the simple aperture photometry (SAP) light curve, which we corrected for the flux contamination (between 0.0 and 1.0\% depending on the quarter) using the value estimated by the \textit{Kepler} team. Only the data spanning three transit durations around each transit were modelled, after normalization with a parabola. The observed transits for Kepler-278b/c and Kepler-391b/c are presented in Figs \ref{transit-fit-Kepler278b}, \ref{transit-fit-Kepler278c}, \ref{transit-fit-Kepler391b}, and \ref{transit-fit-Kepler391c}.

\section{Stellar parameters}
\label{stellar-parameters}
\subsection{Fundamental atmospheric parameters} \label{fundamental}
The atmospheric fundamental parameters (T$_{\mathrm{eff}}$, $\log g$, [Fe/H], and microturbulent velocity, $v_{micro}$) of Kepler-278 and Kepler-391 were derived by imposing spectroscopic equilibrium of neutral and singly-ionized iron lines (\ion{Fe}{I} and \ion{Fe}{II}). Using the average of literature values for each star as the initial set of atmospheric parameters, all the parameters are iteratively modified until the correlations of [Fe/H] with excitation potential (EP=$\chi$) and reduced EW (REW = $\log \mathrm{EW}/\lambda$) are minimized while simultaneously minimizing the difference between the iron abundances obtained from \ion{Fe}{I} lines and those from \ion{Fe}{II} lines (see Fig. \ref{spectroscopic.equilibrium}). For this work we employed the spectral analysis code \texttt{MOOG} \citep{Sneden1973} and linearly interpolated one-dimensional local thermodynamic equilibrium (LTE) Kurucz ODFNEW model atmospheres \citep{Castelli2003} via the \texttt{Qoyllur-quipu} (or $q^{2}$) Python package\footnote{\url{https://github.com/astroChasqui/q2}} \citep{Ramirez2014}.

The iron line list, as well as the atomic parameters (EP and oscillator strengths, $\log gf$), were compiled from \citet{daSilva2011}. The list includes 72 \ion{Fe}{I} and 12 \ion{Fe}{II} lines in the range $\approx$4080--6752 \AA, and EWs were manually measured using the \texttt{splot} task in IRAF. Since weak lines are the most sensitive to changes in the fundamental parameters, we selected only those lines with EWs $<$ 100 m\AA. Lines giving abundances departing $\pm$3$\sigma$ from the average were removed, and the fundamental parameters were recomputed\footnote{Final parameters for Kepler-278 were computed from 50 lines of \ion{Fe}{I} and 9 of \ion{Fe}{II}, whilst for Kepler-391 we employed  56 lines of \ion{Fe}{I} and 8 lines of \ion{Fe}{II}.}. We adopted the solar abundances from \citet{Asplund2009}.

The resulting fundamental parameters, in both cases consistent with those of evolved stars (see \ref{discussion.evolved.stas}), were T$_{\mathrm{eff}}$= 4965 $\pm$ 48 K, $\log g$ = 3.58 $\pm$ 0.08 dex, [Fe/H]= 0.22 $\pm$ 0.04 dex, and $v_{\mathrm{micro}}$ = 1.12 $\pm$ 0.09 km s$^{-1}$ for Kepler-278 and T$_{\mathrm{eff}}$= 5038 $\pm$ 24 K, $\log g$ = 3.62 $\pm$ 0.05 dex, [Fe/H] =  0.04 $\pm$ 0.02 dex, and $v_{\mathrm{micro}}$ = 0.93 $\pm$ 0.06 km s$^{-1}$ for Kepler-391. Here, the errors correspond to intrinsic uncertainties of the technique which are based on the scatter of the individual iron abundances from each individual line, the standard deviations in the slopes of the least-squares fits of iron abundances with REW, excitation, and ionization potential \citep[see, e.g.,][]{Gonzalez1998}. 

We checked for possible differences between the atmospheric parameters derived from the interpolated Kurucz's atmosphere models, obtained via $q^{2}$, and those explicitly calculated (i.e., non-interpolated). For the latter, we employed the \texttt{FUNDPAR}  program\footnote{\url{https://sites.google.com/site/saffecarlos/fundpar}} \citep{Saffe2011}, that also derives atmospheric parameters based on the spectroscopic equilibrium procedure via the \texttt{MOOG} code but uses explicitly 1D LTE Kurucz’s model atmospheres computed with \texttt{ATLAS9} and ODFNEW opacities \citep{Castelli2003}. We find that the differences in the atmospheric parameters derived from these two set of models (in the sense interpolated $-$ calculated) are very small: $\Delta$T$_{\mathrm{eff}}$ = 27 K, $\Delta \log g$ = $-$0.02 dex, $\Delta$[Fe/H] = $-$0.03 dex, $\Delta v_{\mathrm{micro}}$ = $-$0.01  km s$^{-1}$,  and  $\Delta$T$_{\mathrm{eff}}$ = 14 K, $\Delta \log g$ = 0.02 dex, $\Delta$[Fe/H] = 0.03 dex, $\Delta v_{\mathrm{micro}}$ = 0.01 km s$^{-1}$, for Kepler-278 and Kepler-391, respectively.  

For consistency, we also computed the atmospheric parameters of both \textit{Kepler} stars using MARCS model atmospheres \citep{Gustafsson2008} instead of those from Kurucz's ODFNEW grid via the $q^{2}$ program. Although the MARCS grid that $q^2$ employs includes spherically-symmetric models, which are somewhat more realistic representations of evolved star atmospheres, we find that the dependency on the choice of model atmosphere grid is very weak. For both stars, the parameters derived from MARCS models are fully consistent with those computed with Kurucz model atmospheres, obtaining the following differences in the fundamental parameters (Kurucz$-$MARCS): $\Delta$T$_{\mathrm{eff}}$ = 20 K, $\Delta \log g$ = 0.01 dex, $\Delta$[Fe/H] = $-$0.02 dex, $\Delta v_{\mathrm{micro}}$ = $-$0.01  km s$^{-1}$,  and   $\Delta$T$_{\mathrm{eff}}$ = 26 K, $\Delta \log g$ = 0.02 dex, $\Delta$[Fe/H] = 0.02 dex, $\Delta v_{\mathrm{micro}}$ = $-$0.01 km s$^{-1}$, for Kepler-278 and Kepler-391, respectively. These results are in good agreement with those found in other evolved stars studies \citep[e.g.,][]{Ramirez2011a, Carlberg2012, Jofre2015a}.

Regarding 3D or non-LTE effects, several studies show that these effects are noticeable only in warm (T$_{\mathrm{eff}}$ $>$ 6000 K)  and for very metal-poor evolved stars \citep[e.g.,][]{Mashonkina2010, Lind2012}. Since our stars have cooler temperatures and they are not very metal-poor, 3D and non-LTE effects should not compromise our results.

Additionally, we derived projected rotational velocities ($v\sin i$) based on the spectral synthesis of six relatively isolated iron lines, following the procedure of \citet{Carlberg2012} and we adopted the calibration of \citet{Hekker2007} to determine the macroturbulence velocity, $v_{\mathrm{macro}}$. We find both objects are slow rotators: $v\sin i$ = 2.50 $\pm$ 0.65 km s$^{-1}$ and $v\sin i$ = 2.70 $\pm$ 0.70 km s$^{-1}$ for Kepler-278 and Kepler-391, respectively. These results are in excellent agreement with previous estimations computed from TRES and HIRES spectra \citep{Buchhave2012, Huber2013, Petigura2017}, and also with the expected velocities for similar evolved stars \citep[e.g.,][]{Medeiros1996}.

 \renewcommand{\arraystretch}{1.1} 

   \begin{table*}
  \small
      \caption[]{Stellar properties of Kepler-278 and Kepler-391.}
         \label{tableparameters}
     \centering
         \begin{tabular}{l c c c}
            \hline\hline
            
Parameter	&	Kepler-278			&	Kepler-391			& 	Source	\\
\hline											
\multicolumn{4}{c}{Astrometry} \\											
\hline											
Right Ascension, RA 	&	19:20:25.73			&	19:22:29.23			&	\textit{Gaia} DR2	\\
Declination, DEC 	&	+38:42:08.03			&	+51:03:26.32			&	\textit{Gaia} DR2	\\
Proper motion in RA, $\mu_{\alpha}$ [mas yr$^{-1}$]	&	2.637	$\pm$	0.047	&	10.3	$\pm$	0.034	&	\textit{Gaia} DR2	\\
Proper motion in DEC, $\mu_{\delta}$ [mas yr$^{-1}$]	&	8.95	$\pm$	0.052	&	7.585	$\pm$	0.036	&	\textit{Gaia} DR2	\\
Parallax, $\pi$ [mas] \tablefootmark{a}	&	2.146	$\pm$	0.119	&	1.040	$\pm$	0.112	&	\textit{Gaia} DR2	\\
\hline											
\multicolumn{4}{c}{Kinematics and Position} \\											
\hline											
Barycentric radial velocity, RV [km s$^{-1}$]	&	$-$46.98	$\pm$	0.04	&	21.59	$\pm$	0.05	&	This work	\\
Distance, d [pc] \tablefootmark{b}	&	465.98	$\pm$	25	&	963	$\pm$	103	&	This work	\\
Space velocity component, U [km s$^{-1}$]	&	24.56	$\pm$	0.60	&	36.33	$\pm$	2.36	&	This work	\\
Space velocity component, V [km s$^{-1}$]	&	$-$32.24	$\pm$	0.19	&	40.43	$\pm$	0.71	&	This work	\\
Space velocity component, W [km s$^{-1}$]	&	1.26	$\pm$	0.15	&	$-$14.67	$\pm$	1.35	&	This work	\\
Cartesian Galactic coordinate, X [pc]	&	430.4	$\pm$	4.9	&	916.8	$\pm$	10.5	&	This work	\\
Cartesian Galactic coordinate, Y [pc]	&	8349.9	$\pm$	96.1	&	8379.3	$\pm$	96.4	&	This work	\\
Cartesian Galactic coordinate, Z [pc]	&	91.8	$\pm$	1.1	&	268.8	$\pm$	3.1	&	This work	\\
\hline											
\multicolumn{4}{c}{Photometry\tablefootmark{c}} \\											
\hline											
Kep [mag]	&	11.584			&	13.007			&	KIC	\\
G [mag]	&	11.530	$\pm$	0.003	&	12.984	$\pm$	0.001	&	\textit{Gaia} DR2	\\
B$_{P}$ [mag]	&	12.068	$\pm$	0.001	&	13.492	$\pm$	0.002	&	\textit{Gaia} DR2	\\
R$_{P}$ [mag]	&	10.873	$\pm$	0.001	&	12.342	$\pm$	0.001	&	\textit{Gaia} DR2	\\
B [mag]	&	12.79	$\pm$	0.24	&	14.075	$\pm$	0.018	&	APASS	\\
V [mag]	&	11.818	$\pm$	0.14	&	13.203	$\pm$	0.015	&	APASS	\\
g [mag]	&	12.328			&	13.599	$\pm$	0.003	&	PAN-STARRS	\\
r [mag]	&	11.644			&	13.206			&	PAN-STARRS	\\
i [mag]	&	11.338			&	12.892			&	PAN-STARRS	\\
z [mag]	&	11.163			&	12.713			&	PAN-STARRS	\\
y [mag]	&	11.081			&	12.535	$\pm$	0.005	&	PAN-STARRS	\\
J [mag]	&	10.047	$\pm$	0.021	&	11.564	$\pm$	0.025	&	2MASS	\\
H [mag]	&	9.581	$\pm$	0.018	&	11.072	$\pm$	0.027	&	2MASS	\\
Ks [mag]	&	9.465	$\pm$	0.018	&	10.97	$\pm$	0.025	&	2MASS	\\
W1 [mag]	&	9.37	$\pm$	0.024	&	10.916	$\pm$	0.024	&	ALLWISE	\\
W2 [mag]	&	9.456	$\pm$	0.02	&	11.001	$\pm$	0.02	&	ALLWISE	\\
W3 [mag]	&	9.285	$\pm$	0.025	&	10.991	$\pm$	0.057	&	ALLWISE	\\
W4 [mag]	&	8.57	$\pm$	0.207	&	$<$ 9.365			&	ALLWISE	\\
\hline											
\multicolumn{4}{c}{Fundamental and physical parameters} \\											
\hline											
Effective temperature, T$_{\mathrm{eff}}$ [K]	&	4965	$\pm$	54	&	5038	$\pm$	57	&	This work	\\
Surface gravity, $\log g$ [cgs]	&	3.58	$\pm$	0.095	&	3.62	$\pm$	0.05	&	This work	\\
Metallicity, $\mathrm{[Fe/H]}$	&	0.22	$\pm$	0.04	&	0.04	$\pm$	0.02	&	This work	\\
Microturbulent velocity, $v_{\mathrm{micro}}$ [km s$^{-1}$]	&	1.12	$\pm$	0.09	&	0.92	$\pm$	0.06	&	This work	\\
Macroturbulent velocity, $v_{\mathrm{macro}}$ [km s$^{-1}$]	&	3.54	$\pm$	1.32	&	3.72	$\pm$	1.70	&	This work	\\
Rotational velocity, v$\sin i$ [km s$^{-1}$]	&	2.50	$\pm$	0.67	&	2.60	$\pm$	0.70	&	This work	\\
Spectral type, ST	&	K2 III-IV			&	K2 III-IV			&	This work	\\
Activity index $\log (R'_{HK})_{Ca II-IRT}$	&	$-$5.31 $\pm$ 0.04	&	$-$5.14 	$\pm$ 0.04	&	This work	\\
Mass, $M_{\mathrm{\star}}$ [$M_{\mathrm{\odot}}$]	&	1.227	$\pm$	0.061	&	1.270	$\pm$	0.081	&	This work	\\
Radius, $R_{\mathrm{\star}}$ [$R_{\mathrm{\odot}}$]	&	2.861	$\pm$	0.060	&	2.879	$\pm$	0.318	&	This work	\\
Age, $\tau_{\star}$ [Gyr]	&	5.761	$\pm$	1.019	&	4.365	$\pm$	0.899	&	This work	\\
Density, $\rho_{\star}$ [g cm$^{-3}$]\tablefootmark{d}	&	0.074	$\pm$	0.005	&	0.073	$\pm$	0.006	&	This work	\\
Luminosity, $L_{\mathrm{\star}}$ [$L_{\mathrm{\odot}}$]\tablefootmark{e}	&	4.46	$\pm$	0.57	&	4.60	$\pm$	1.34	&	This work	\\
\hline											
\multicolumn{4}{c}{Asteroseismic properties} \\											
\hline											
Frequency separation, $\Delta \nu$ [$\mu$Hz]	&	30.63	$\pm$	0.20	&	-			&	\citet{Huber2013}	\\
Frequency of maximum oscillation power, $\nu_{max}$ [$\mu$Hz]	&	500.7	$\pm$	7.0	&	-			&	\citet{Huber2013}	\\
\hline											

            \hline
         \end{tabular}

 \tablefoot{        
\tablefoottext{a}{Values have been corrected for the $-$82 $\pm$ 33 $\mu$arcsec offset found by \citet{Stassun2018}.}\\
\tablefoottext{b}{Derived from the corrected \textit{Gaia} DR2 parallaxes \citep{Gaia2018}. To compute the errors on the distances, we have quadratically included 0.1 mas to the parallaxes uncertainties to account for systematic errors of \textit{Gaia}’s astrometry \citep{Luri2018}.} \\
\tablefoottext{c}{We assume 1\% of error when no errors are available in the catalogs.}\\
\tablefoottext{d}{Calculated from the derived stellar radius and mass.} \\
  \tablefoottext{e}{Calculated using the relation $\left( \dfrac{L_{\mathrm{\star}}}{L_{\mathrm{\odot}}} \right) = \left( \dfrac{R_{\mathrm{\star}}}{R_{\mathrm{\odot}}} \right)^{2} \left( \dfrac{T_{\mathrm{eff, \star}}}{T_{\mathrm{eff, \odot}}} \right)^{4} $.}  

}  
        \end{table*}

\subsubsection{Consistency checks on T$_{\mathrm{eff}}$ and $\log g$} 
\label{consistency}

Fundamental atmospheric parameters, especially T$_{\mathrm{eff}}$ and $\log g$, have a great impact on the stellar mass and radius determination and, ultimately, on the resulting planetary properties. Therefore, we present a set of consistency checks on our spectroscopically established T$_{\mathrm{eff}}$ and $\log g$ values to determine their reliability and external accuracy \citep{Sousa2011}. 
 
We performed the following checks on the spectroscopic effective temperatures:

\textit{Photometric estimate}. We computed photometric effective temperatures using the metallicity-dependent T$_{\mathrm{eff}}$-color calibrations from \citet{Casagrande2010}. Based on available photometric data (see Table \ref{tableparameters}), we calculated ($B-V$), ($V-J$), ($V-H$), ($V-K_{S}$), and ($J-K_{S }$) colors. Magnitudes were corrected for extinction using the tables from \citet{Arenou1992} and adopting the extinction ratios, $k = E(color)/ E(B-V)$, from \citet{Ramirez2005}. Using our [Fe/H] values, we obtained average effective temperatures of T$_{\mathrm{eff}}$ =  4939 $\pm$ 60 K and T$_{\mathrm{eff}}$ =  4997 $\pm$ 80 K for Kepler-278 and Kepler-391, respectively, which in both cases are in good agreement with our spectroscopic T$_{\mathrm{eff}}$ determinations.

\textit{Equivalent width line strength ratios}. We used the \texttt{T$_{\mathrm{eff}}$-LR}\footnote{\url{http://www.astro.up.pt/~sousasag/ares/line_ratiopick2.php}} code that relies on the calibration between T$_{\mathrm{eff}}$ and 433 line EWs ratios obtained by \citet{Sousa2010} analyzing 451 FGK dwarf stars. The EWs line ratios are built from 171 spectral lines of different chemical elements including  Fe, Na, Si, Sc, Cr, Co, Ti, V, Ni, and Co. In Fig. \ref{spectra} we mark some of the doublets used for both stars. The 433 line ratios are build from a list of 171 lines \citep{Sousa2010} whose EWs were measured automatically using the upgraded version of the code \texttt{ARES}\footnote{\url{http://www.astro.up.pt/~sousasag/ares/}} \citep{Sousa2015}. Of the 433 EWs ratios included in the list of Sousa et al. to build the T$_{\mathrm{eff}}$-LR calibration, our final T$_{\mathrm{eff}}$ values were computed from 376 EWs ratios for Kepler-278 and from 368 for Kepler-391. This is because some of the EWs ratios (57 for Kepler-278 and 65 for Kepler-391) provided a temperature departing more than 2$\sigma$ from the average T$_{\mathrm{eff}}$. For the final number of EWs ratios, we obtained  T$_{\mathrm{eff}}$ = 4950 $\pm$ 55 K for Kepler-278 and T$_{\mathrm{eff}}$ = 5070 $\pm$ 50 K for Kepler-391. Even though the T$_{\mathrm{eff}}$-LR calibration was built using only dwarfs, the resulting values for our evolved stars are consistent, within the error bars, with our spectroscopic determinations.   

    \begin{figure*}[th!]
   \centering
   \includegraphics[width=.43\textwidth]{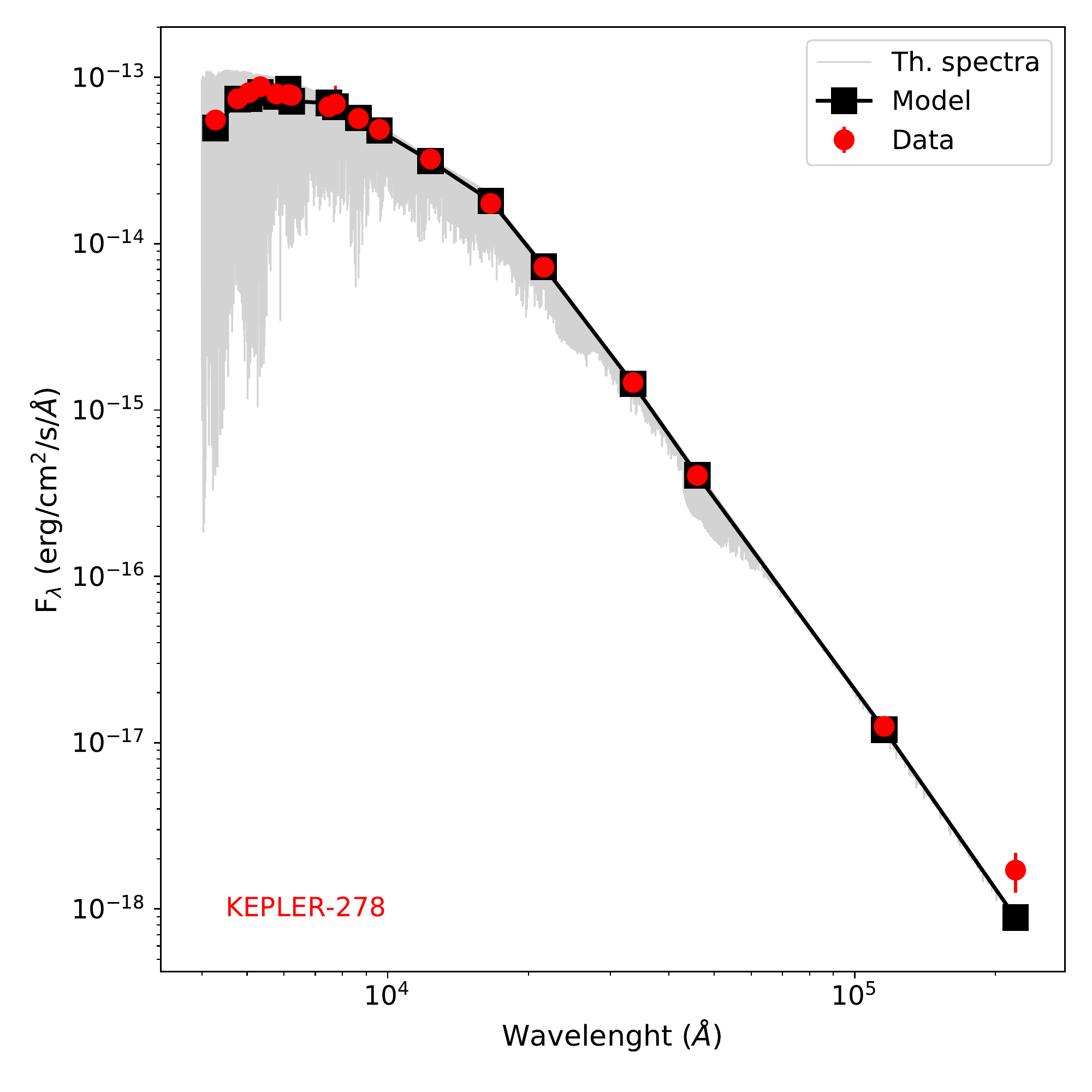}
      \includegraphics[width=.43\textwidth]{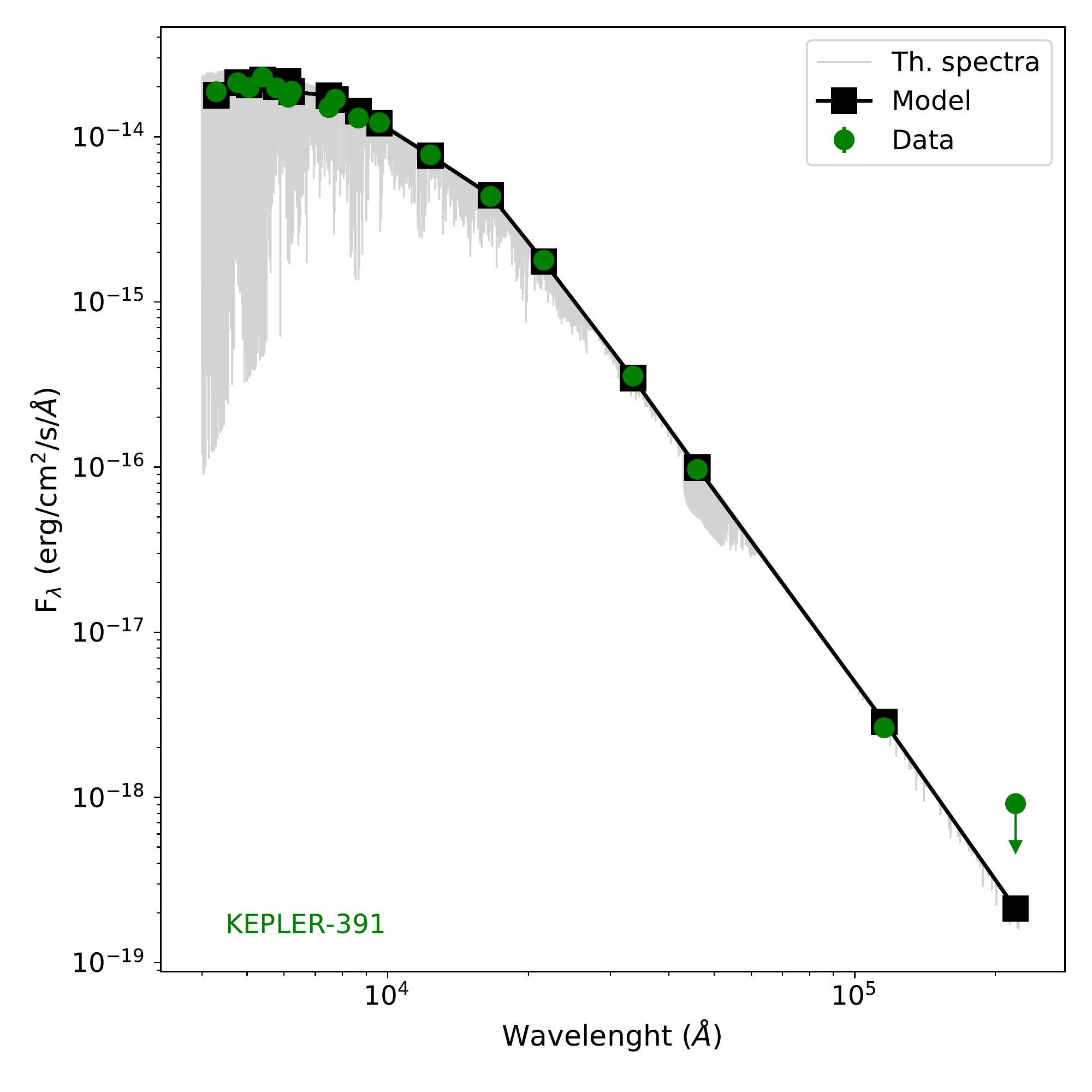}

   \caption{Spectral energy distributions of Kepler-278 (\textit{left}) and Kepler-391 (\textit{right}). The black squares and black solid lines are the best fitting models, circles (red and green) mark the observed fluxes from the optical and infrared magnitudes listed in Table \ref{tableparameters}. Grey solid lines are the best-fit synthetic spectra.}
              \label{seds}%
    \end{figure*}

\textit{Spectral energy distribution}. We obtained T$_{\mathrm{eff}}$ from the analysis of the Spectral Energy Distribution (SED) based on the latest version of the \texttt{Virtual Observatory SED Analyzer}\footnote{\url{http://svo2.cab.inta-csic.es/theory/vosa/}} \citep[\texttt{VOSA},][]{Bayo2008}. We constructed the SEDs of Kepler-278 and Kepler-391 from broadband photometry, including magnitudes from AAVSO Photometric All-Sky Survey \citep[APASS\footnote{\url{http://aavso.org/apass
}};][]{Henden2016}, Panoramic Survey Telescope and Rapid Response System \citep[PAN-STARRS;][]{Hodapp2004, Chambers2016}, 2-Micron All-Sky Survey \citep[2MASS;][]{Skrutskie2006}, \textit{Gaia} DR2  \citep{Gaia2018}, \textit{Kepler} \citep{Borucki2010}, and Wide-field Infrared Survey Explorer \citep[WISE;][]{Wright2010}. Table \ref{tableparameters} summarizes this information. We modeled the data using BT-NextGen stellar model atmospheres \citep{Allard2012} from a Chi-square fit in order to compute the expected values of T$_{\mathrm{eff}}$, surface gravity, metallicity, extinction, and a proportionality factor. The effective temperatures computed from VOSA are T$_{\mathrm{eff}}$ = 5000 $\pm$ 50 K and T$_{\mathrm{eff}}$ = 5100 $\pm$ 60 K for Kepler-278 and Kepler-391, respectively, which once again agree reasonably well with our spectroscopic estimates. The SEDs for both stars are shown in Fig. \ref{seds}. From these figures, it can be noticed that Kepler-278 might exhibit a small infrared excess which is particularly notable at the 22 $\mu$m WISE W4 bandpass. However, using the test presented in \citet{Rebull2015}, we find that the W4-excess would not be significant. Although Kepler-391 also seems to show an IR excess, in this case, the W4 magnitude is only an upper limit value.

\begin{figure*}[th!]
   \centering
   \includegraphics[width=.495\textwidth]{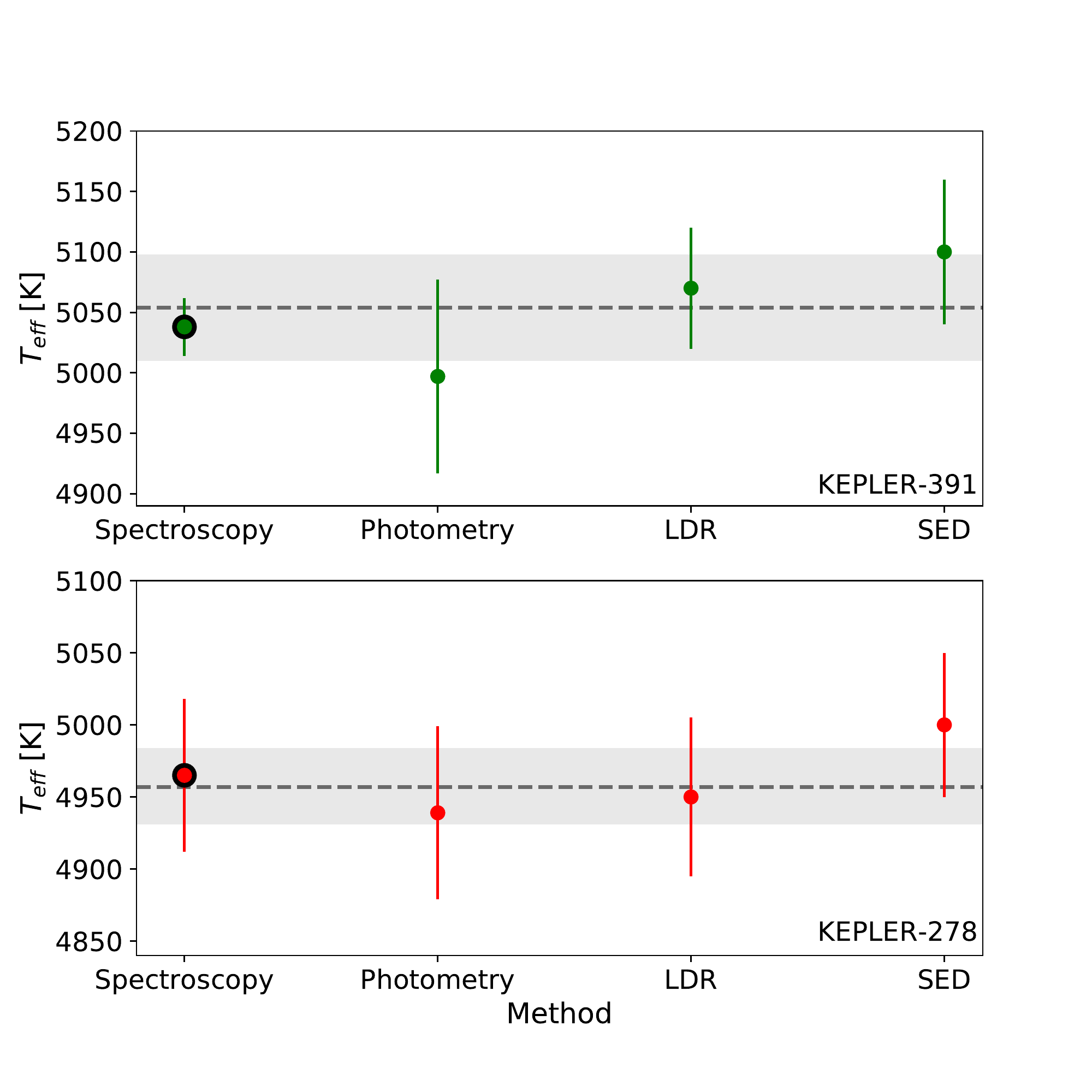}
      \includegraphics[width=.495\textwidth]{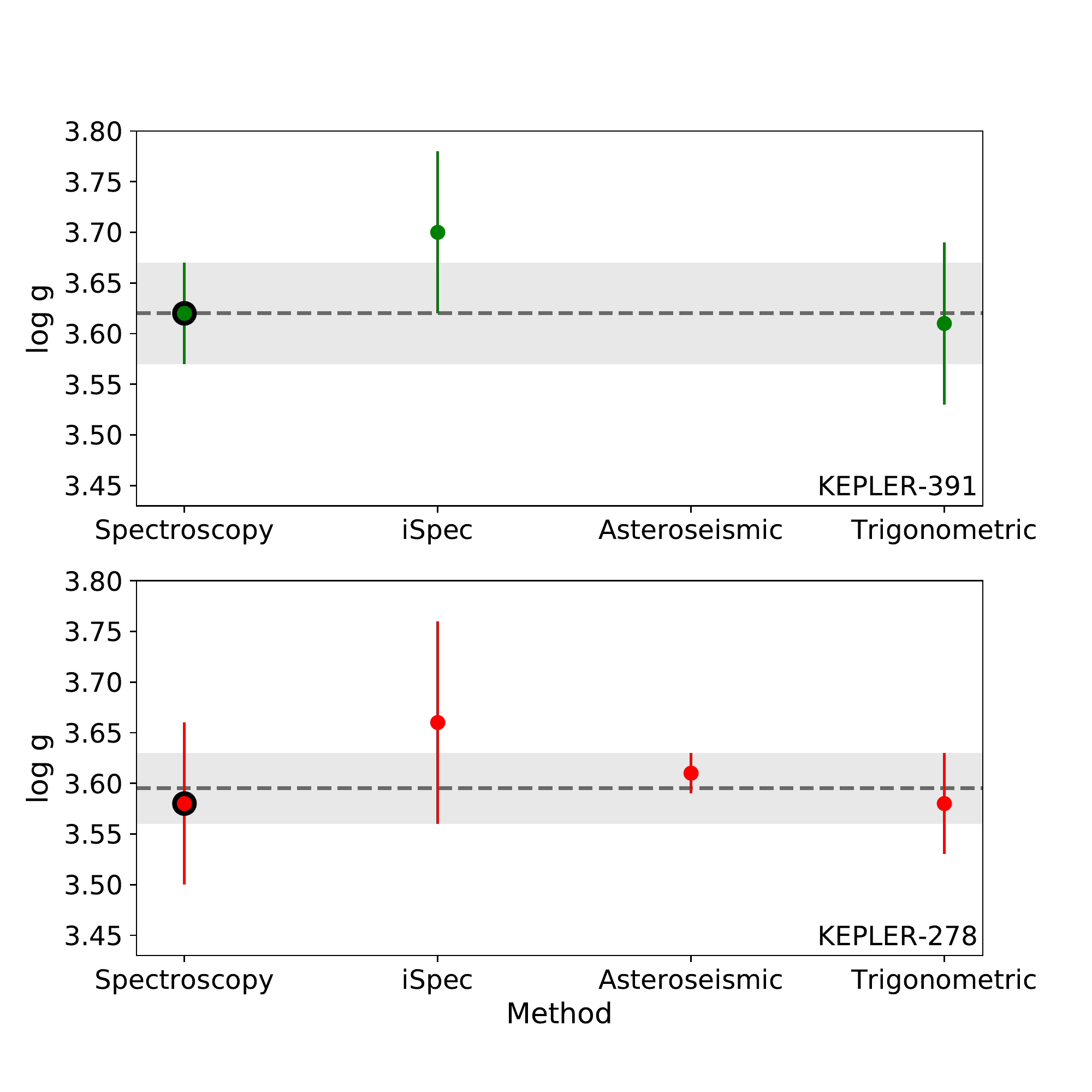}

   \caption{Stellar temperatures (\textit{left}) and surface gravity values (\textit{right}) obtained by the different consistency checks detailed in Section \ref{consistency}. Dashed lines mark the median values and the shaded areas indicate the standard deviations. The adopted spectroscopic T$_{\mathrm{eff}}$ and $\log g$ values for Kepler-278 and Kepler-391 are showed with black edge-color circles.}
              \label{checks}%
    \end{figure*}

Given the good agreement between the effective temperatures of Kepler-278 and Kepler-391 estimated from GRACES spectra and those derived from the approaches detailed above (see left panel of Fig. \ref{checks}), we can conclude that the first ones are reliable and, therefore, we adopt them as our final values to compute the chemical abundances and physical stellar and planetary parameters in the next sections.

On the other hand, for the surface gravities, we carried out the following tests: 

\textit{Surface gravity from \ion{Mg}{I} b, \ion{Na}{I} D, and Ca lines}. We obtained $\log g$ values from the spectral synthesis of the wings of several strong features using the \texttt{iSpec} spectral analysis tool\footnote{\url{https://www.blancocuaresma.com/s/iSpec}} \citep{Blanco-Cuaresma2014}, following a similar method as described in \citet{Bruntt2010}, \citet{Ramirez2011b}, and \citet{Doyle2017}. The lines analyzed include \ion{Mg}{I} b ($\lambda$5180 \AA), \ion{Na}{I} D ($\lambda$5889, 5895 \AA), and  \ion{Ca}{I} ($\lambda$6122, 6162, 6439 \AA) from which we obtained an average value of $\log g$ = 3.66 $\pm$ 0.10 dex for Kepler-278 and $\log g$ =  3.70 $\pm$ 0.08 dex for Kepler-391. Both values are in agreement, within the errors, with the estimates found from the spectroscopic equilibrium.        

 \textit{Asteroseismic gravity}. Asteroseismic parameters (such as the maximum frequency, $\nu_{max}$), obtained from high-precision photometry (e.g., \textit{Kepler} or \textit{CoRoT}), can be combined with a T$_{\mathrm{eff}}$ estimation to provide very accurate surface gravities \citep[e.g.,][]{Gai2011, Morel2012, Hekker2013}. Using asteroseismic data from \citet{Huber2013} and the scaling relation of \citet{Brown1991}:

\begin{equation}
\log g_{seis} = \log g_{\odot} + \log \left(\dfrac{\nu_{max}}{\nu_{max,\odot}}\right)+ \dfrac{1}{2} \log \left( \dfrac{T_{\mathrm{eff}}}{T_{\mathrm{eff, \odot}}}\right) ,
\end{equation}
where $\nu_{max,\odot}$ = 3090 $\mu$Hz, $T_{\mathrm{eff, \odot}}$ = 5777 K and $\log g_{\odot}$ = 4.44 dex, we find $\log g_{seis}$ = 3.61 $\pm$ 0.02 dex for Kepler-278, which is in good agreement with the spectroscopic value. 
We could not perform this check on Kepler-391 because, unfortunately, there is no asteroseismic information available for this star. The reason for this is that Kepler-391 does not have \textit{Kepler} short-cadence data and, moreover given its $\log g$, the $\nu_{max}$ value for this star is expected to be above the Nyquist frequency for the long-cadence data \citep[e.g.,][]{Chaplin2014}.

\textit{Trigonometric gravities}. We employed the 1.3 version of the \texttt{PARAM} web interface\footnote{\url{http://stev.oapd.inaf.it/cgi-bin/param_1.3}} that performs a Bayesian estimation of stellar parameters \citep{daSilva2006, Miglio2013} based on PARSEC isochrones \citep{Bressan2012}. As input we used V magnitudes, \textit{Gaia} DR2 parallaxes \citep{Gaia2018} and our spectroscopic T$_{\mathrm{eff}}$ and [Fe/H] values. In perfect agreement with our previous estimations, \texttt{PARAM} returned  $\log g$ = 3.59 $\pm$ 0.05 dex for Kepler-278 and $\log g$ = 3.61 $\pm$ 0.08 dex for Kepler-391. Similar results are obtained using the $q^{2}$ program from which we computed $\log g$ = 3.63 $\pm$ 0.06 dex for Kepler-278 and $\log g$ = 3.60 $\pm$ 0.05 dex for Kepler-391. 
 
Similar to the results obtained for T$_{\mathrm{eff}}$, we find that the $\log g$ values determined with other independent methods are in good agreement with the ones determined in this work via ionization equilibrium for both stars (see right panel of Fig. \ref{checks}), and hence, ensuring their reliability.

We also tested the effects on T$_{\mathrm{eff}}$ and [Fe/H] of fixing the surface gravity to an external accurate value. Recently, it has been suggested the existence of a strong degeneracy between T$_{\mathrm{eff}}$, [Fe/H], and $\log g$ when solving for all three quantities simultaneously, especially when methods based on spectral synthesis are used \citep{Torres2012}. To avoid this degeneracy and improve stellar parameters, it has been pointed out that fixing $\log g$ to an external accurate value (e.g., from asteroseismic data or transits) improves the final stellar parameters. However, it has also been noticed that fixing $\log g$ to an external value has no significant effect on the temperatures or metallicities when working with methods based on EWs, and then it is sufficient to use the unconstrained values \citep{Mortier2014, Doyle2017}.

We checked these results for Kepler-278 by recomputing their fundamental parameters with $\log g$ fixed to the accurate value estimated from asteroseismology\footnote{$\log g$ from transits might be less accurate for shallower transits, low S/N light curves or planets in eccentric orbits \citep{Huber2013}.}, obtaining T$_{\mathrm{eff}}$ = 4967 $\pm$ 49 K and [Fe/H]= 0.24 $\pm$ 0.04 K, which represent a difference of only 2 K and 0.02 dex when compared to the parameters obtained with no constraints on $\log g$. Therefore, in agreement with the results of \citet{Mortier2014} and \citet{Doyle2017}, we find that T$_{\mathrm{eff}}$ and [Fe/H] are not significantly altered when $\log g$ is fixed and, hence, in the next sections we adopt the unconstrained set of fundamental stellar parameters. 

\subsubsection{Formal errors} 
\label{formal}

In addition to the internal precision errors for T$_{\mathrm{eff}}$ and $\log g$ provided in Section \ref{fundamental}, we also computed systematic (or accuracy) errors for these parameters following \citet{Sousa2011}. We derived an estimation of the systematic error in T$_{\mathrm{eff}}$ and $\log g$ by comparing the results obtained from the excitation and ionization equilibrium with those derived with the other independent methods in Section \ref{consistency}. For T$_{\mathrm{eff}}$, we obtained a mean difference of 1 $\pm$ 25 K and 22 $\pm$ 51 K for Kepler-278 and Kepler-391. Therefore, we can assume a systematic error in T$_{\mathrm{eff}}$ of 25 K and 51 K for Kepler-278 and Kepler-391, respectively, which are consistent with the average systematic error obtained recently for a large sample of subgiant stars with planets by \citet{Ghezzi2018}.

In the case of surface gravity, the comparison between the $\log g$ values obtained via ionization balance and those derived with the other techniques revealed a mean difference of 0.04 $\pm$ 0.04 dex and 0.05 $\pm$ 0.04 dex for Kepler-278 and Kepler-391, respectively. Then, we adopted 0.04 dex as the systematic error in $\log g$ for Kepler-278 and Kepler-391.

The formal errors in the spectroscopic T$_{\mathrm{eff}}$ and $\log g$ are taken as the quadratic sum of the intrinsic and systematic errors. The formal error for [Fe/H] is computed by adding in quadrature the intrinsic error, given by the scatter of the individual line-to-line iron abundances, and the errors introduced by propagating our formal uncertainties in the other atmospheric parameters. The final atmospheric parameters along with their formal errors are summarized in the fourth block of Table \ref{tableparameters}. An additional source of error in the spectroscopic parameters, not taken into account in this work, could come from the use of classical solar-scaled opacities instead of non-solar-scaled opacities that could amount up to 26 K, 0.05 dex, and 0.02 dex  in T$_{\mathrm{eff}}$, $\log g$, and [Fe/H], respectively \citep{Saffe2018}.

    \begin{figure*}[th!]
   \centering
   \includegraphics[width=.485\textwidth]{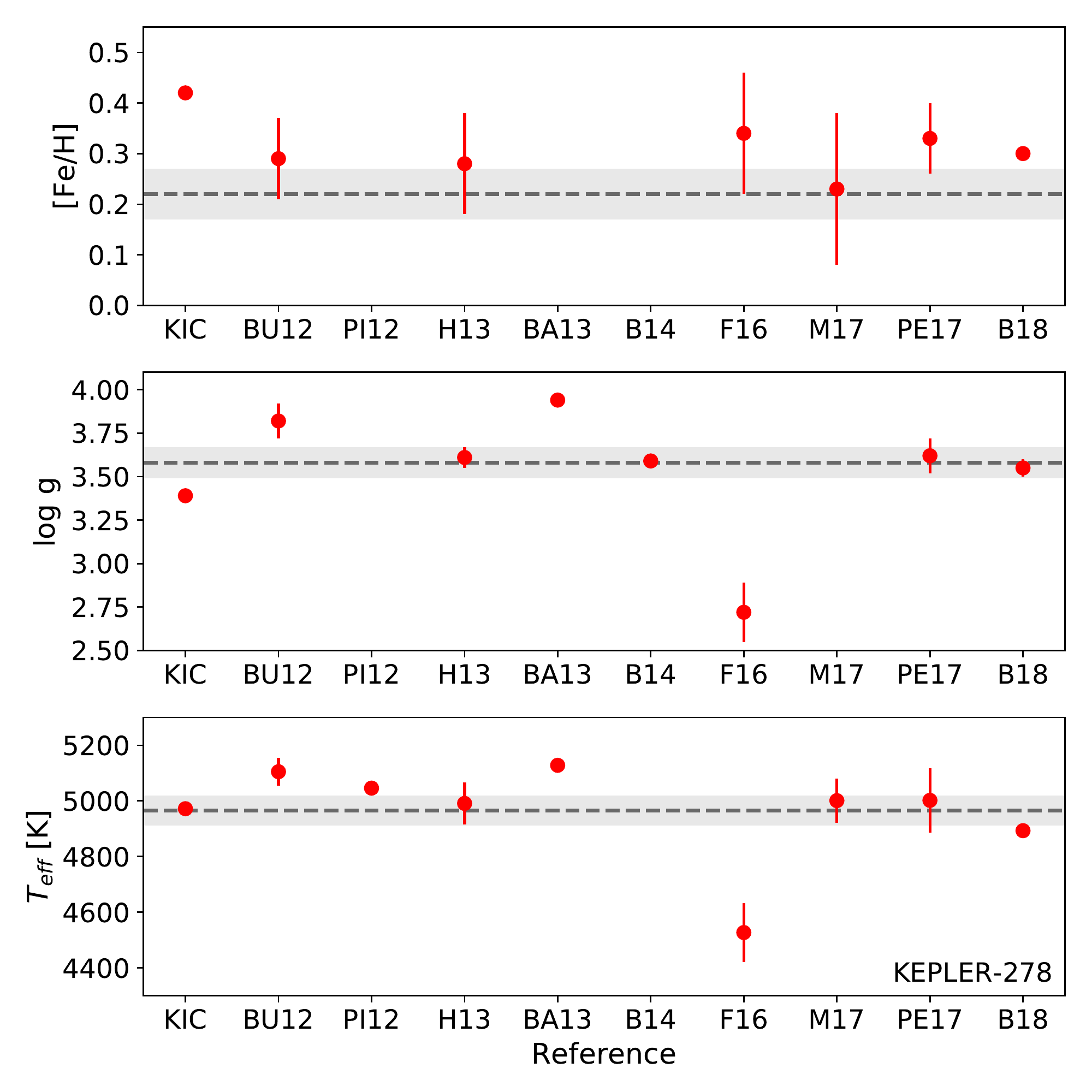}
      \includegraphics[width=.485\textwidth]{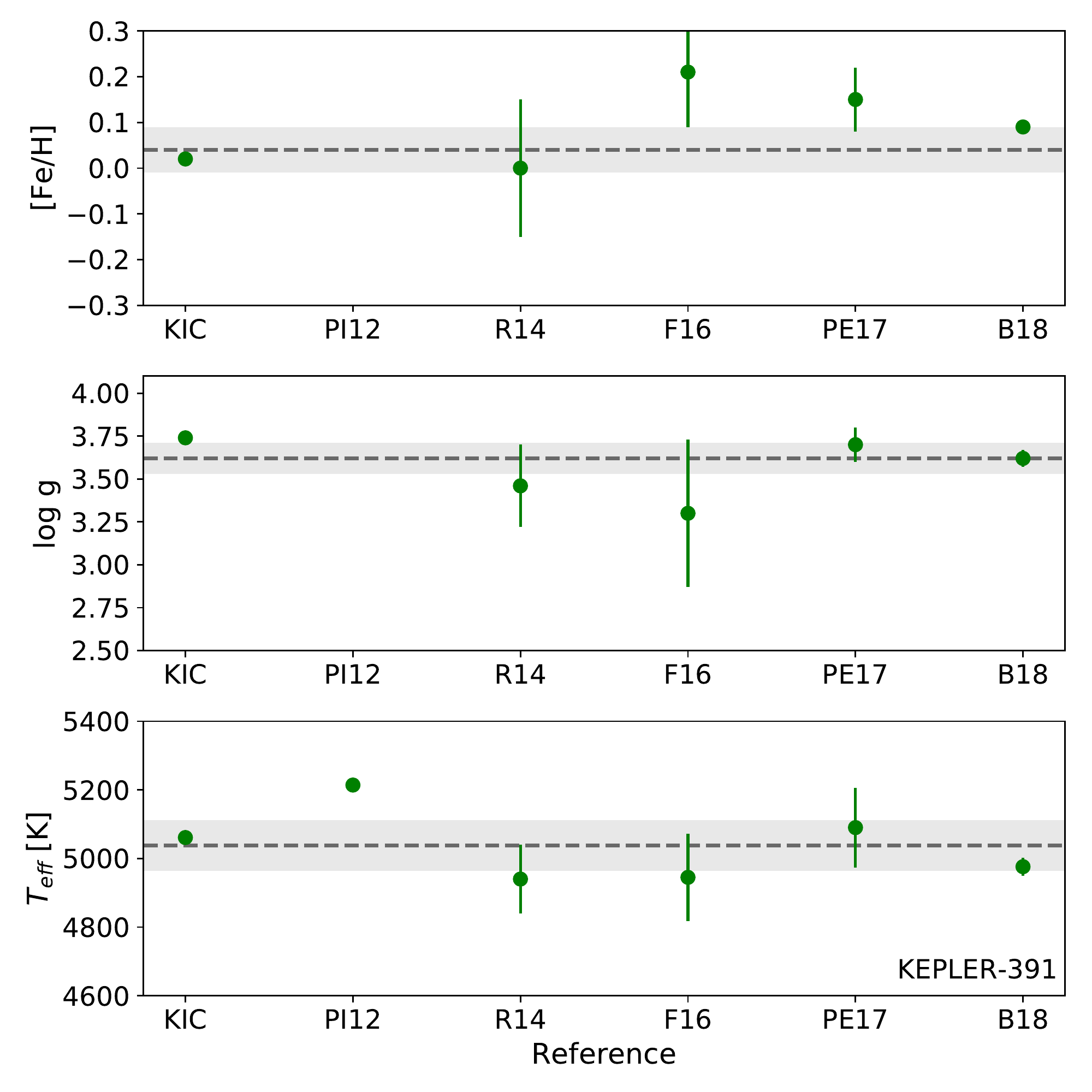}

   \caption{Comparison of our spectroscopic T$_{\mathrm{eff}}$, $\log g$, and [Fe/H] values (dashed lines) for Kepler-278 (\textit{left}) and Kepler-391 (\textit{right}) with those reported by the Kepler Input Catalogue \citep[][KIC]{Kepler2009}, \citet[][BU12]{Buchhave2012}, \citet[][PI12]{Pinsonneault2012}, \citet[][H13]{Huber2013}, \citet[][BA13]{Batalha2013}, \citet[][R14]{Rowe2014}, \citet{Bastien2014}, \citet[][F16]{Frasca2016}, \citet[][M17]{Mathur2017}, \citet[][PE17]{Petigura2017}, and \citet[][BF18]{Brewer2018a}. Grey shaded areas indicate the formal error in our results as detailed in Section \ref{formal}.}
              \label{lite-fundamental}%
    \end{figure*}

\subsubsection{Comparison with previous works}
\label{previous.atmospheric}
As a final external validation test on the reliability of our estimations, we compared our results with those previously derived based on different methods and data quality. Fig. \ref{lite-fundamental} shows our T$_{\mathrm{eff}}$, $\log g$, and [Fe/H] estimations in comparison with those of the Kepler Input Catalog \citep[][KIC hereafter]{Kepler2009}, \citet[][BU12 hereafter]{Buchhave2012}, \citet[][PI12 hereafter]{Pinsonneault2012}, \citet[][H13 hereafter]{Huber2013}, \citet[][BA13 hereafter]{Batalha2013}, \citet[][R14 hereafter]{Rowe2014}, \citet[][B14 hereafter]{Bastien2014}, \citet[][F16 hereafter]{Frasca2016}, \citet[][M17 hereafter]{Mathur2017}, \citet[][PE17 hereafter]{Petigura2017}, and \citet[][BF18 hereafter]{Brewer2018a}.
In general, our estimations for both stars are in fair agreement with those from previous studies, although a few measurements deviate $\sim$2$\sigma$ from our values. Next we discuss the potential source of these discrepancies.

Although for Kepler-391 our results are in excellent agreement with those from the KIC\footnote{No errors are provided in this catalog.}, for Kepler-278 the discrepancies are particularly significant for the surface gravity and [Fe/H] values ($\Delta \log g$ = 0.19 dex, and $\Delta$[Fe/H] =$-$ 0.2 dex). As we mentioned in the introduction, the limited accuracy of atmospheric parameters (in particular $\log g$ and [Fe/H]) based on broadband photometric calibrations only \citep{Huber2014, Bruntt2012} might explain the discrepancies with our spectroscopic values. This also could be the origin for the differences with the effective temperatures derived by PI12, which are also based on photometric calibrations at fixed [Fe/H] = $-$0.2 dex. Their temperatures are larger than our values by 84 K and 176 K for Kepler-278 and Kepler-391, respectively. 
 
From high-resolution reconnaissance spectra (i.e., with low S/N) obtained by the \textit{Kepler} Follow-up Observing Program (FOP), BU12 reported fundamental parameters of the host stars of 226 small exoplanet candidates discovered by \textit{Kepler}, including Kepler-278. Their results are obtained using the Stellar Parameter Classification (\texttt{SPC}) technique, which fits synthetic spectra to the observed data. Their results for Kepler-278\footnote{Obtained using a spectrum with S/N $\approx$ 45.}, although within the error, are systematically larger than ours: $\Delta$T$_{\mathrm{eff}}$ = $-$140 K, $\Delta \log g$ = $-$0.24 dex, and $\Delta$[Fe/H] = $-$0.07 dex. Potential sources of this discrepancy include the substantial difference in the S/N of the spectra used and the distinct analysis techniques. Moreover, as we mentioned before, techniques based on spectral synthesis are more sensitive to show strong systematic biases on T$_{\mathrm{eff}}$ and [Fe/H] when $\log g$ is unconstrained \citep{Torres2012}. This also could explain the large differences obtained in T$_{\mathrm{eff}}$ ($\Delta$T$_{\mathrm{eff}}$ = 163 K) and $\log g$ ($\Delta \log g$ = 0.16 dex) with the parameters determined by BA13\footnote{No estimation of [Fe/H] nor errors in T$_{\mathrm{eff}}$ and $\log g$ are reported.} from spectral synthesis based on low S/N and high resolution spectroscopy \citep{Gautier2010}.

The largest discrepancies, in both stars, are obtained with the results of F16 who determined stellar parameters from the spectral synthesis of LAMOST low resolution spectra \citep{Wang1996}. For Kepler-278 the differences are $\Delta$T$_{\mathrm{eff}}$ = 438 K, $\Delta \log g$ = 0.86 dex, and  $\Delta$[Fe/H] = $-$0.12 dex. For Kepler-391 the differences are a little bit smaller, but still the largest in comparison with any of the other studies: $\Delta$T$_{\mathrm{eff}}$ = 93 K, $\Delta \log g$ = 0.32 dex, and  $\Delta$[Fe/H] = $-$0.17 dex. The most probable sources for these discrepancies are the quality of the spectra, especially resolution, and the analysis technique, in which $\log g$ values are not constrained \citep[e.g.,][]{Doyle2017}.

    \begin{figure*}[th!]
   \centering
   \includegraphics[width=.495\textwidth]{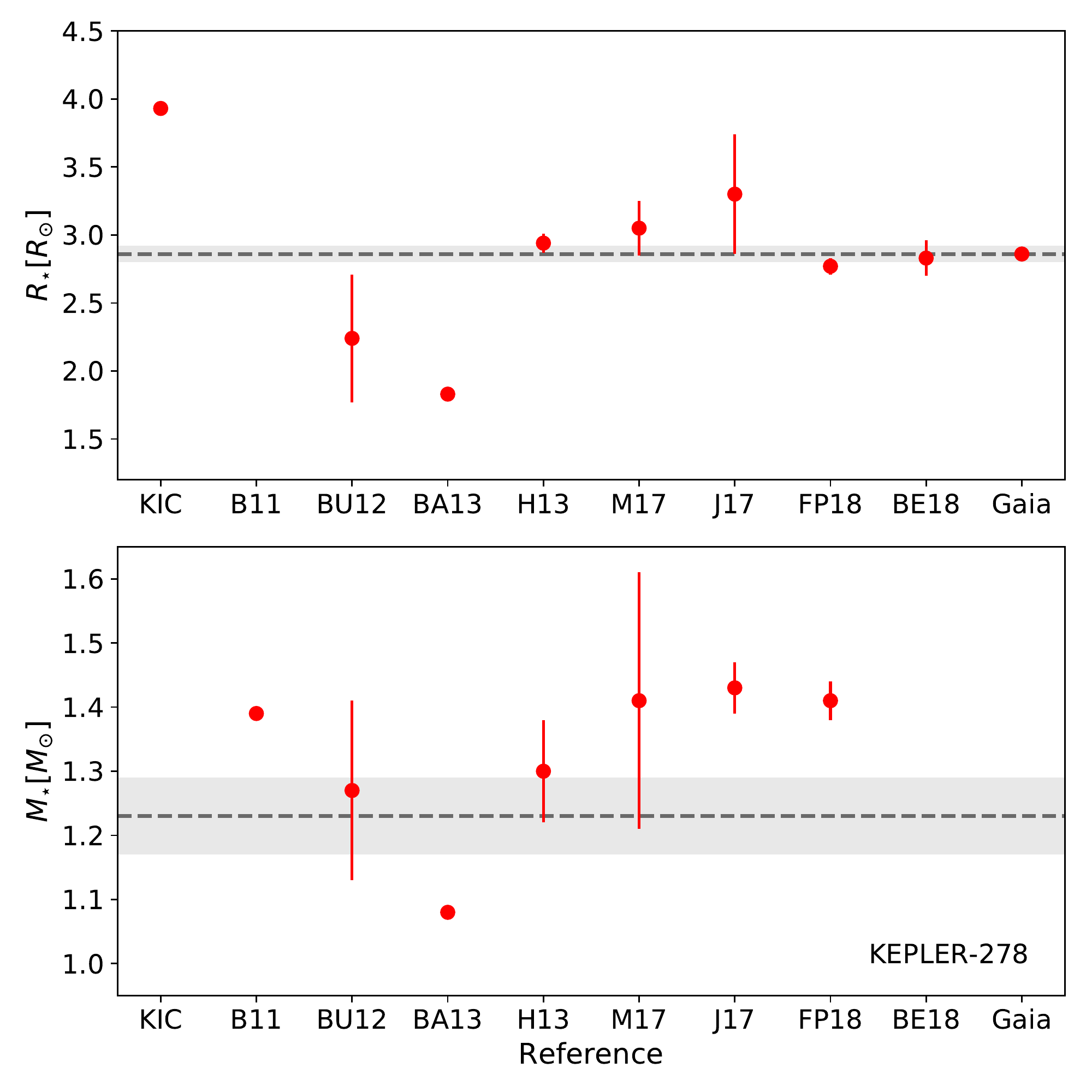}
      \includegraphics[width=.495\textwidth]{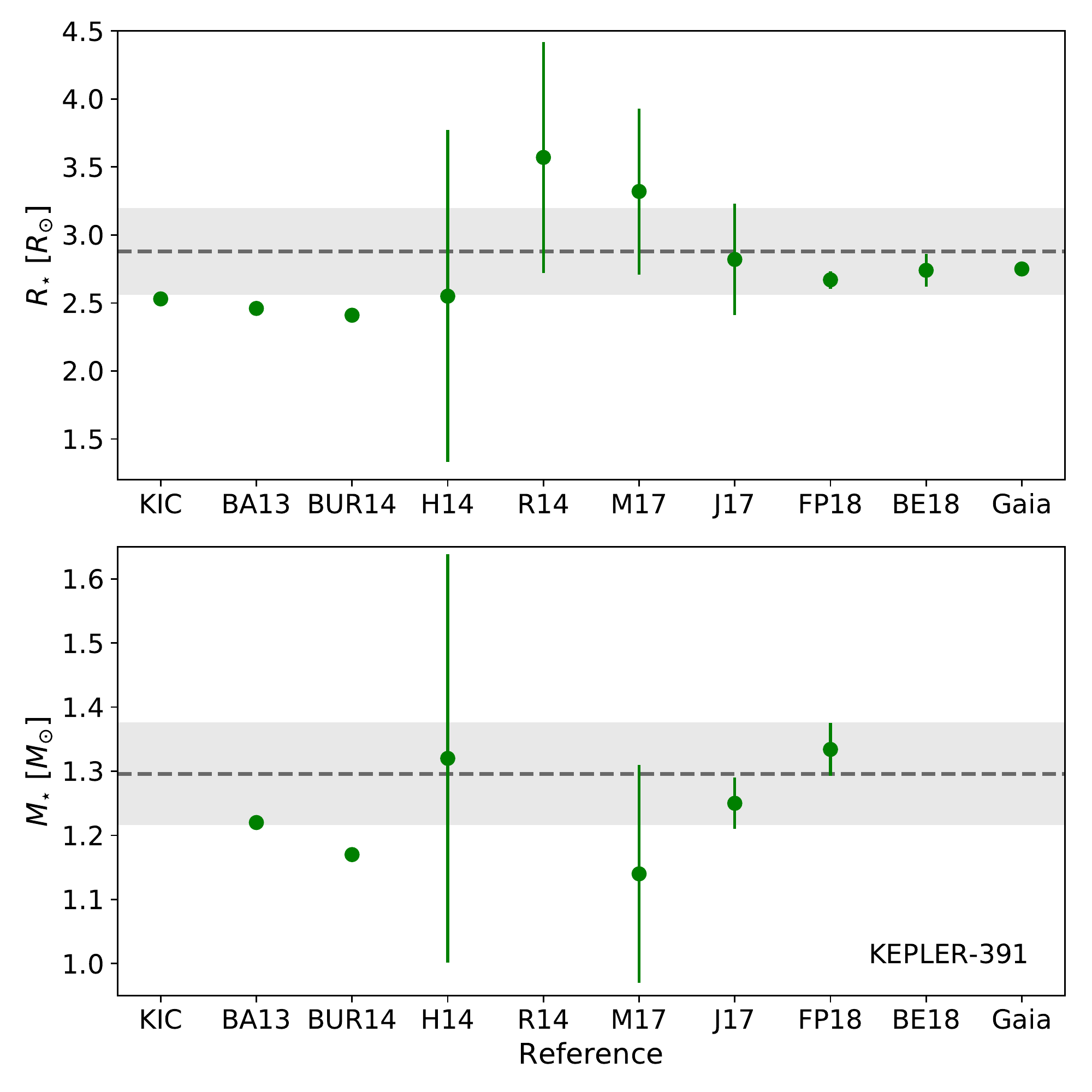}

   \caption{Stellar masses and radii determined by different authors (filled circles) in comparison with those derived in this study (dashed lines) for Kepler-278 (\textit{left}) and Kepler-391 (\textit{right}). Grey shaded areas indicate 1$\sigma$ uncertainty in our results.}
              \label{comp-lite-physical}%
    \end{figure*}

\subsection{Stellar activity indicators}
Following the procedure described in \citet{Mittag2013} for evolved stars, we characterized the chromospheric stellar activity of Kepler-278 and Kepler-391 from the fluxes of the \ion{Ca}{II} H \& K lines centered at $\lambda$3968 and $\lambda$3934 \AA\ via the $S$ and $\log(R'_{HK})$ indicators \citep[e.g.,][]{Baliunas1995, Lovis2011, Egeland2017}. Since the wavelength coverage of the GRACES spectra does not include these lines, we used public Keck-HIRES spectra obtained by the California-Kepler Survey\footnote{Observations were taken on 2014 July 14 UT for Kepler-278 and on 2012 June 20 UT for Kepler-391.}  \citep[CKS,][]{Petigura2017}. We obtained $S$ = 0.11 and $\log(R'_{HK}) = -5.10$ for Kepler-278, whilst for Kepler-391 we derived $S$ = 0.12 and $\log(R'_{HK}) = -5.14$. These values are in line with those found for the vast majority of subgiants, indicating low chromospheric activity \citep{Isaacson2010}. Moreover, our derived $S$ and $\log(R'_{HK})$ values are in good agreement with those computed recently by \citet{Brewer2018a}, although their $\log(R'_{HK})$ values were based on the definition of \citet{Noyes1984}, which is only valid for main-sequence stars. Additionaly, following \citet{Isaacson2010}, we estimated a jitter of 4.33 m s$^{-1}$ for Kepler-278 and 4.11 m s$^{-1}$ for Kepler-391. 

On the other hand,  given that GRACES spectra include the infrared triplet lines of ionized calcium (\ion{Ca}{II} - IRT) at 8498, 8542 and 8662 \AA\/, we obtained information about stellar activity from these lines following the method outlined in \citet{Busa2007}. In particular, we estimated  $\log(R'_{HK})$ = $-$5.31 $\pm$ 0.04 for Kepler-278 and $\log(R'_{HK})$ = $-$5.14 $\pm$ 0.04 for Kepler-391, considering a 10\%-error in the stellar flux.  Given that for Kepler-391 the $\log(R'_{HK})$ values derived from HIRES and GRACES data are in good agreement, we consider that the difference in the results for this activity indicator for Kepler-278 could be caused mainly by a difference in the level of stellar activity. It would be worthwhile to analyze additional spectroscopic data in order to confirm the stellar variability that might be caused by stellar spots which already have been detected in their photometry \citep{VanEylen2015}.  

\subsection{Stellar mass, radius, and age}  
\label{sec.stellar.parameters}

We derived the stellar mass $M_{\mathrm{\star}}$, radius $R_{\mathrm{\star}}$, and age $\tau_{\star}$ of Kepler-278 and Kepler-391 through the 1.3 version of the Bayesian interface \texttt{PARAM} \citep{daSilva2006} using PARSEC isochrones \citep{Bressan2012}. This applet also gives $\log g$, as mentioned in Section \ref{consistency}. As input, we provided our spectroscopic T$_{\mathrm{eff}}$ and [Fe/H], \textit{Gaia} DR2 parallaxes \citep{Gaia2018} corrected for the 82 $\mu$arcsec offset found by \citet{Stassun2018}, and V dereddened magnitudes along with all the parameters' uncertainties. For Kepler-278, \texttt{PARAM} returned $M_{\mathrm{\star}}$ = 1.325 $\pm$ 0.060 $M_{\mathrm{\odot}}$, $R_{\mathrm{\star}}$ = 2.973 $\pm$ 0.152 $R_{\mathrm{\odot}}$, and $\tau_{\star}$= 4.466 $\pm$ 0.630 Gyr, whilst for Kepler-391, we found $M_{\mathrm{\star}}$ = 1.296 $\pm$ 0.080 $M_{\mathrm{\odot}}$, $R_{\mathrm{\star}}$ = 2.879 $\pm$ 0.318 $R_{\mathrm{\odot}}$, and $\tau_{\star}$= 4.365 $\pm$ 0.899 Gyr. The computed stellar masses and radii imply stellar densities of $\rho_{\star}$= 0.071 $\pm$ 0.006 g cm$^{-3}$ and $\rho_{\star}$= 0.077 $\pm$ 0.011 g cm$^{-3}$ for Kepler-278 and Kepler-391, respectively. 

As another option, the asteroseismic quantities ($\Delta \nu$: large frequency separation and $\nu_{max}$: frequency of maximum oscillation power) can be used in \texttt{PARAM} as input parameters in replacement of V magnitude and parallax. Using this option for Kepler-278, for which seismic information is available \citep[$\nu_{max}$ = 500.7 $\pm$ 7  $\mu$Hz, $\Delta \nu$ = 30.63 $\pm$ 0.20 $\mu$Hz;][]{Huber2013}, \texttt{PARAM} returned $M_{\mathrm{\star}}$ = 1.227 $\pm$ 0.061 $M_{\mathrm{\odot}}$, $R_{\mathrm{\star}}$ = 2.861 $\pm$ 0.057 $R_{\mathrm{\odot}}$, $\tau_{\star}$= 5.761 $\pm$ 1.019 Gyr, $\log g$= 3.606 $\pm$ 0.006 dex, and from the values of $M_{\mathrm{\star}}$ and $R_{\mathrm{\star}}$ we derived  $\rho_{\star}$ = 0.074 $\pm$ 0.005 g cm$^{-3}$. All values are in excellent agreement, within 1$\sigma$, with those obtained with the first input. As final results, for Kepler-278 we adopted these values based on asteroseismic information and those based on DR2 \textit{Gaia} parallaxes for Kepler-391. Independent techniques produced consistent results as can be seen in Appendix \ref{appendix-1}.

\subsubsection*{Comparison with previous works}
\label{comparison-literature-stellar-radii-mass}
As a further check on our computed stellar physical parameters, we compared them with those reported in the literature. In Fig. \ref{comp-lite-physical}, we show the comparison of our stellar masses and radii of Kepler-278 and Kepler-391 with those obtained by KIC, \citet[][B11 hereafter]{Borucki2011}, BU12, \citet[][BA13 hereafter]{Batalha2013}, H13, \citet[][H14 hereafter]{Huber2014}, R14, M17, \citet[][J17 hereafter]{Johnson2017}, \citet[][FP18 hereafter]{Fulton2018}, and \citet[][BE18 hereafter]{Berger2018}. As can be noticed, there are significant discrepancies with some of the results of previous studies, mainly for Kepler-278.

The stellar radius reported in the KIC, $R_{\mathrm{\star}}$ = 3.93  $R_{\mathrm{\odot}}$, is $\sim$37\% larger than our value. As we mentioned before, the origin of this discrepancy is likely caused by the non-negligible difference between fundamental parameters (see Section \ref{previous.atmospheric}). In line with this difference in the stellar radius, \citet{Johnson2017} found that stellar radii in the KIC, based on broadband photometry only, have fractional uncertainties of 40\%.

Based on the $R_{\mathrm{\star}}$ and $\log g$ values reported by KIC, B11 derived a stellar mass of $M_{\mathrm{\star}}$ = 1.39 $M_{\mathrm{\odot}}$ (no error provided) for Kepler-278, which is 13\% larger than our value. Again, this is likely due to the limited accuracy of stellar parameters in the KIC. 

BU12 used the T$_{\mathrm{eff}}$, $\log g$, and [Fe/H] determined via the SPC technique from the low S/N reconnaissance spectra, in combination with a grid of Yonsei-Yale models to infer the stellar mass and radius of Kepler-278. Although the stellar mass reported by BU12 is in good agreement with our value, their radius is $\sim$22\% smaller than our estimation. This discrepancy is likely originated by the large difference between our T$_{\mathrm{eff}}$ and $\log g$ values and those from BU12 (see Section \ref{previous.atmospheric}).

H13 determined $M_{\mathrm{\star}}$ and $R_{\mathrm{\star}}$ using the so called grid-based modeling, where atmospheric parameters (T$_{\mathrm{eff}}$, [Fe/H]) and asteroseismic contraints are fitted to a grid of isochrones\footnote{H13 used six different models:  ASTEC \citep{Christensen2008}, BaSTI \citep{Pietrinferni2004}, DSEP \citep{Dotter2008}, Padova \citep{Marigo2008}, Yonsei-Yale
\citep{Demarque2004}, and YREC \citep{Demarque2008}.}. For Kepler-278, their estimations agree, within the errors, with our results. 

BA13 employed $T_{\mathrm{eff}}$, $\log g$, and [Fe/H] derived from spectroscopy (for Kepler-278) or values compiled from KIC (for Kepler-391) as initial constraints to obtain stellar masses and radii, through Yonsei--Yale stellar evolution models. For Kepler-391 they reported $M_{\mathrm{\star}}$ = 1.22 $M_{\mathrm{\odot}}$ and $R_{\mathrm{\star}}$ = 2.46 $R_{\mathrm{\odot}}$ (no errors provided), which are just below our 1$\sigma$ range. For Kepler-278, BA13 reported $M_{\mathrm{\star}}$ = 1.08 $M_{\mathrm{\odot}}$ and $R_{\mathrm{\star}}$ = 1.80 $R_{\mathrm{\odot}}$, which are $\sim$12\% and $\sim$36\% smaller than our estimations. The origin of these discrepancies is probably related to differences between our fundamental parameters and those from BA13, especially for Kepler-278 (see Fig. \ref{lite-fundamental}).

BUR14 employed photometric T$_{\mathrm{eff}}$ from PI12 and $\log g$ and [Fe/H] values from the  KIC as initial constraints to obtain stellar masses and radii via stellar evolution models as in BA13. For Kepler-391, they determined $M_{\mathrm{\star}}$ = 1.17 $M_{\mathrm{\odot}}$ and $R_{\mathrm{\star}}$ = 2.41 $R_{\mathrm{\odot}}$ (no errors provided), which are very similar to those from BA13 and are just outside the 1$\sigma$ region in our results. As before, the cause of these discrepancies is likely related to differences between our fundamental parameters, especially for the  T$_{\mathrm{eff}}$ (see Fig. \ref{lite-fundamental}).

For Kepler-391, R14 derived stellar parameters based on Yonsei-Yale models matching, using the atmospheric parameters derived from low S/N HIRES spectra as initial constraints. They estimated $R_{\mathrm{\star}}$ = 3.57 $\pm$ 0.85 $R_{\mathrm{\odot}}$, which is $\sim$24\% larger than our estimation. Curiously, they do not report the value for the stellar mass nor the age. However, they report an estimation for the stellar density of $\rho_{\mathrm{\star}}$ = 0.042 g cm$^{-3}$, which combined with the stellar radius would imply a mass of $M_{\mathrm{\star}}$ = 1.35 $M_{\mathrm{\odot}}$. All the parameters agree with our results within the errors. 

J17 used Dartmouth stellar evolution models to convert the spectroscopic properties T$_{\mathrm{eff}}$, $\log g$, and [Fe/H], compiled from PE17, into mass, radius, and age via the \texttt{isochrones} code \citep{Morton2015}. For Kepler-391, their estimations agree well with our results. However, for Kepler-278, their derived stellar radius and mass values are $\sim$15\%  and $\sim$16\% larger than our estimations. Recently, FP18 derived stellar radii for the CKS sample from the Stefan-Boltzmann law based on \textit{Gaia} parallaxes, \textit{Kepler} photometry, and spectroscopic temperatures compiled from PE17. In parallel, they also determined masses, radii, and ages via MIST isochrone grids \citep{Choi2016} by providing T$_{\mathrm{eff}}$, $\log g$, and [Fe/H] from PE17, along with $m_{K}$ 2MASS constraints through the \texttt{isoclassify} package \citep{Huber2017}. The agreement between the stellar radii they derived by different approaches is excellent both for Kepler-278 and Kepler-391. However, similarly to J17, we found a significant discrepancy in the mass values for Kepler-278 of $\sim$15\%. Given that there is no significant difference between our fundamental parameters against those of PE17 and J17, the origin of the discrepancies in the stellar mass of Kepler-278 is not clear. A possibility could be the use of different stellar models (e.g., MESA vs. PARSEC) to obtain the stellar parameters, although it is not evident why there are no differences for Kepler-391. Additionally, as demonstrated by \citet{Huber2019} for TOI-197, uncertainties in the stellar parameters can be underestimated by the use of single, instead of multiple, model grids that do not allow to take systematic errors into account.

Finally, our stellar radii are in perfect agreement with those provided by \textit{Gaia} DR2 \citep{Gaia2018}, based on SED modeling, and those from BE18 derived by combining \textit{Gaia} DR2 parallaxes with the DR25 \textit{Kepler} Stellar Properties Catalog.

\begin{figure}[]
   \centering
   \includegraphics[width=0.50 \textwidth]{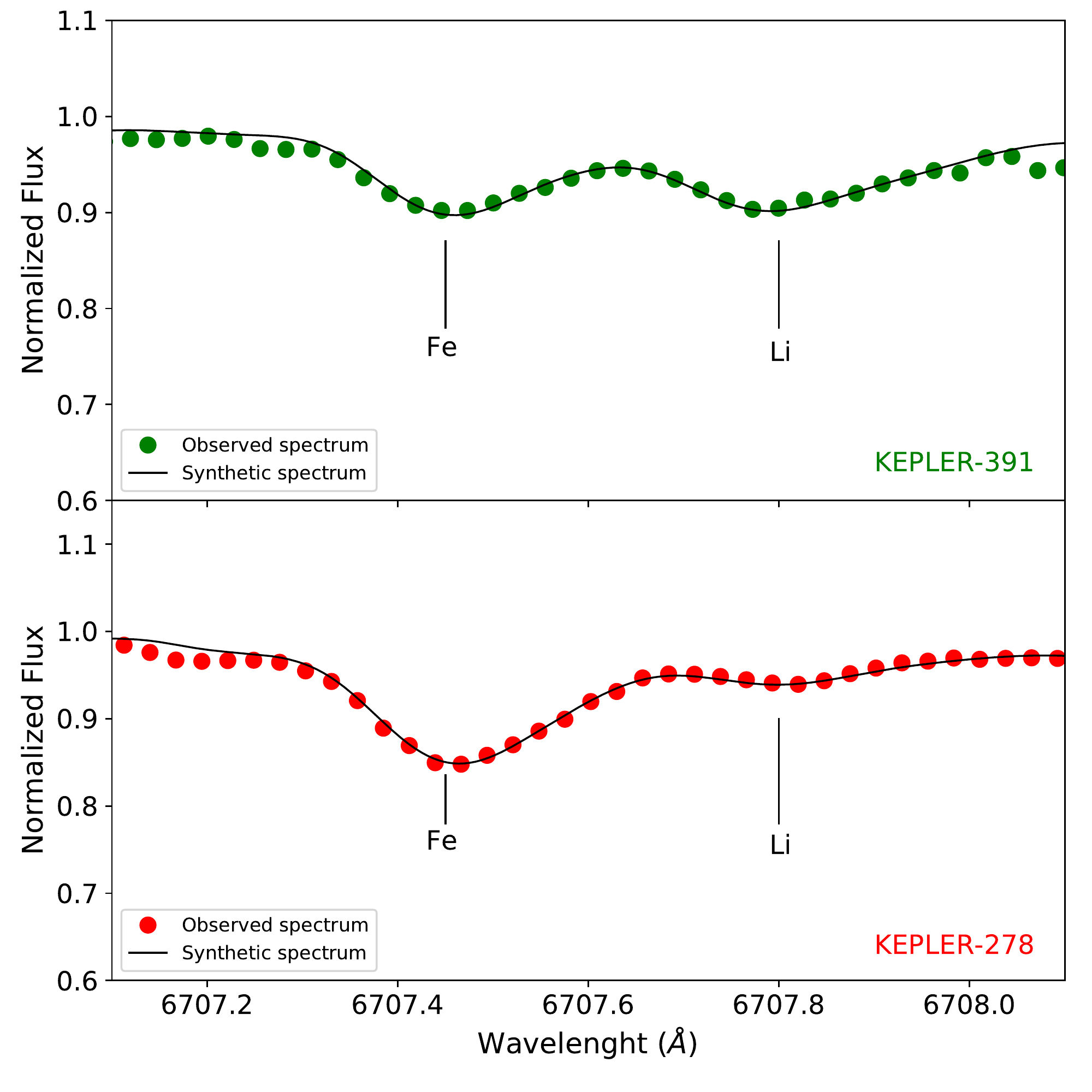}

   \caption{Best fit obtained between the synthetic and the observed
GRACES spectra of Kepler-278 and Kepler-391 around the \ion{Li}{I} $\lambda$6708 feature.}
              \label{sintesis}%
    \end{figure}

\subsection{Detailed chemical abundances}
\subsubsection{Analysis}
From the high-quality GRACES spectra obtained for Kepler-278 and Kepler-391, in addition to iron, we derived chemical abundances of five light (Li, C, N, Na, Al), eight iron-peak (Sc, V, Cr, Mn, Co, Ni, Cu, Zn), six alpha (O, Mg, Si, S, Ca, Ti), and five heavy elements (Sr, Y, Zr, Ba, Ce). Abundances were computed using both EWs and spectrum synthesis analysis in combination with the LTE Kurucz model atmospheres previously calculated.
 
The abundances of O, Na, Mg, Al, Si, Ca, Sc, Ti, V, Cr, Mn, Co, Ni, Cu, Zn, Sr, Y, Zr, Ba, and Ce were derived from the curve-of-growth approach employing the \texttt{MOOG} code (abfind driver) through the \texttt{$q^{2}$} code. The EWs were measured carefully by hand using Gaussian profile fits with the task \texttt{splot} in IRAF. The line-list and atomic parameters for most of the elements were compiled from \citet{Neves2009}, \citet{Adibekyan2015}, \citet{Takeda2005}, \citet{Chavero2010}, \citet{Ramirez2014}, \citet{Saffe2015}, and \citet{Delgado2017}. For Sc, Ti, and Cr our final list includes both neutral and singly-ionized lines, whilst for Y, Ba, and Ce only singly-ionized lines are available. For the rest of the elements, only neutral lines were used. Hyperfine splitting (HFS) was taken into account for V, Mn, Co, Cu, and Ba using the \textit{blends} driver and the HFS constants of \citet{Kurucz1995}.

We applied non-local thermodynamic equilibrium (NLTE) corrections to the oxygen abundances, obtained from the $\lambda$7771-5 {\AA} infrared triplet, using the grid of \citet{Ramirez2007}. We also obtained NLTE abundances for Na, using the corrections interpolated from the tables of \citet{Lind2009} through the INSPECT database\footnote{\url{www.inspect-stars.com}}.

The abundances of C, N, Li, S, and also the carbon $^{12}$C/$^{13}$C isotopic ratio were derived by fitting synthetic spectra to the data using the \textit{synth} driver of the MOOG code. Carbon abundances were derived from the C$_{2}$ Swan band at $\lambda$5086 and $\lambda$5135 using a line list from the VALD line database \citep{Kupka1999}. Abundances of N and the $^{12}$C/$^{13}$C isotopic ratio were computed by fitting $^{12}$CN and $^{13}$C features in the range 8002--8004 {\AA} using the line list of \citet{Carlberg2012}. We employed the following molecular dissociation energies: D$_{0}$ = 6.21 \citep[C2;][]{Huber1979}, and D$_{0}$ = 7.65 \citep[CN;][]{Bauschlicher1988}. For lithium, we analyzed the \ion{Li}{I} feature at $\lambda$6707.8 {\AA} adopting the line list of \citet{Carlberg2012}, which includes blends from atomic and
molecular (CN) lines. We obtained NLTE lithium abundances using the corrections by \citet{Lind2009} via the INSPECT database. In the case of sulphur, we fitted the \ion{S}{I} features at $\lambda$6757.15 {\AA}, and $\lambda$8694.62 {\AA} using the atomic data of \citet{Takeda2016}. Fig. \ref{sintesis} shows the best fits to the Li features for both stars. 

The final computed abundances, relative to the solar values from \citet{Asplund2009}, together with the errors are listed in Table \ref{abundances}. Here, $\sigma_{lines}$ corresponds to the line-by-line abundance dispersion for each element whilst $\sigma_{pars}$ indicates the error introduced by propagating our formal uncertainties in the atmospheric parameters \citep[see, e.g.,][]{Ramirez2011a, Morel2014}. As usual, the formal total error given in the last column of Table \ref{abundances} is obtained by adding quadratically all the error contributions. 

\begin{figure}[t!]
   \centering
   \includegraphics[width=0.50 \textwidth]{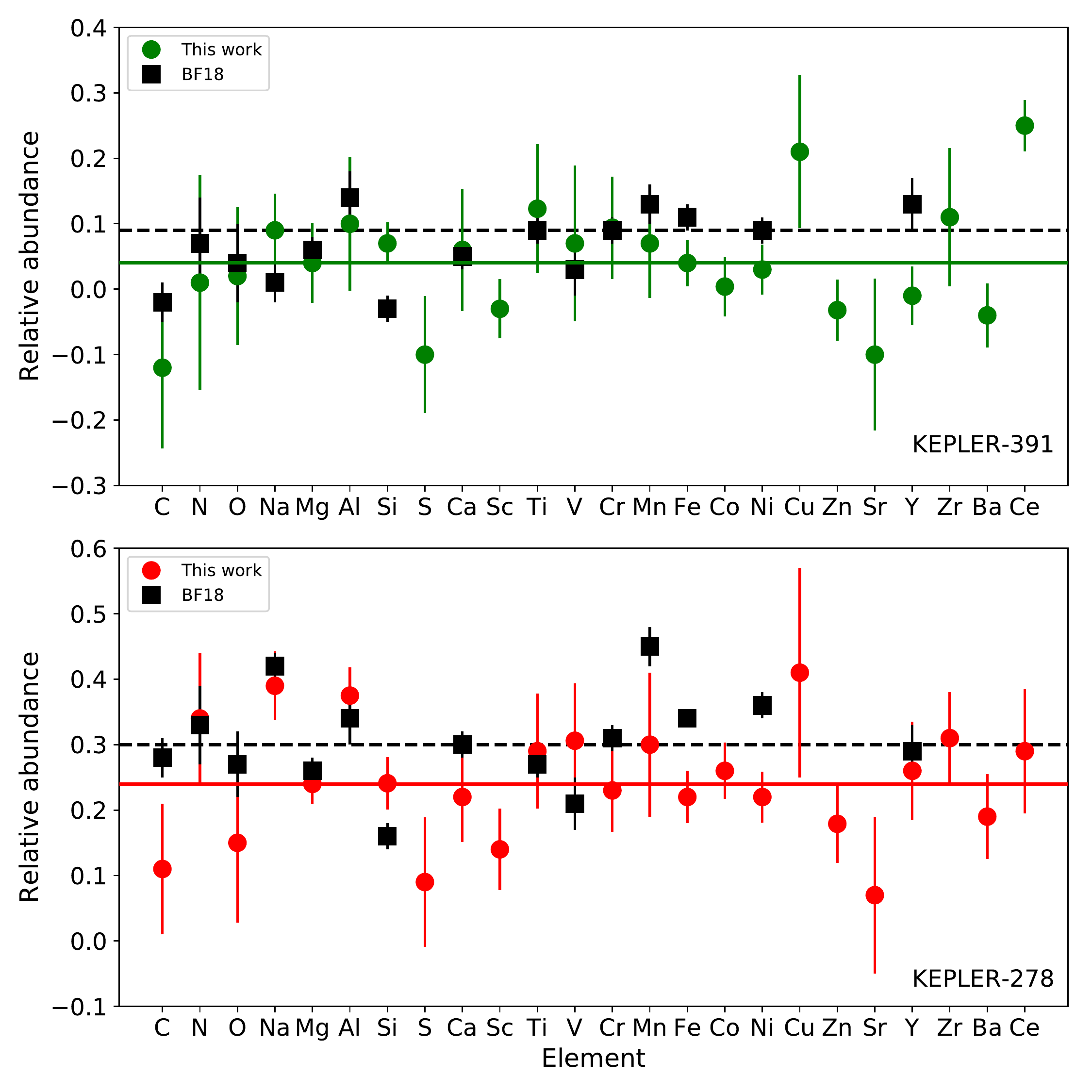}

   \caption{Elemental abundances derived in this work (filled circles) in comparison to those reported by BF18 (filled squares) for Kepler-278 (\textit{bottom}) and Kepler-391 (\textit{top}). Continuous and dashed lines indicates the mean metallicities obtained in this work and in BF18, respectively.}
              \label{abun.literature}%
    \end{figure}

\subsubsection{Comparison with literature}

In addition to the atmospheric parameters, BF18 also derived chemical abundances for 15 elements (C, N, O, Na, Mg, Al, Si, Ca, Ti, V, Cr, Mn, Fe, Ni, and Y) via spectral synthesis of low S/N HIRES spectra through the \texttt{SME} code \citep{Valenti1996}. In Fig. \ref{abun.literature} we show the abundances of Kepler-278 and Kepler-391 derived in this work in comparison with those obtained by BF18. In general, there is a good agreement between both measurements for most elements in common. Our mean metallicities (continuous lines) are slightly lower than the ones of BF18 (dashed lines) by $-$0.05 $\pm$ 0.08 dex for Kepler-278 and $-$0.02 $\pm$ 0.07 dex for Kepler-391. For some elements such as C, Mn, Fe, and Ni, however, the abundances obtained by BF18 are systematically larger than our values, and the differences are even larger for Kepler-278 for which the agreement is only within 2$\sigma$. On the other hand, the abundances of Si determined by BF18 are lower than our estimations. Also, for Kepler-391, their abundance of Y is $\sim$0.15 dex ($>$ $1\sigma$) larger than our estimation. The systematic differences are likely related to the use of different techniques to compute abundances, atmospheric parameters (see Section \ref{previous.atmospheric}), models of stellar atmospheres, and line-lists \citep{Hinkel2016, Jofre2017}. Moreover, the large difference in the S/N between their HIRES and our GRACES spectra ($\Delta S/N \gtrsim$ 300 for Kepler-278 and $\Delta S/N \gtrsim$ 240 for Kepler-391) likely produces a non-negligible effect in the computed chemical abundances. An additional reason for the discrepancies could be the existence of a possible systematic flaw in the abundance analysis that, according to BF18, might be affecting their results for the evolved stars.    

Another feature visible in Fig. \ref{abun.literature} is that, similar to the uncertainties in the atmospheric parameters, error bars in the chemical abundances provided by BF18 are significantly smaller (up to $\sim$ 80 \%) than those computed in this work. The origin of such differences is that our formal quoted errors, as explained in the previous section, include the uncertainties in the atmospheric parameters (considering both internal and external errors) and the dispersion in the line-by-line measurements, whilst those reported by BF18 correspond to statistical uncertainties from fitting models to observations only, and likely underestimate the true uncertainties.  

Recently, \citet{Berger2018} measured Li abundances from EWs for 1305 Kepler targets, also based on HIRES spectra taken by the CKS. They obtain A(Li)$_{LTE}$ = 0.52 $\pm$ 0.16 dex and A(Li)$_{LTE}$ = 0.96 $\pm$ 0.27 dex for Kepler-278 and Kepler-391, respectively. These values are 0.17 dex and 0.18 dex smaller than our estimations for Kepler-278 and Kepler-391, respectively. As before, the discrepancies, are probably related to the use of different techniques employed (EWs vs. spectral synthesis), the difference in the adopted T$_{\mathrm{eff}}$ (which they compiled from PE17, see Section \ref{previous.atmospheric}), and the use of different data quality.

\begin{figure}[t!]
   \centering
   \includegraphics[width=.51\textwidth]{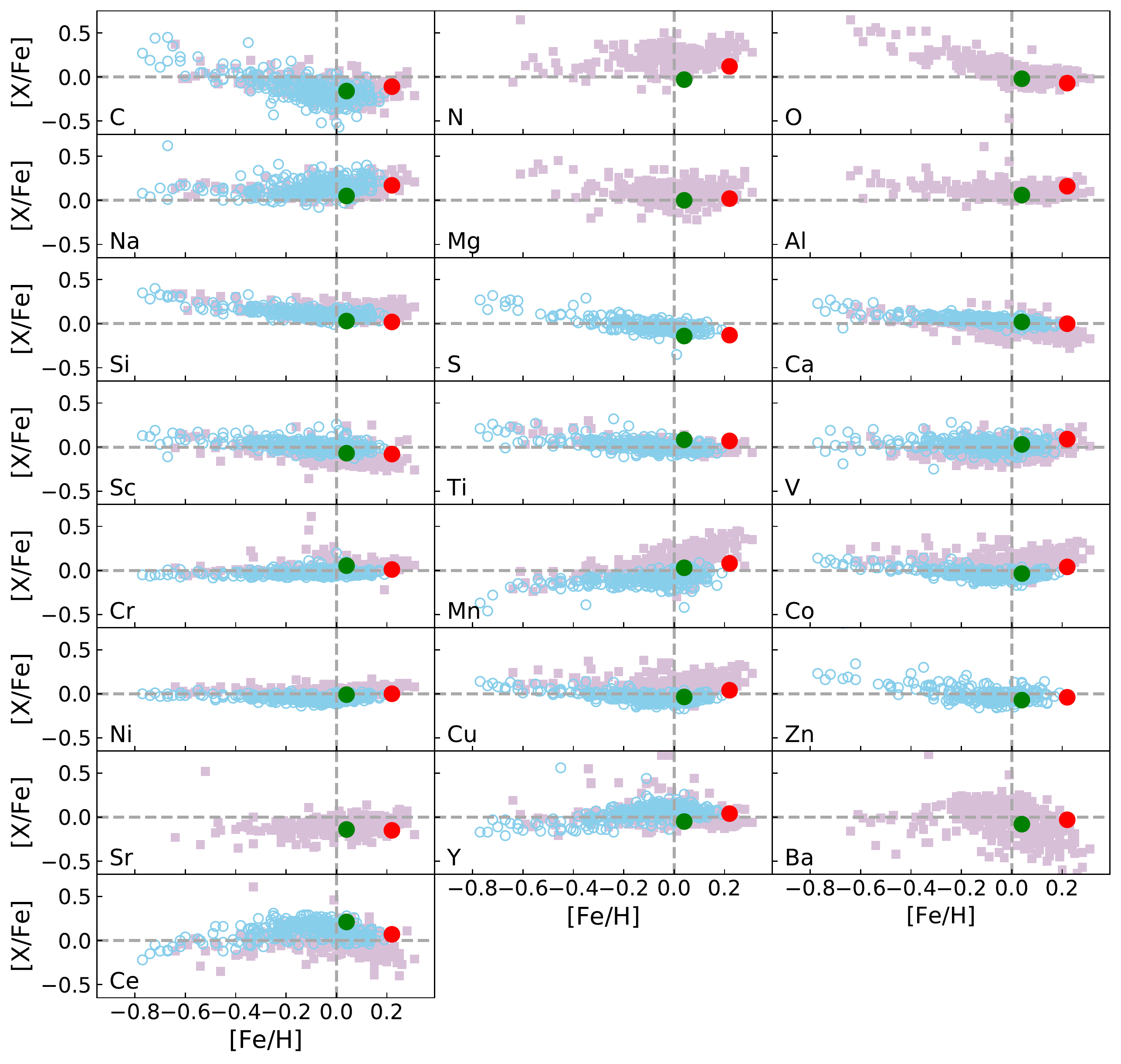}
   \caption{Elemental abundance ratios of Kepler-278 (red circle) and Kepler-391 (green circle) compared to the Galactic chemical evolution trends by \citet[][squares]{Luck2007} and \citet[][empty circles]{Takeda2008, Takeda2016}. Dashed lines indicate the solar values.
}
              \label{galactic}%
    \end{figure}

\begin{figure*}[th!]
   \centering
   \includegraphics[width=.495\textwidth]{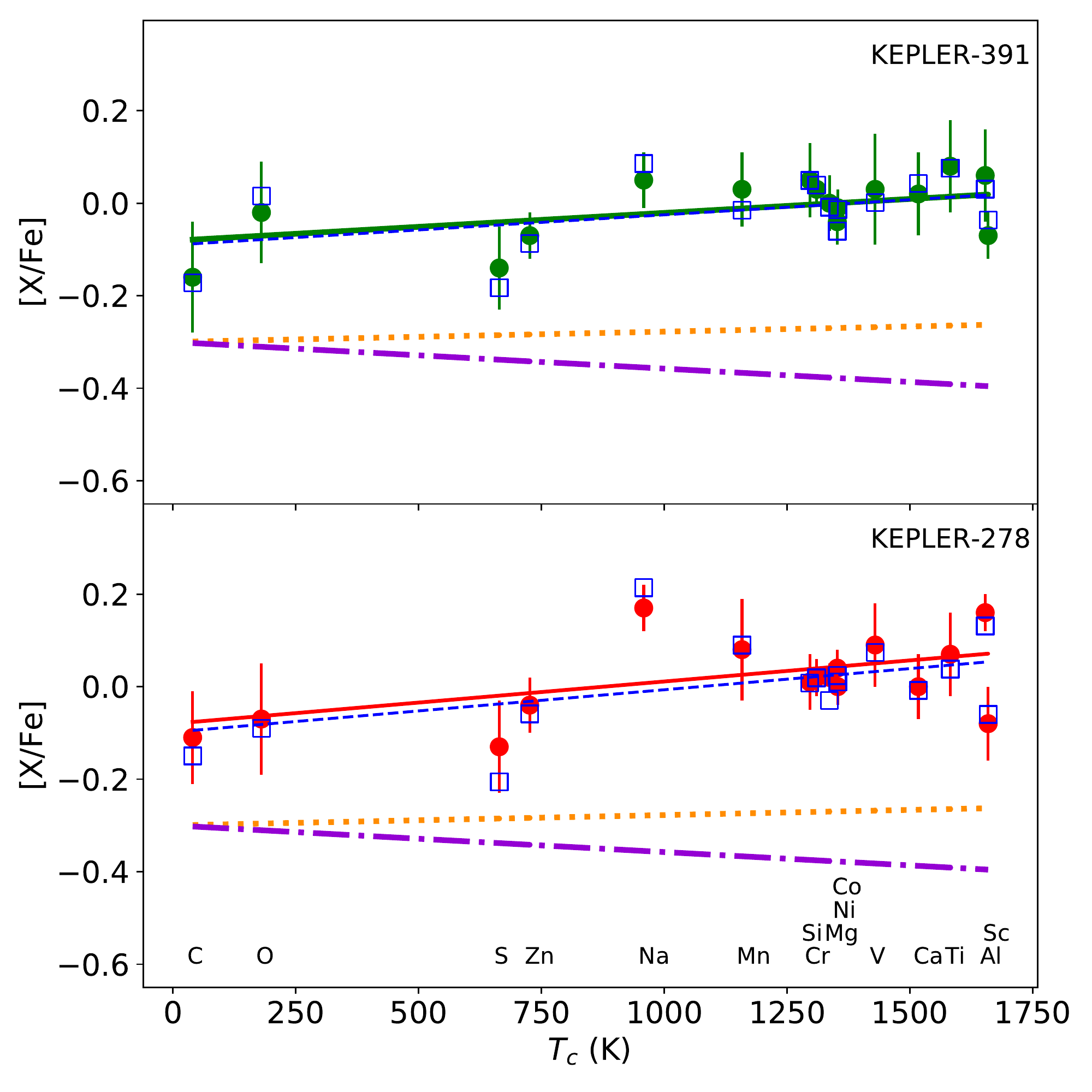}
   \includegraphics[width=.495\textwidth]{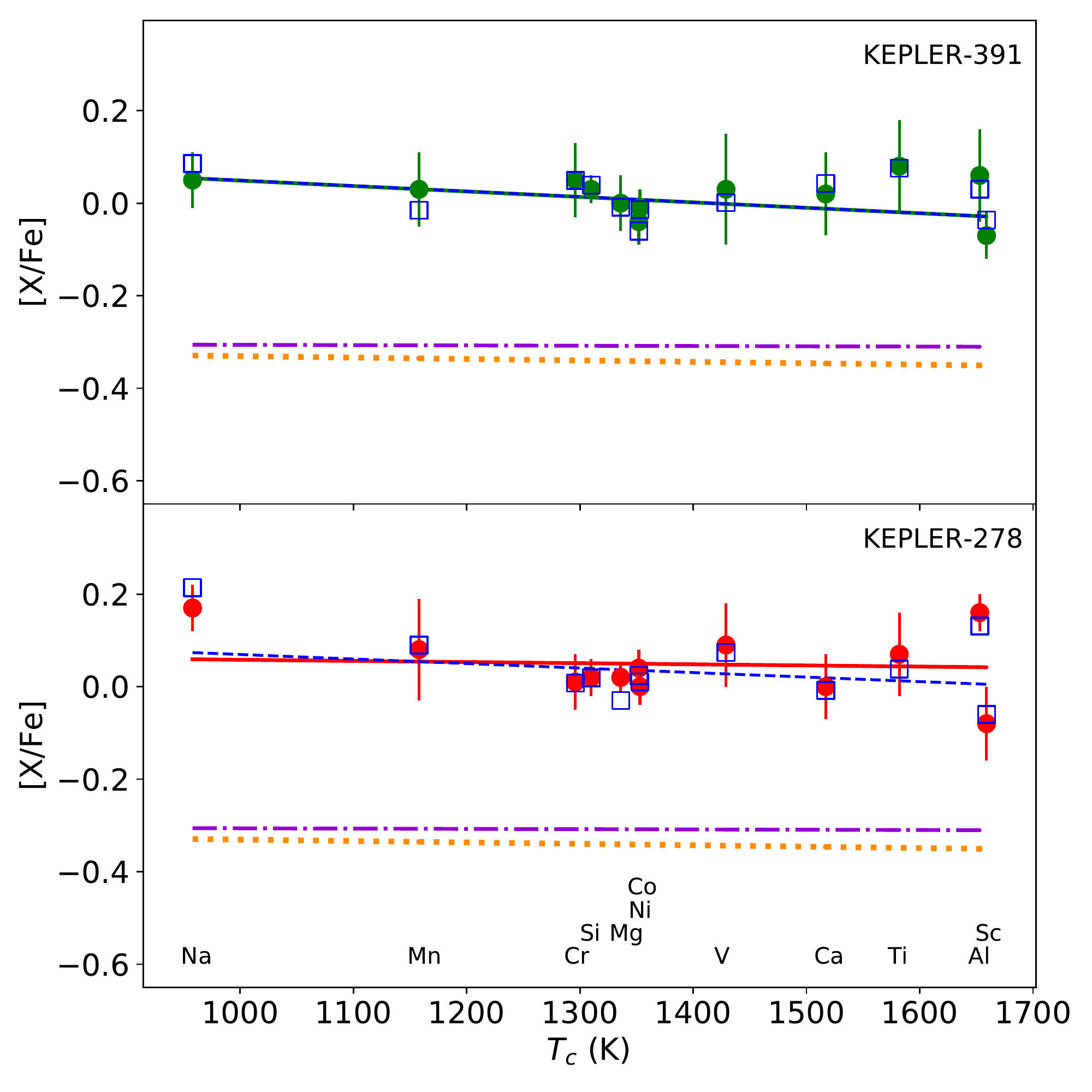}

   \caption{\textit{Left}: Abundances of volatile and refractory elements for Kepler-278 (bottom) and Kepler-391 (top) as a function of dust condensation temperature; filled circles and blue squares represent [X/Fe] without and with Galactic chemical evolution (GCE) corrections, respectively. Solid and dashed lines show the weighted linear fits to the abundance values  without and with GCE corrections, respectively. Orange dotted and purple dot-dashed lines show the mean trend found by M16 for subgiants and giants with planets, respectively. \textit{Right}: Same as left panel, but for refractory elements only.}
              \label{tcond}%
    \end{figure*}

\subsubsection{Location on the [X/Fe]--[Fe/H] plane and kinematic membership} 
\label{sec.kinematic}

To check unusual chemical compositions, we investigated abundance trends with metallicity for all the 
elements. In Fig. \ref{galactic}, we show the [X/Fe] abundances as a function of the
[Fe/H] of Kepler-278 and Kepler-391 compared to Galactic chemical evolution trends of solar neighborhood evolved stars. The comparison data were taken from \citet[][purple squares]{Luck2007} for all elements except Zn, and from \citet[][light blue circles]{Takeda2008, Takeda2016} for all elements with the exception of N, Mg, O, and Ba. In both samples, abundances have been computed from high-resolution spectra using similar techniques to ours and are dominated by stars of the thin disk. In general, the chemical composition of both Kepler stars does not appear to be anomalous but rather consistent with the abundances of the thin-disk nearby evolved stars with similar metallicities. In particular, no important $\alpha$-element enhancement is observed neither for Kepler-278 nor Kepler-391. Averaging the abundances of O, Mg, Si, Ca, and Ti we obtain [$\alpha$/Fe]= 0.04 dex for both stars. The Ca abundance for Kepler-278 and Ti abundances for both stars appear slightly overabundant compared to the mean trend of both reference samples. However, they are still consistent with the abundances of the comparison stars considering the 1$\sigma$ scatter. Also, we do not observe any evident signs of anomalies in the abundance of Fe-peak or \textit{s}-process elements.  

Using proper motions and parallaxes\footnote{We corrected the Gaia parallaxes for the 82 $\mu$arcsec suggested by \citet{Stassun2018} and derived the distances from these values (see Table \ref{tableparameters}).} from \textit{Gaia} DR2 \citep{Gaia2018}, we derived Galactic space-velocity UVW components for both stars following the procedure detailed in \citet{Jofre2015a}. We find (U, V, W) = (24.56 $\pm$ 0.60, $-$32.24 $\pm$ 0.19, 1.26 $\pm$ 0.15) km s$^{-1}$ for Kepler-278 and (U, V, W) = (36.33 $\pm$ 2.33, 40.43 $\pm$ 0.71, $-$14.67 $\pm$ 1.31) km s$^{-1}$ for Kepler-391. Then, from these space-velocity components and using the membership formulation by \citet{Reddy2006}, we found that the probability of belonging to the thin disk population is $\sim$98\% for both stars, which is consistent  with the results from the chemical analysis. 


\subsubsection{Lithium abundance}

The lithium feature at $\lambda$6707.8 is detectable in the spectra of Kepler-278 and it is even stronger on the spectra of Kepler-391 as can be noticed in Fig. \ref{sintesis}. For Kepler-278, we determined A(Li)$_{NLTE}$ = 0.87 $\pm$ 0.10 dex and A(Li)$_{NLTE}$ = 1.29 $\pm$ 0.09 dex for Kepler-391. The Li abundance of Kepler-391 is just below the standard limit, A(Li) $\approx$ 1.5 dex, from which evolved stars are considered to have an anomalous abundance of Li and termed as Li-rich stars \citep[e.g.,][]{Kumar2011}. According to their T$_{\mathrm{eff}}$ and $\log g$ values, Kepler-391, is near the base of the RGB (see Fig. \ref{hr-todos}) and then only recently started the first dredge-up (FDU). Therefore, the relatively high Li abundance of Kepler-391 is most likely a remnant from the main-sequence phase as in other similar cases \citep[e.g.,][]{Adamow2014, Adamow2015}. In agreement with this scenario, we do not find evidence of increased stellar rotation or other chemical anomalies (see next section) that could indicate planet engulfment events \citep[][]{Siess1999, Carlberg2012, Adamow2014, Jofre2015b}. 

Another possibility to explain high lithium content in evolved stars is a fresh lithium production phase via the Cameron-Fowler mechanism \citep{Cameron1971}. This scenario is generally associated to stars near the luminosity bump \citep[LB, see, e.g., Fig. 2 of][]{Jofre2015b} on the RGB, where most of low-mass lithium-enhanced evolved stars tend to cluster. However, from its position on the HR-diagram (see Fig. \ref{hr-todos}), Kepler-391 it is relatively far away from the LB position.

 \renewcommand{\arraystretch}{1.2} 
 
\begin{table*} [t!]
 \footnotesize
 
        \caption[]{Planetary parameters of the Kepler-278 and Kepler-391 systems.}
         \label{planetary-parameters-table}
     \centering
         \begin{tabular}{  m{14em}  m{3cm} m{3.5cm}  m{3cm}  m{3cm} } 
            \hline\hline

Parameter	&	Kepler-278b								&	Kepler-278c								&	Kepler-391b								&	Kepler-391c								\\
\hline																																					
\multicolumn{5}{c}{Model parameters} \\																																					
\hline																																					
\textit{Kepler} long-cadence additional white noise level	&	\multicolumn{2}{c}{2.495	$\pm$		0.040}										&	\multicolumn{2}{c}{1.3993	$\pm$	0.0095					}										\\
\textit{Kepler} short-cadence additional white noise level	&	\multicolumn{2}{c}{1.1317	$\pm$		0.0032				}										&			\multicolumn{2}{c}{N/A}									\\
Stellar mean density, $\rho_{\star}$ [g cm$^{-3}$] \tablefootmark{a}	&	\multicolumn{2}{c}{0.07237	$\pm$		0.00096				}										&	\multicolumn{2}{c}{0.0822	$\substack{	+	0.0099	\\	-	0.012	}$}										\\
Linear limb darkening, $u_{\mathrm{a}}$	&	\multicolumn{2}{c}{0.74	$\substack{	+	0.13	\\	-	0.16	}$}										&	\multicolumn{2}{c}{0.30	$\substack{	+	0.32	\\	-	0.21	}$}										\\
Quadratic limb darkening, $u_{\mathrm{b}}$	&	\multicolumn{2}{c}{$-$0.27	$\substack{	+	0.22	\\	-	0.12	}$}										&	\multicolumn{2}{c}{0.31	$\substack{	+	0.31	\\	-	0.40	}$}										\\
Scaled semi-major axis, $a/R_{\mathrm{\star}}$ 	&	15.152	$\pm$		0.067					&	21.541	$\pm$		0.096					&	6.20	$\substack{	+	0.24	\\	-	0.31	}$	&	12.23	$\substack{	+	0.47	\\	-	0.62	}$	\\
Eccentricity, $e$\tablefootmark{b}	&	0.696	$\substack{	+	0.017	\\	-	0.026	}$	&	0.616	$\substack{	+	0.015	\\	-	0.023	}$	&	$<$ 0.46\tablefootmark{c}								&	$<$ 0.27								\\
Inclination, i [$^{\circ}$]\tablefootmark{d}	&	85.11	$\substack{	+	0.72	\\	-	0.52	}$	&	88.58	$\substack{	+	1.2	\\	-	0.51	}$	&	85.5	$\substack{	+	2.9	\\	-	1.7	}$	&	87.27	$\substack{	+	0.72	\\	-	0.48	}$	\\
Argument of pericentre, $\omega$ [$^{\circ}$]	&	21.1	$\pm$		7.7					&	20.80	$\pm$		7.9					&	103	$\substack{	+	53	\\	-	46	}$	&	126	$\substack{	+	100	\\	-	63	}$	\\
Difference of the longitude of the ascending node, $\Omega_{1}-\Omega_{2}$ [$^{\circ}$] 	&	\multicolumn{2}{c}{10	$\substack{	+	10	\\	-	32	}$}										&	\multicolumn{2}{c}{0	$\pm$	74					}										\\
$\sqrt{e} \cos \omega$	&	0.778	$\substack{	+	0.039	\\	-	0.056	}$	&	0.735	$\substack{	+	0.037	\\	-	0.055	}$	&	$-$0.06	$\pm$		0.32					&	$-$0.07	$\substack{	+	0.25	\\	-	0.22	}$	\\
$\sqrt{e} \sin \omega$	&	0.295	$\pm$		0.098					&	0.274	$\pm$		0.093					&	0.35	$\substack{	+	0.14	\\	-	0.21	}$	&	0.16	$\pm$		0.22					\\
Mean anomaly, $M_{\mathrm{0}}$	&	11.2	$\substack{	+	1.4	\\	-	1.1	}$	&	340.7	$\pm$ 1.7		&	321	$\substack{	+	28	\\	-	300	}$	&	227	$\pm$		63					\\
Impact parameter, b	&	0.544	$\substack{	+	0.066	\\	-	0.085	}$	&	0.27	$\substack{	+	0.11	\\	-	0.23	}$	&	0.38	$\substack{	+	0.23	\\	-	0.26	}$	&	0.561	$\substack{	+	0.087	\\	-	0.019	}$	\\
Epoch of transit, $T_{\mathrm{0}}$ [${\rm BJD}_{\rm TDB}$]	&	2455695.3727	$\substack{	+	0.0036	\\	-	0.0095	}$	&	2455700.4118	$\pm$		0.0052					&	2455690.3959	$\substack{	+	0.013	\\	-	0.0089	}$	&	2455696.9123	$\substack{	+	0.0085	\\	-	0.0074	}$	\\
Period, P [days]\tablefootmark{e}	&	30.1556	$\substack{	+	0.0041	\\	-	0.0020	}$	&	51.111	$\substack{	+	0.023	\\	-	0.011	}$	&	7.4175	$\substack{	+	0.0027	\\	-	0.0037	}$	&	20.501	$\substack{	+	0.032	\\	-	0.010	}$	\\
Radius ratio, $R_{\mathrm{p}}$ /$R_{\mathrm{\star}}$ 	&	0.01266	$\substack{	+	0.00023	\\	-	0.00027	}$	&	0.01046	$\pm$ 0.00025	&	0.00795	$\substack{	+	0.00034	\\	-	0.00027	}$	&	0.00915	$\pm$		0.00034					\\
Mass ratio, $M_{\mathrm{p}}$ /$M_{\mathrm{\star}}$ 	&	0.000142	$\substack{	+	0.000095	\\	-	0.000033	}$	&	0.000087	$\substack{	+	0.000026	\\	-	0.000050	}$	&	$<$ 0.0026								&	$<$ 0.0036								\\
Planet mean density, $\rho_{\mathrm{p}}$ [g cm$^{-3}$]	&	5.0	$\substack{	+	3.2	\\	-	1.1	}$	&	5.5	$\substack{	+	1.5	\\	-	3.2	}$	&	$<$ 380								&	$<$ 400								\\
 \hline																																					
\multicolumn{5}{c}{Derived properties} \\																																					
 \hline																																					
Semi-major axis, $a$ [AU]	&	0.202	$\pm$		0.005					&	0.287	$\pm$		0.007					&	0.083	$\substack{	+	0.012	\\	-	0.013	}$	&	0.164	$\substack{	+	0.026	\\	-	0.022	}$	\\
Radius, $R_{\mathrm{p}}$ [$R_{\mathrm{\oplus}}$]	&	3.955	$\pm$		0.080					&	3.268	$\pm$		0.067					&	2.50	$\pm$ 0.29	&	2.88	$\pm$		0.33					\\
Mass, $M_{\mathrm{p}}$ [$M_{\mathrm{\oplus}}$]	&	56	$\substack{	+	37	\\	-	13	}$	&	34.9	$\substack{	+	9.9	\\	-	21	}$	&	$<$ 1100								&	$<$ 450								\\
Incident Flux, $S_{\mathrm{inc}}$ [$S_{\mathrm{\oplus}}$]	&	110	$\pm$		14					&	54.3	$\pm$		7.2					&	690	$\substack{	+	390	\\	-	410	}$	&	179	$\substack{	+	96	\\	-	110	}$	\\
Equilibrium temperature, $T_{\mathrm{eq}}$ [K] \tablefootmark{f}	&	586	$\pm$		19					&	492	$\pm$		16					&	930	$\substack{	+	37	\\	-	140	}$	&	662	$\substack{	+	89	\\	-	97	}$	\\
Semiamplitude, $K_{\mathrm{\star}}$ [m $s^{-1}$]	&	14.0	$\substack{	+	21	\\	-	4.7	}$	&	6.7	$\substack{	+	2.4	\\	-	5.1	}$	&	$<$ 340								&	$<$ 93								\\
          \hline																																					
         \end{tabular} 																																					
 \tablefoot{				
   \tablefoottext{a}{For Kepler-278 we used as prior the stellar density  derived from asteroseismology by H13 and the one determined from stellar models in Section \ref{sec.stellar.parameters} for Kepler-391.} \\
   \tablefoottext{b}{The orbital parameters correspond to Jacobi elements computed at the reference time t$_{\rm ref}$=2455695.37632 BJD$_{\rm TBD}$ for Kepler-278 and at t$_{\rm ref}$=2455696.9123 BJD$_{\rm TBD}$ for Kepler-391.}  \\
   \tablefoottext{c}{All upper limits correspond to the 95\% confidence interval.}  \\
   \tablefoottext{d}{The prior for the inclination of planets b is uniform between [0, 180]$^{\circ}$, and [0, 90]$^{\circ}$ for planets c. The value for the inclination in the Table for planets b is reflected with respect to i=90$^{\circ}$, the supplementary angle is equally probable.}  \\
   \tablefoottext{e}{Defined as $P \equiv \sqrt{\frac{3\pi}{\mathcal G \rho_{\star}}\left(\frac{a}{R_{\star}} \right)^3}$}\\
   \tablefoottext{f}{Calculated with equation 4 from \citet{Johnson2017} assuming a Bond albedo $\alpha$ of 0.3.} \\

 }
\end{table*}

\subsubsection{[X/Fe] versus T$_{c}$}
Beyond the overall metallicity enhancement of main-sequence stars with planets \citep[e.g.,][]{Ghezzi2010a}, it has been suggested that trends in elemental abundances with condensation temperature (T$_{c}$) are possible signatures of planet formation and evolution processes. A deficit of refractory elements (T$_{c}$ $\gtrsim$ 900 K; $\alpha$-group and iron-peak elements) relative to volatiles (T$_{c}$ $<$ 900 K, e.g., C, O, S) on the stellar atmosphere might indicate this material was sequestered in the formation of planetesimals, rocky planets, or the cores of giant planets \citep{Melendez2009, Ramirez2010}. On the other hand, an overabundance of refractory elements in comparison to volatiles might indicate the accretion of hydrogen-depleted rocky material onto the star \citep{Gonzalez1997, Murray2002, Melendez2017, Saffe2017}.

Recently, \citet[][M16 hereafter]{Maldonado2016} analyzed the abundances of different groups of stars (dwarfs, subgiants, and giants), with and without planets, as a function of T$_{c}$ in order to search for differences that could be linked to the planet formation process. Considering all elements (volatiles and refractories), these authors found no significant difference between the slopes of evolved stars with and without planets. When restricting the analysis to refractory elements, however, they found differences  between stars with and without known planets for the samples of main-sequence and subgiant stars but no for the sample of giants. Given that the sample of main-sequence and subgiant stars contain less massive and older stars than the sample of giants, M16 consider that Galactic radial mixing offers a more suitable scenario for explaining the observed trends in main-sequence and subgiant stars rather than planet formation.

In the left panel of Fig. \ref{tcond} we show our [X/Fe] values as function of T$_{c}$ for Kepler-278 and Kepler-391, considering both volatile and refractory elements in common with those analyzed by M16 (16 elements). For T$_{c}$, we have used the 50\% values from \citet{Lodders2003}. No significant trend is observed for Kepler-278 nor for Kepler-391. A weighted linear fit reveals small positive slopes of (9.12 $\pm$ 6.03) $\times$ $10^{-5}$ dex K$^{-1}$ for Kepler-278 and (6.02 $\pm$ 4.14) $\times$ $10^{-5}$ dex K$^{-1}$ with standard fit deviations of 0.07 and 0.05 dex for Kepler-278 and Kepler-391, respectively. In both cases the slopes are consistent, within the errors, with the average value presented by M16 for the subgiants with planets sample: (2.24 $\pm$ 1.17) $\times$ $10^{-5}$ dex K$^{-1}$, rather than the one obtained for giants with planets sample: ($-$5.76 $\pm$ 1.58) $\times$ $10^{-5}$ dex K$^{-1}$. 

We also applied the Galactic chemical evolution (GCE) corrections to our [X/Fe] values based on the studies of \citet{Gonzalez2013} following the same procedure as \citet{Saffe2015} and \citet{Saffe2017}. These corrections are very small for both stars and therefore there are no significant changes in the slopes values. In the right panel of Fig. \ref{tcond} we show the [X/Fe] vs. T$_{c}$ trends when only the refractory elements common to the study of M16 are considered (12 elements). As before, no evident trend in the abundances as a function of T$_{c}$ is present for any of the stars. Weighted linear fits to the data result in negative slopes for both Kepler-278 ($-$2.45 $\pm$ 8.91 $\times$ $10^{-5}$ dex K$^{-1}$) and Kepler-391 ($-$11.8 $\pm$ 7.08 $\times$ $10^{-5}$ dex K$^{-1}$) in agreement with the negative average slopes found by M16 for subgiants with planets ($-$3.06 $\pm$ 2.32 $\times$ $10^{-5}$ dex K$^{-1}$) and giants with planets ($-$0.62 $\pm$ 2.35 $\times$ $10^{-5}$ dex K$^{-1}$). Again, similar negative slopes are observed when GCE corrections are considered.


    \begin{figure*}[th!]
   \centering
      \includegraphics[width=.497\textwidth]{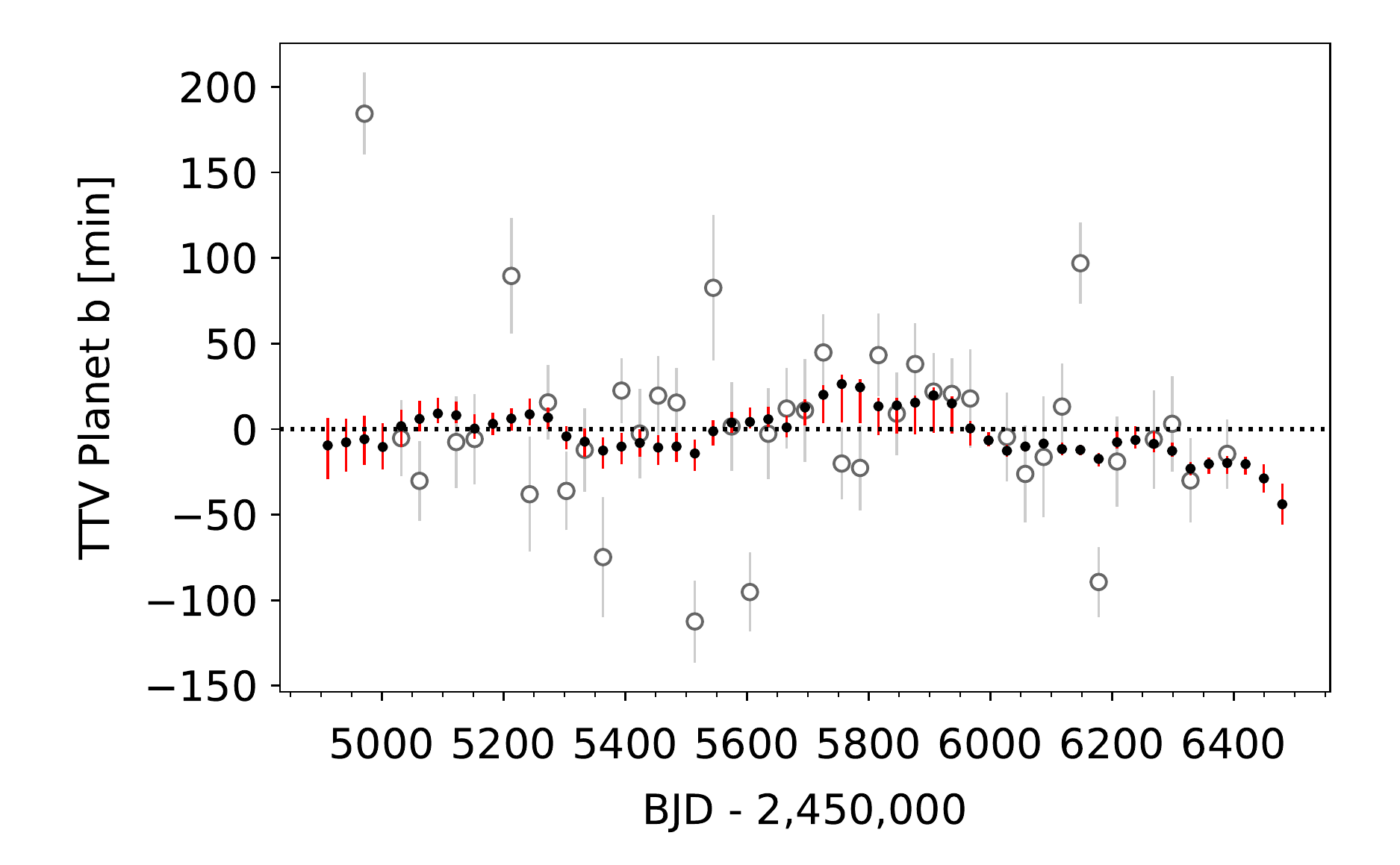}
      \includegraphics[width=.497\textwidth]{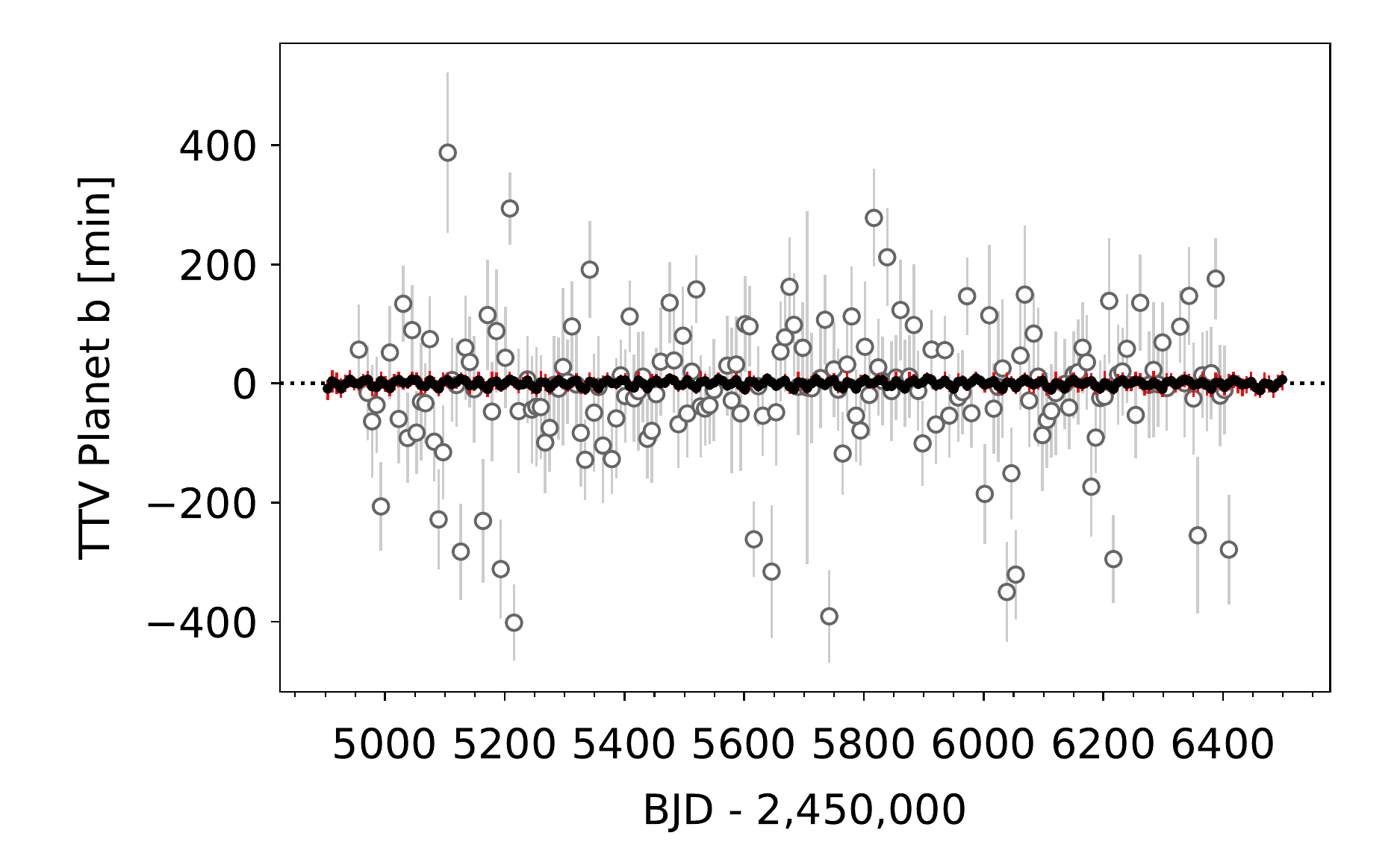}

      \includegraphics[width=.497\textwidth]{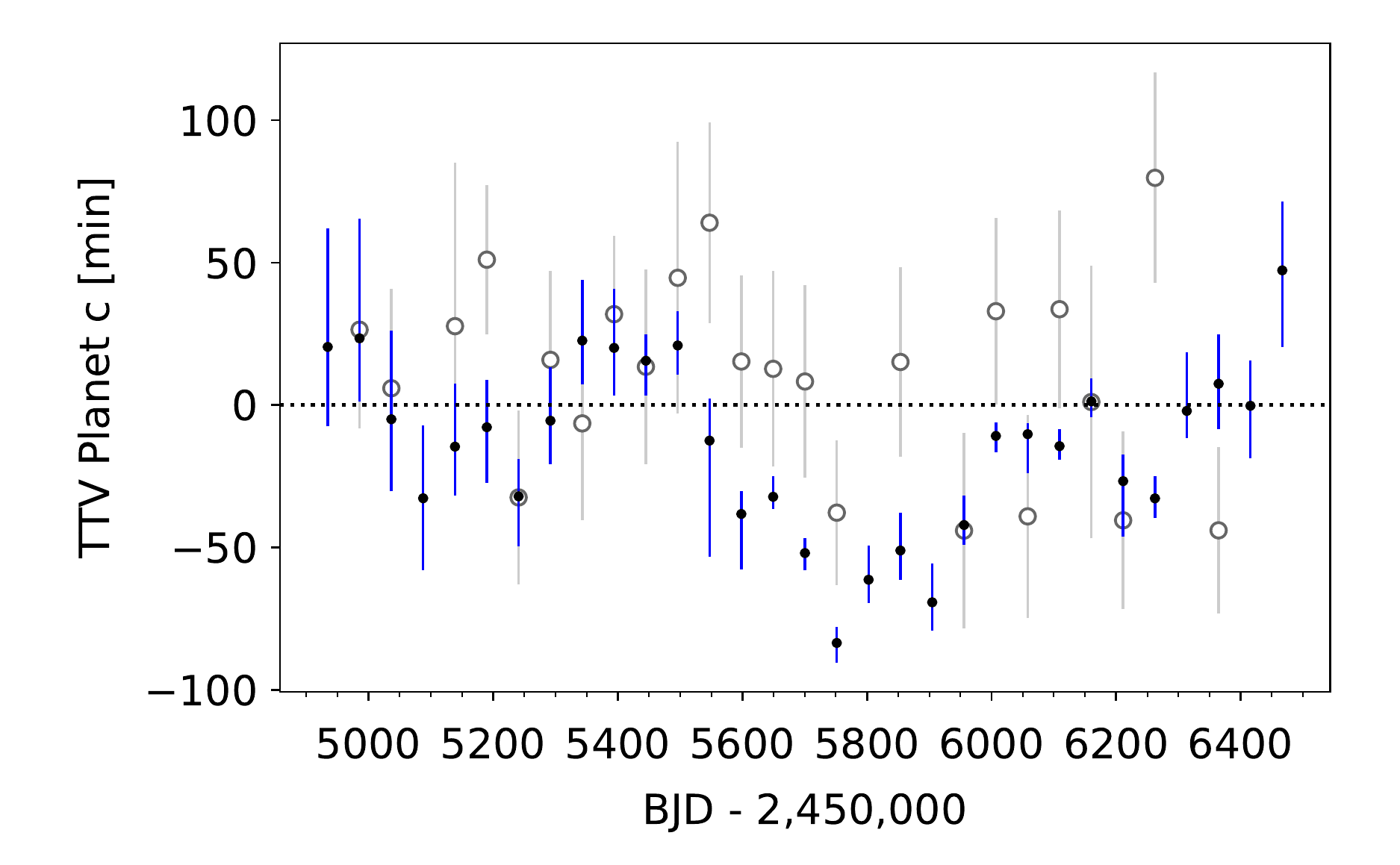}
      \includegraphics[width=.497\textwidth]{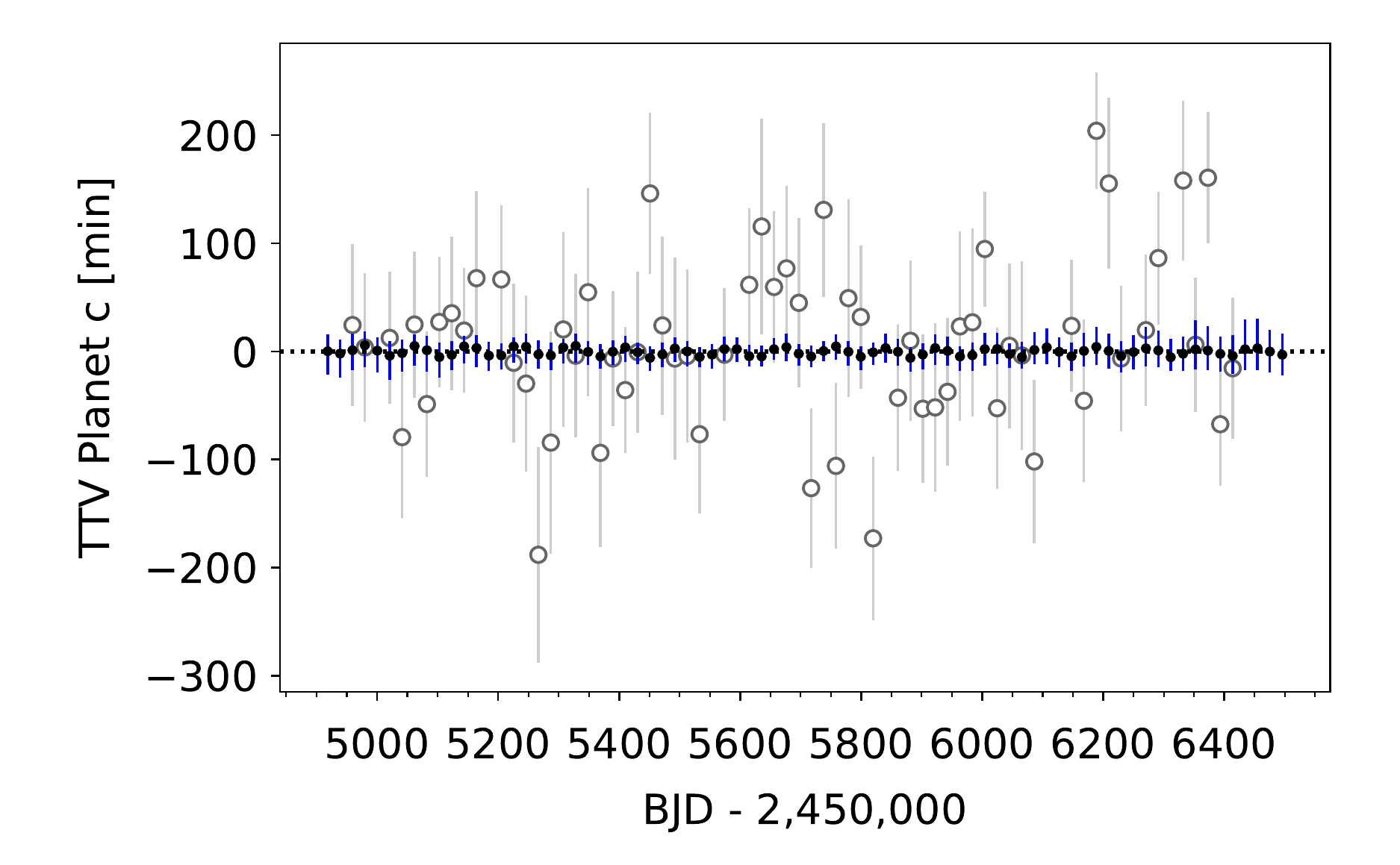}

   \caption{Posterior TTV of Kepler-278 b (top left panel, red), Kepler-278c (bottom left panel, blue), Kepler-391 b (top right panel, red), and Kepler-391 c (bottom right panel, blue) from the photodynamical modeling. For comparison the TTV of \citet{Rowe2015} measured on individual transits are shown as empty circles with grey errorbars.}
              \label{TTV-plot}%
    \end{figure*}

\section{Planetary properties}
\label{planetary-parameters}
\subsection{Transit photodynamical modeling}


We analyzed the \textit{Kepler} data using a photodynamical model \citep{carter2011} described in \citet{Almenara2018}. Briefly, we used the {\sc \tt REBOUND} code \citep{rein2012}, with the {{\sc \tt WHFast} {\it N}-body integrator \citep{rein2015}, to compute the positions of the star and the planets during the \textit{Kepler} observations. The latter is used to compute the light curve model with the analytic description of \citet{mandelagol2002} using a quadratic limb-darkening law \citep{manduca1977} and the parameterization of \citet{kipping2013} to consider only physical values. To model the long-cadence data, we oversampled the model by a factor of 10 and then binned back to the observed cadence. This accounts for the deformation of the signal due to the duration of the exposure \citep{kipping2010}. With an {\it N}-body time-step of 0.05~d we estimate the error of the model to be lower than 1~ppm following \citet{Almenara2018}. 

For Kepler-278, the model has 22 free parameters: the stellar density, two limb-darkening coefficients, five orbital elements, and a mass and radius ratio per planet, the difference in longitudes of the ascending nodes, the amplitude of an additional multiplicative white noise term for the \textit{Kepler} long and short-cadence data, and a free normalization factor for each dataset, corresponding to the out-of-transit flux. The five orbital parameters of each planet and the difference in longitudes of the ascending node are set at the reference time $2\;455\;695.37632$~BJD$_{\mathrm{TDB}}$ for Kepler-278, and $2\;455\;690.39125$~BJD$_{\mathrm{TDB}}$ for Kepler-391. For Kepler-391 there is only long-cadence data so the model has 20 parameters. 

We used a normal prior for the stellar density ($\rho_{\star}=0.07240\pm0.00094~\mathrm{g\;cm^{-3}}$ for Kepler-278 from H13, and $\rho_{\star}=0.077\pm0.011~\mathrm{g\;cm^{-3}}$ for Kepler-391, Section ~\ref{sec.stellar.parameters}), and uniform prior distributions for the remaining parameters. We limit the inclination of the inner planet to the range [0, 180] degrees and the inclination of the outer planet to the range [0, 90] degrees due to the symmetry of the problem.

We used the {{\sc \tt emcee}\xspace} algorithm \citep{goodmanweare2010, Foreman2013} to sample from the posterior distributions of the parameter models. We ran 100 walkers for 1.2$\times10^{6}$ steps for Kepler-278, and 0.7$\times10^{6}$ steps for Kepler-391. Only the last 100\;000 steps were used for the final inference.

    \begin{figure*}[]
   \centering
   \includegraphics[width=.495\textwidth]{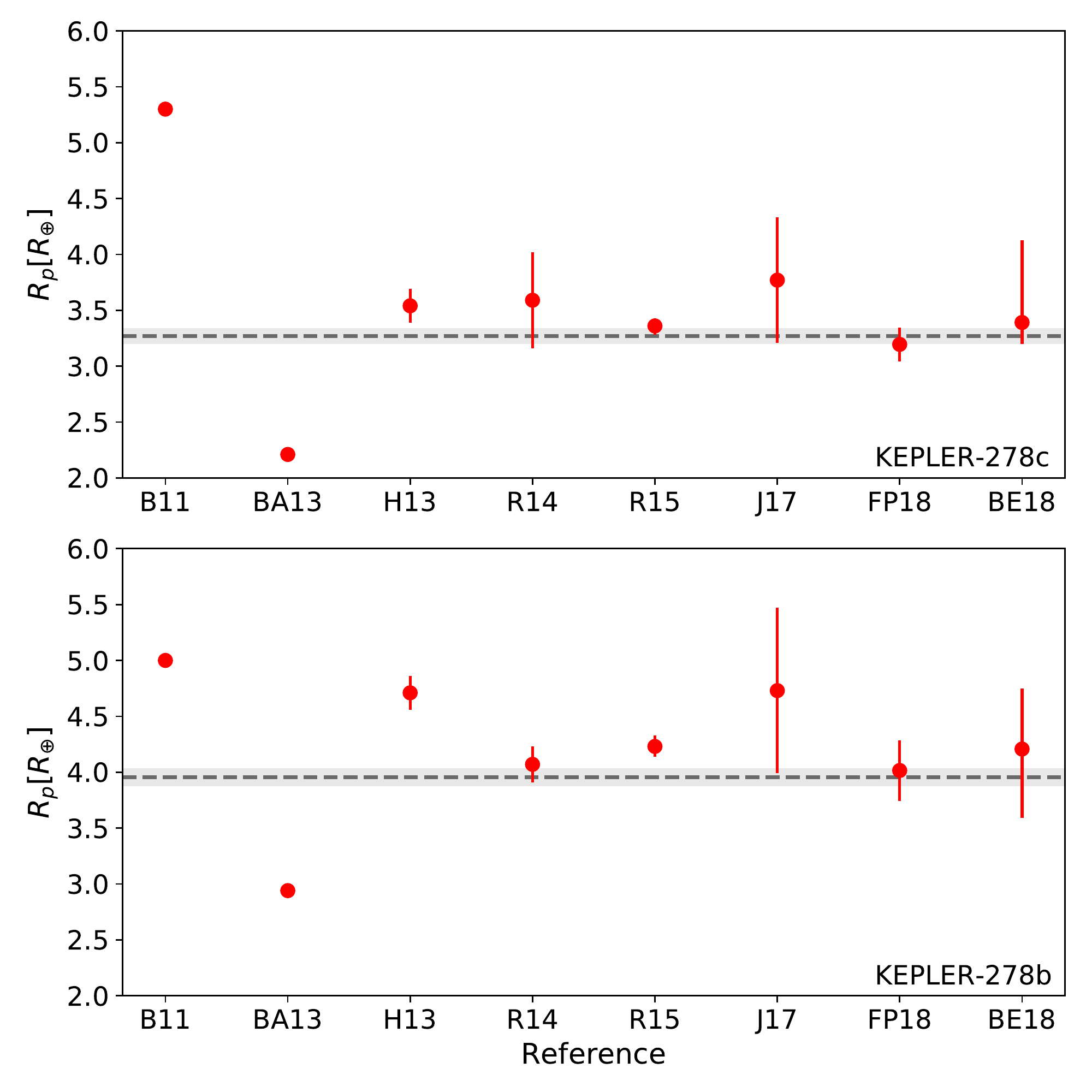}
      \includegraphics[width=.495\textwidth]{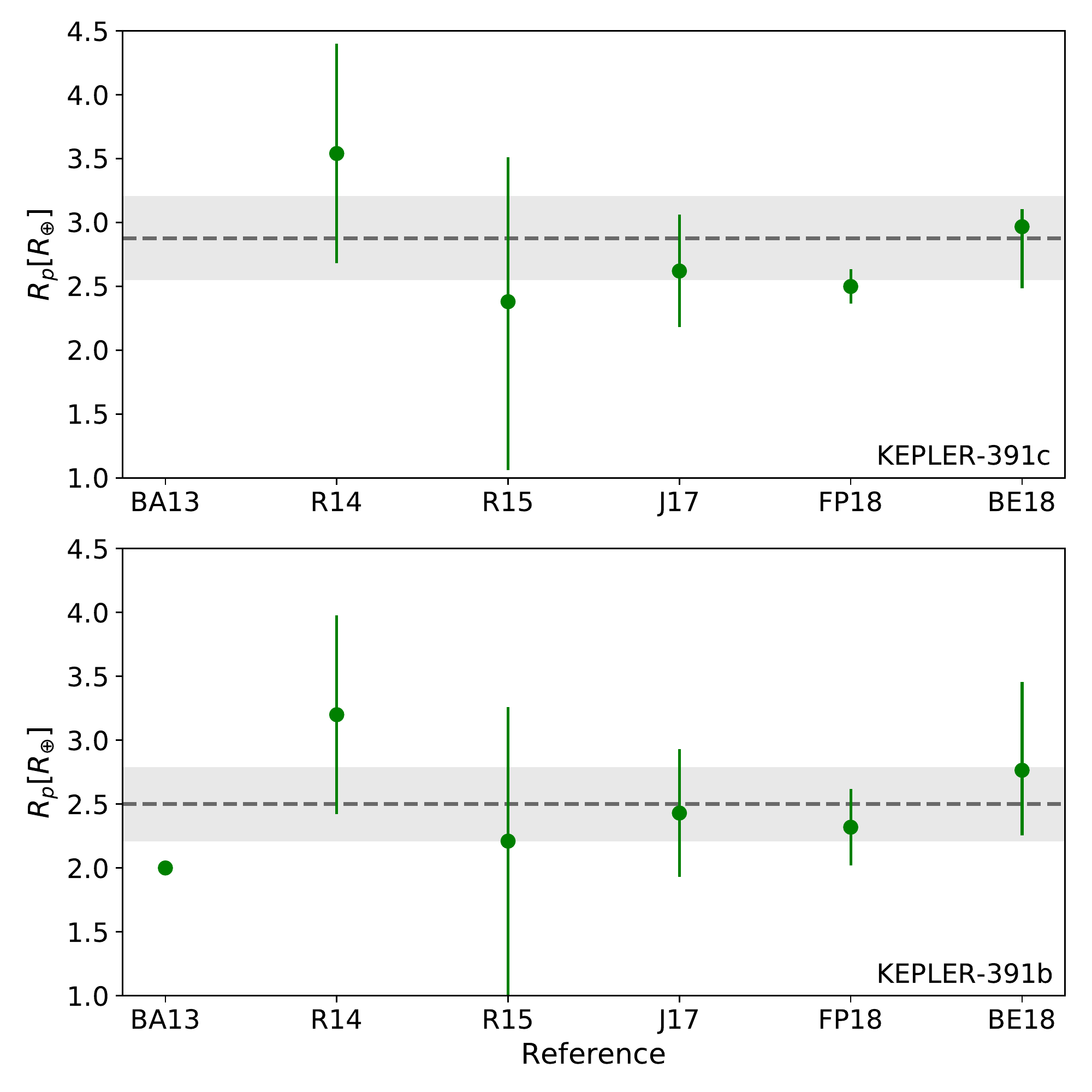}

   \caption{Planetary radii derived in this study (dashed lines) in comparison with those reported by 
   \citet[][BA13]{Batalha2013}, \citet[][H13]{Huber2013}, \citet[][R14]{Rowe2014}, \citet[][R15]{Rowe2015},   \citet[][J17]{Johnson2017}, \citet[][FP18]{Fulton2018}, \citet[][BE18]{Berger2018} indicated by red and green filled circles for Kepler-278b/c (\textit{left}) and Kepler-391b/c (\textit{right}), respectively. Grey shaded areas indicate 1$\sigma$ uncertainty in our results.}
              \label{comp-lite-planetary-radii}%
    \end{figure*}

\subsection{Results}

The photodynamical modeling allows improving the precision of the planetary parameters significantly with respect to an analysis correcting the light curves using individually-measured transit times. This can be particularly important in low S/N regimes, such as is the case for Kepler-278 and Kepler-391, where the transit times measured on individual transits are plagued by biases from spot crossing and other stellar variability issues \citep[see, e.g.,][]{Barros2013}. One of the largest advantages of the photodynamical modeling is minimizing the impact of these effects by using the entire set of transits to constrain each transit time and analyzing the dynamics of the system \citep{Almenara2015}. However, even with a photodynamical analysis, the small S/N of the transits of Kepler-278 and Kepler-391 calls for caution. For example, the mean stellar density can be inferred with a precision of around 50\% from the transit light curve analysis for Kepler-278, but the asteroseismic analysis provides a precision of 1.3\% (H13). The inference may be biased by unmodelled or unknown systematics effects, that at this level of signal dominate the error budget. The two-dimensional projections of the posterior sample are shown in Fig. \ref{pyramidmodel278} and \ref{pyramidmodel391} whilst the summary statistics of the marginal posterior distributions of each parameter is presented in Table \ref{planetary-parameters-table}. For each value, we report the median and 68.3\% confidence interval. For certain parameters, such as the eccentricities and mass ratios of the planets around Kepler-391, only upper limits are available. In this case, we report only the upper limit of the 95-\% Highest Density Interval (HDI)\footnote{The $q$-\% Highest Density Interval on a one-dimensional density is defined by all the points such that their densities are larger than a given value $W$ and such that the integral over all those values is $q/100$, i.e., the inteval $I = \{x : f(x) > W \}$ where $W$ satisfies that $\int_{x: f(x)>W} f(x) = q/100$.}.

Concerning the planets around Kepler-278, the inference based on the photodynamical model indicates that the planets are in eccentric ($e\sim0.7$) aligned (similar arguments of the pericentre, inclinations and longitudes of the ascending nodes) orbits. The posterior of the mean stellar density is dominated by the asteroseismic prior. The planet/star radius ratio values, $R_p/R_*$,  are known with precisions of around 2.3\%, and combined with the stellar radius computed in Section ~\ref{sec.stellar.parameters}, they yield planetary radii of $3.96\pm 0.08\;R_\oplus$, and $3.27\pm0.07\;R_\oplus$, for planets b and c, respectively. Because the planets exhibit transit timing variations (Fig. \ref{TTV-plot}, left; see also Sec. \ref{TTV-278}), the mass ratios $M_p/M_*$ can also be determined, albeit with a precision of around 40\% for both planets. The posterior distributions of the mass ratios exhibit two or three not fully separated modes, indicating that the solution is not unique, and highlighting the difficulty of exploring the parameter space of the photodynamical model when transits have low signal-to-noise ratio \citep[see also][]{Almenara2018}. The 95\% upper limits for the masses are 127.12 $\;M_{\oplus}$ (0.4$\;M_\mathrm{Jup}$), and 54 $\;M_{\oplus}$ (0.17$\;M_\mathrm{Jup}$), for planets b and c, respectively. The planets are, therefore, likely to be Neptune-like, but their density is very poorly constrained. 

The orbits of the planets around Kepler-391 are compatible with circular orbits, according to the inference using the photodynamical model, but their eccentricities are badly constrained, with 95\% upper limits of 0.46 and 0.27 for the inner and outer planet, respectively. The difference of the longitudes of the ascending nodes is compatible with zero, but the value is again, poorly constrained. In this case, the stellar density changes slightly from the prior, but the precision is not much improved by the inclusion of the data. Because no TTV are observed (Fig.~\ref{TTV-plot}, right) only upper limits on the masses can be measured. Assuming the stellar mass from Section~\ref{sec.stellar.parameters}, we find that the upper limits of the 95\%-HDI are 4.4 $M_\mathrm{Jup}$ and 4.8 $M_\mathrm{Jup}$, for planets $b$ and $c$, respectively. In fact, the posterior distribution of planet $b$ is slightly bimodal, and the 95-\% HDI is disjointed: $[0.0, 2.3] \cup [3.2, 4.4]$ $M_\mathrm{Jup}$. In any case, the data do not provide strong constraints on the planetary masses. On the other hand, the radii are determined with a precision of around 10\%, and indicate the planets are sub-Neptunes.

Finally, we combined the stellar parameters derived earlier in Section \ref{sec.stellar.parameters} with the relative parameters from the photodynamical analysis to compute other physical properties for all the planets. Our final planet properties are listed in Table \ref{planetary-parameters-table}. 

In Fig. \ref{comp-lite-planetary-radii} we show the comparison between our planetary radii and those from B11, BA13, H13, R14, \citet[][R15, hereafter]{Rowe2015}, J17, FP18, and BE18. In general, as can be noticed, our radii agree fairly well with those obtained previously, particularly for Kepler-391b/c. Alternatively, for Kepler-278b/c although most of the results are in relatively good agreement (within 1$\sigma$), we note that a few estimations from literature (e.g., B11, BA13, H13) agree with our values only within 2$\sigma$. The discrepancies with B11 and BA13 mainly arise owing to their stellar radii are $\sim$37\% larger and $\sim$36\% smaller, respectively, than our estimation (see Fig. \ref{comp-lite-physical}). Additionally, although within 1$\sigma$, the large discrepancies with the estimations of J17 are also originated from the differences in the stellar radius. The disagreement with the planetary radii derived by H13 can be explained from significant differences between our derived radius ratio values $R_{\mathrm{p}}$ /$R_{\mathrm{\star}}$ and those computed by BA13 that are adopted by H13. The radius ratio values determined by BA13 are 16\% and 6\% larger than ours for Kepler-278b and Kepler-278c, respectively. The disagreement is likely related to different methods employed to analyze the \textit{Kepler} light curves \citep[i.e., photodynamical vs. individual transit curves;][]{Almenara2018}, which also includes significant differences in the assumption for excentricity, limb-darkening, and normalization, among others. A similar scenario possibly explains the small discrepancy with the radius of Kepler-278b derived by R15.

Our estimations of semi-major axis and incident flux are in good agreement with the available results computed by H13, R14, R15, J17, and FP18. Finally, the masses of the planets around Kepler-278 presented here are independent of the measured stellar mass given that they are computed from the planetary radii and densities derived from the photodynamical model using the asteroseismic stellar density.


\section{Discussion}
\label{discussion}
\subsection{Host stars ascending the red giant branch}
\label{discussion.evolved.stas}
Fig. \ref{hr-todos} shows the location of Kepler-278 and Kepler-391 in the $T_{\mathrm{eff}}- \log g$ plane in comparison with other confirmed exoplanet hosts detected via RVs and transits. Both stars, with similar early K spectral types, lie close to the base of the RGB at the boundary between the luminosity classes of giants and subgiants. Their derived surface gravities are more compatible with their classification as subgiants \citep[3.5 $<$ $\log g$ $<$ 4.1,][]{Bastien2016}. Based on the criteria to distinguish between subgiants and giants, that relies on the bolometric magnitude $M_{\mathrm{bol}}$, both Kepler-278 ($M_{\mathrm{bol}}$ = 3.12) and Kepler-391 ($M_{\mathrm{bol}}$ = 3.36) would be also classified as subgiants\footnote{ $M_{\mathrm{bol}} >$ 2.82: subgiants; $M_{\mathrm{bol}} <$ 2.82: giants \citep{Ghezzi2010b, Maldonado2013, Jofre2015a}}. However,  the two stars are classified as giants according to physically motivated boundaries from solar metallicity interior models \citep{Huber2017, Berger2018}, using the \texttt{evolstate} code\footnote{see \url{https://github.com/danxhuber/evolstate}}. In any case, consistent with their location on the HR-diagram, we obtained a relatively high carbon isotopic ratio, $^{12}\mathrm{C}/^{13}\mathrm{C}$ $>$ 40 (no detection) for both stars, indicating that CN-cycled material is little or not yet well mixed \citep[e.g.,][]{Gilroy1991, Thoren2004, Afsar2012} and therefore confirming that both stars are just starting their ascent on the RGB.

As can be noticed in Fig. \ref{hr-todos}, there are several stars all along the RGB with planets detected from RV surveys. However, only a few late subgiants and early red giants ($\sim$15) are known to host transiting planets, which highlight the difficulty to detect planetary transits at this evolutionary stage. Recently, \citet{Huber2019} presented the first oscillating late subgiant star, \object{TOI-197}, with a transiting planet discoverd by TESS. The stellar parameters of TOI-197 (T$_{\mathrm{eff}}$ = 5080 $\pm$ 90 K, $\log g$ = 3.60 $\pm$ 0.08 dex, $M_{\mathrm{\star}}$ = 1.21 $\pm$ 0.07 $M_{\mathrm{\odot}}$, and $R_{\mathrm{\star}}$ = 2.94 $\pm$ 0.06 $R_{\mathrm{\odot}}$) are very similar to those of Kepler-278 and Kepler-391. Moreover, the measured $\Delta \nu$ value of 28.94 $\pm$ 0.15 $\mu$Hz in TOI-197 \citep{Huber2019}, indicating that the star has just started its ascent on the RGB \citep{Mosser2014}, is also very similar to that of Kepler-278. 

\begin{figure}[]
   \centering
   \includegraphics[width=0.495 \textwidth]{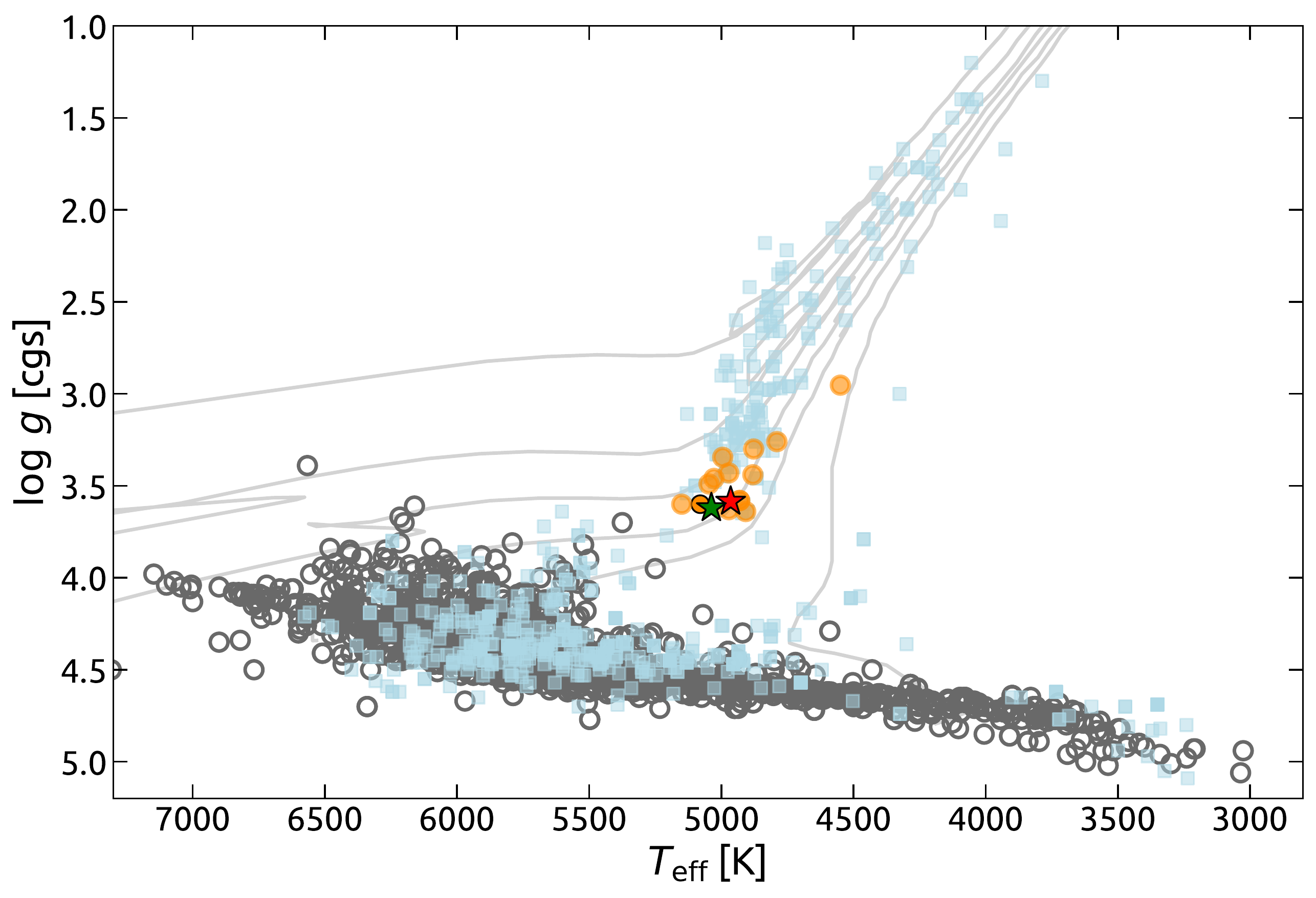}

   \caption{Location of Kepler-278 (red star) and Kepler-391 (green star) in the HR diagram, based on the spectroscopic T$_{\mathrm{eff}}$ and $\log g$ measured in this work, in comparison with other confirmed exoplanet hosts \citep[taken from the NASA Exoplanet Archive on 2019 March 19;][]{Akeson2013}: light blue squares represent stars with confirmed planets detected via RVs, dark gray circles are stars with planets found by transits, and orange circles represent the small population of RGB stars with transiting planets. The star TOI-197, with values taken from \citet{Huber2019}, is indicated by an orange circle with black edge-color. Error bars are omitted for clarity. Evolutionary tracks, corresponding to masses of 3, 2, 1.6, 1.3, 1.0, and 0.6 $M_{\mathrm{\odot}}$  (left to right) for [Fe/H]=+0.0 dex, from \citet{Girardi2000} are overplotted with continuous lines.}
              \label{hr-todos}%
    \end{figure}

\subsection{TTV in the system Kepler-278}    
\label{TTV-278}
The first sinusoidal TTV signals for Kepler-278c were reported by \citet{VanEylen2015}, and later by \citet{Holczer2016}. Also, R15 measured TTV in the Kepler-278 system and included their effect in the transit models. However, none of these works reported long-term TTV signals for the inner planet Kepler-278b. The increased precision in the transit times obtained from the photodynamical analysis allowed us not only to confirm the presence of a TTV signal in the outer planet but also to show, for the first time, a TTV signal in Kepler-278b (upper left panel in Fig. \ref{TTV-plot}), and hence estimate the mass of the outer planet. For comparison, in Fig. \ref{TTV-plot} we also show the TTV of R15 measured on individual transits (empty circles). As can be seen, the error bars of the transit times based on individual measurements are considerably larger than those from the photodynamical analysis \citep[e.g.,][]{Almenara2018}, which in the case of the planets around Kepler-278 hinder the detection of a TTV signal in the inner planet.

Using the Bayesian information criterion \citep[BIC\footnote{BIC$ = -2\ln \mathcal{L}_{max} + d \ln n$ ; where $\mathcal{L}_{max}$ is the maximum value of the Likelihood function, $d$ is the number of free parameters in the model, and $n$ is the number of data-points.},][]{Schwarz1978}, we found that the photodynamical model with the planetary masses as free parameters (model with TTV), as derived in Sec 4.1 for the system Kepler-278, provides a better fit to the data than a model where the planetary masses are fixed to zero (model without TTV). We obtained BIC$_{\rm TTV}$= $-$813267.0 and BIC$_{\rm no-TTV}$ = $-$813226.8 for the first and second case, respectively. Considering these values we derived $\Delta$BIC $\sim$ 40, which indicates a very strong case for TTV \citep{Kass1995}.

It is interesting to note also that the posterior TTV of Kepler-278b (top) and Kepler-278c (bottom) are anticorrelated. The presence of anticorrelated TTV signals among planet candidates on a single target provides strong evidence that the objects are true interacting planets \citep[][and references therein]{Steffen2013}. However, our posterior TTV are conditional to the model hypothesis that both planets are in the same system and interact gravitationally. Therefore, the anti-correlation of our TTV is simply a consequence of this hypothesis.

        \begin{figure}[]
   \centering
      \includegraphics[width=.497\textwidth]{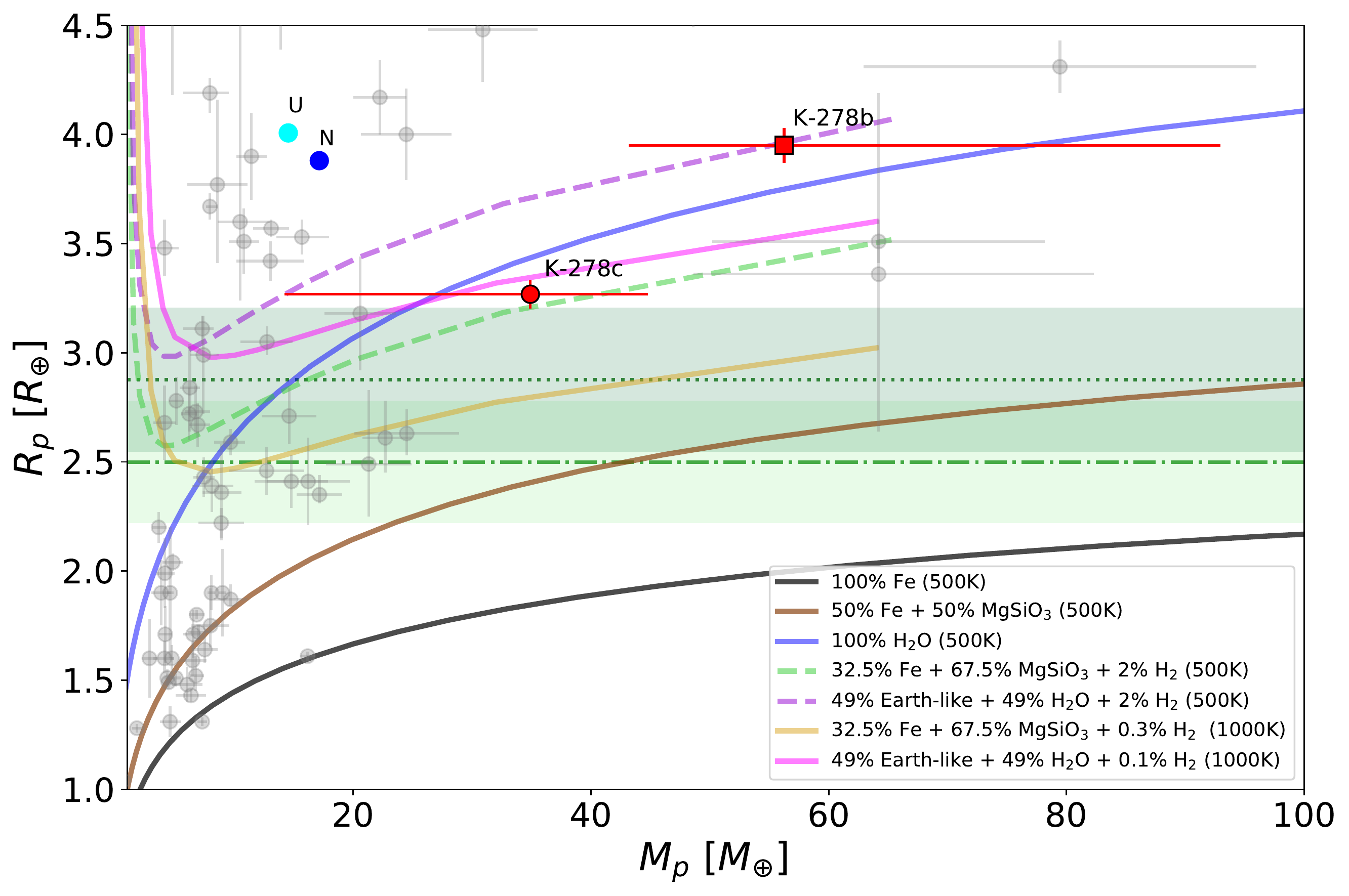}

   \caption{Mass--radius diagram for all confirmed planets with masses between 1--100 $M_{\mathrm{\oplus}}$ and radii 1--4.5 $R_{\mathrm{\oplus}}$ determined with a precision better than 30\% (data taken from the NASA Exoplanet Archive on 2019 June 10). Theoretical composition models from \citet{Zeng2016} are displayed with different lines and colors. Kepler-278b and Kepler-278c are indicated with a red square and circle. Kepler-391b and Kepler-391c are indicated with dark and light green bands, respectively, that represent the 1$\sigma$ uncertainty in our derived planetary radii. For reference, the solar system planets Uranus and Neptune are marked with a cyan and blue circle, respectively.}
              \label{mass-radius-diagram}%
    \end{figure}

\subsection{Planets in the mass-radius diagram}
\label{sec-mass-radius}
Figure \ref{mass-radius-diagram} shows the location of Kepler-278b and Kepler-278c in a mass-radius diagram compared to other relatively small transiting planets ($R_{\mathrm{p}}$ $<$ 4.5 $R_{\mathrm{\oplus}}$) with measured masses and radii better than 30\%. Planet structure models from \citet{Zeng2016} are also overplotted. In terms of mass and radius, Kepler-278c and especially Kepler-278b fall in relatively unpopulated regions of the mass-radius diagram. Although the large uncertainties in the planetary masses of Kepler-278b/c ($\sim$50 \%) prevent us from performing a detailed analysis of their composition, their radii are precise enough to locate both planets completely above the regime of solid planets with pure-iron or Earth-like rocky (32.5\% Fe + 67.5\% MgSiO$_{3}$) composition. Instead, both planets present bulk structures consistent with significant H$_{2}$O or H$_{2}$ content. In particular, Kepler-278b/c span a regime in mass in which they may either have a water-rich core (49.95\% Earth-like rocky core + 49.95\% H$_{2}$O layer) with a H$_{2}$-dominated gas envelope ($\sim$2\% for both planets or 0.1\% for Kepler-278c) or being pure H$_{2}$O planets. For Kepler-278c given our uncertainties, the mass and radius also match the regime of an Earth-like rocky core with a 2\% H$_{2}$ gaseous envelope. Improved measurements of the planetary masses for Kepler-278b/c would better constrain their interior compositions. Future studies, beyond the scope of this paper, like a full probabilistic Bayesian inference analysis would require precise RVs combined with the precise planetary radii and the stellar abundances obtained here \citep[see, e.g.,][]{Dorn2017a, Dorn2017b, Almenara2018}.

For the Kepler-391 system, given the lack of TTV signals, we cannot estimate the planetary masses and therefore both planets are indicated in Fig. \ref{mass-radius-diagram} with bands corresponding to their radii. However, considering the predicted masses from the mass--radius relation of \citet[][6.74 $\substack{+	4.35\\-3.27}$ $M_{\mathrm{\oplus}}$ for Kepler-391b and 8.92 $\substack{+	6.22\\-4.47}$ $M_{\mathrm{\oplus}}$ for Kepler-391c]{Chen2017}, these planets also might fall into the composition regimes consistent with a significant amount of water content or Earth-like rocky core and the presence of H$_{2}$ gaseous envelopes.

        \begin{figure}[]
   \centering
      \includegraphics[width=.497\textwidth]{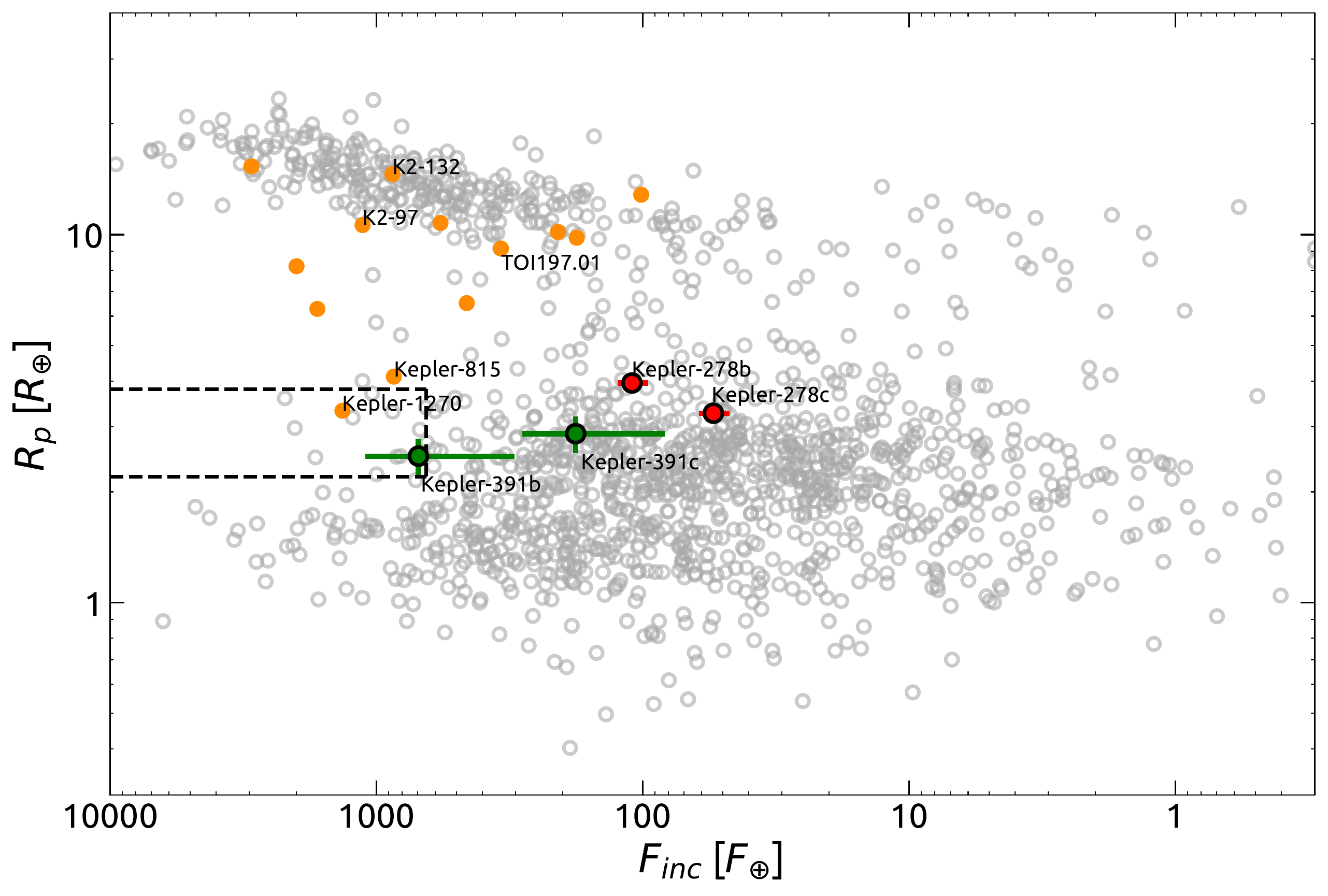}

   \caption{Planet radius versus incident flux for transiting confirmed exoplanets (grey circles; data taken from the NASA Exoplanet Archive on 2019 March 19). Planets around Kepler-278 and Kepler-391 are highlighted in red and green circles, respectively. Planets around RGB stars are indicated with filled orange circles. The black dashed lines delimit the super-Earth desert as derived in \citet{Lundkvist2016}. Other planets around RGB stars are also labelled (see text for more details). }
              \label{planet-context}%
    \end{figure}

\subsection{Binary companion candidate to Kepler-278}  
\label{binary}
It is worth mentioning that, using adaptive-optics imaging with the NIRC2 facility on the Keck II telescope, \citet{Kraus2016} detected a faint visual companion to Kepler-278 with a separation of $1''.984 \pm 0''.050$, projected separation of $\sim$ 860 AU, and position angle PA $10^{\circ}.1 \pm 0^{\circ}.11$. Based on star count models and the computed magnitude difference of $\Delta K$= 9.83 $\pm$ 0.13 relative to Kepler-278, \citet{Kraus2016} determined that this faint object is likely a bound companion given the low chance of background stars alignment. Nevertheless, they caution that multi-epoch imaging and common proper motion analysis are still necessary to conclusively confirm this companion is actually bound. Unfortunately, the visual companion is not resolved by \textit{Gaia}. 

The large $4''$ pixel scale of \textit{Kepler} implies that it is not possible to isolate the photometry of Kepler-278 from that of its potential close secondary companion, whose extra flux might contribute to dilute the observed transit depth, and therefore the derived planetary radii. However, given the large magnitude difference $\Delta K$, we find that transit depths are insignificantly diluted by the light of the stellar companion and therefore the planetary radii do not require any correction\footnote{The corrected planet radius can be obtained as $R_{\mathrm{p, corr}}$ = $R_{\mathrm{p}}\sqrt{1+10^{-0.4\Delta m}}$, where $\Delta m$ is the difference in magnitudes between the secondary and the primary \citep{Furlan2017} .}. This is also in agreement with the results of \citet{Furlan2017}. Moreover, considering the low mass inferred for the stellar companion \citep[$\sim$0.1 $M_{\mathrm{\odot}}$;][]{Kraus2016} and the transit duration of Kepler-278b ($\sim$ 7.5 hr) and Kepler-278c ($\sim$ 11.3 hr), we determine that the planets are actually transiting the late subgiant / early red giant star and not the low-mass visual companion.
  
Considering the angular separation of the companion candidate ($\sim$ 2 arcsec),  the difference in magnitudes between both stars ($\Delta$K = 9.83), and that our spectroscopic observations were taken with a seeing of $\sim$0.8 arcsec\footnote{The on-sky size of the GRACES’ fiber is 1.2 arcsec.}, we expect no significant contamination by the light of the companion candidate on the spectra of Kepler-278. Also, the spectra show no indication of a second set of stellar lines, which rules out additional close companions or background stars. Therefore, the existence of the visual companion should not affect any of the measurements performed from the GRACES spectra of Kepler-278.

\subsection{Planet properties in context}

Using the derived planetary parameters listed in Table \ref{planetary-parameters-table}, in Fig. \ref{planet-context} we located Kepler-278b/c and Kepler-391b/c in the radius-flux diagram. According to the classification scheme of \citet{Petigura2018} based on planetary radius and orbital period,  Kepler-278b ($P$ = 30.1 d, $R_{\mathrm{p}}$ = 3.95 $R_{\mathrm{\oplus}}$), Kepler-278c ($P$ = 51 d,  $R_{\mathrm{p}}$ = 3.26 $R_{\mathrm{\oplus}}$), and  Kepler-391c ($P$ = 20.5 d and $R_{\mathrm{p}}$ = 2.88 $R_{\mathrm{\oplus}}$) would be classified as warm sub-Neptunes, whilst Kepler-391b ($P$ = 7.4 d and $R_{\mathrm{p}}$ = 2.49 $R_{\mathrm{\oplus}}$) would be termed as a hot sub-Neptune. Only other two multi-planet systems have been reported to transit around RGB stars, Kepler-56 and Kepler-432, although they are composed by Jupiter-size planets and only the inner planet transits in the Kepler-432 system. Moreover, as can be seen in Fig. \ref{planet-context}, the planets around Kepler-391 are the smallest planets ($R_{\mathrm{p}}$ $\lesssim$ 3 $R_{\mathrm{\oplus}}$) detected around stars ascending the RGB so far.  

As evidenced on Fig. \ref{planet-context}, Kepler-278b, Kepler-278c, and Kepler-391c fall outside the limits of the hot super-Earth ``desert''  \citep{Lundkvist2016, Berger2018}, whilst Kepler-391b resides within the rightmost boundary of this region. In this area there is a deficit of super-Earth to Neptune-size planets  (2.2 $R_{\mathrm{\oplus}}$ $<$ $R_{\mathrm{p}}$ $<$ 3.8 $R_{\mathrm{\oplus}}$) at high irradiance ($F_{inc}$ $>$ 650 $F_{\oplus}$), that could be a consequence of the photoevaporation of low-mass planets atmospheres \citep[e.g.,][]{Owen2016}. Although the errorbar in the $F_{inc}$  might still locate Kepler-391b slightly outside the rightmost limit, as the host star continues its ascent on the RGB the significant increase of $F_{inc}$ will move the position of Kepler-391b well inside this desert in just a few Myr. During this phase, a potential hydrogen atmosphere in Kepler-391b, which is likely given its radius and predicted mass (see Section \ref{sec-mass-radius}), could be stripped by photoevaporation \citep{Owen2017}. With similar parameters to those of Kepler-391b, the planets around \object{Kepler-1270} and \object{Kepler-815} might experience a similar process as their host stars evolve up the RGB.

On the other hand, planets such as those orbiting Kepler-391 and Kepler-278 are probably too small or too far from their hosts to fall into the regime of significant inflation \citep[e.g.,][]{VanEylen2016}. For example, with $F_{inc}$ $\sim$ 340 $F_{\oplus}$ and $R_{\mathrm{p}}$ = 9.16 $R_{\mathrm{Jup}}$, the planet TOI-197.01 \citep{Huber2019} is located at the base of a ``inflation sequence'', a region in the radius-incident flux diagram (Fig. \ref{planet-context}) from which the planet radius increases with stellar irradiation \citep[see, e.g.,][]{Demory2011}. While TOI-197.01 would just be starting to be re-inflated \citep{Huber2019}, there are other planets around RGB stars on this inflation sequence, including the giant planets \object{K2-97b} and \object{K2-132b} whose radii have already become re-inflated as result of stellar evolution \citep{Grunblatt2016, Grunblatt2017}. 

Planets Kepler-278b/c and Kepler-391b/c are part of the small group of planets ($\sim$ 20), most of them detected via transits, in close-in orbits around evolved stars with a semi-major axis below $\sim$ 0.5 AU (P $\lesssim$ 100 d). One of the main scenarios to explain the paucity of close-in planets considers that these objects might end up engulfed by their host stars as they ascend on the RGB \citep[e.g.,][]{Villaver2009}. Based on our computed T$_{\mathrm{eff}}$, [Fe/H], $R_{\mathrm{\star}}$, and $M_{\mathrm{\star}}$ values, listed on Table \ref{tableparameters}, and evolutionary tracks from \citet{Girardi2000}, we found that the stellar surface of Kepler-391 will reach the inner planet in $\sim$410 Myr ($\approx$17 $R_{\mathrm{\odot}}$) and the outer planet in 428 Myr ($\approx$63 $R_{\mathrm{\odot}}$). Kepler-278, that is slightly more evolved than Kepler-391, will engulf its inner planet in $\sim$408 Myr ($\approx$43 $R_{\mathrm{\odot}}$) and in $\sim$428 Myr ($\approx$63 $R_{\mathrm{\odot}}$) the outer planet. These values represent a conservative upper limit on the remaining lifetimes of the planets, because the engulfment could be accelerated due to tidal interactions in the star-planet systems causing the orbital decay of the planets \citep{Villaver2009, Matsumura2010, Kunitomo2011}.

\subsection{Stellar abundances and planets}
The stellar metallicities that we measured for Kepler-278 ([Fe/H] = 0.22 dex) and Kepler-391 ([Fe/H] = 0.04 dex), from our GRACES spectra, are in line with the tendency that small planets can occur around stars with a wide range of metallicities \citep{Buchhave2012, Buchhave2014, Petigura2018}. Interestingly, the higher metallicity of Kepler-278 in comparison with that of Kepler-391 agrees with the trend of increasing planet radius with the host star metallicity \citep{Petigura2018}. On the other hand, recent studies suggest that [Mg/Si] mineralogical ratio probably plays an important role in the formation of small planets, since they tend to orbit stars with larger [Mg/Si] ratios in comparison with Jovian host stars or control stars without detected planets \citep{Adibekyan2015, Mack2018}. The relatively low mineralogical ratio that we found for both stars ([Mg/Si] = 0.00 for Kepler-278 and [Mg/Si] = $-$0.03 for Kepler-391) seems to point out that these evolved stars do not follow the reported trend, although we note that in \citet{Adibekyan2015} planets are categorized by mass rather than radius and most of them are single systems around dwarf stars. To make a proper comparison, it would be interesting to see how the results for these evolved \textit{Kepler} stars are compared with the [Mg/Si] ratios of a larger sample of evolved stars with planets. Considering that, to date, most of RGB stars host giant RV planets, this sample would represent a good starting point for such a comparison.

\subsection{Eccentricity of the planets orbiting Kepler-278}
\label{eccentricity}
The highly eccentric orbits of Kepler-278b ($e$ = 0.696 $\substack{+0.017\\-	0.026}$) and Kepler-278c ($e$ = 0.616 $\substack{+0.015\\-0.023}$) revealed by our photodynamical analysis are extremely rare among the known multi-planet systems and therefore add further interest to this system. The only previous measurement of eccentricity in this system was performed by \citet{VanEylen2015} based on a technique relying on Kepler's second law, measuring the duration of individual transits from \textit{Kepler} data and using the asteroseismic stellar density. After removing the TTV signal detected for the outer planet, using a sinusoidal model, they also found an eccentric orbit with a mode of $e$ = 0.51 and a 68\% confidence interval [0.39, 0.70]. The authors, however, advise caution about this result due to the large degeneracy with the impact parameter and the poor quality of the transit light curves. For Kepler-278b, on the other hand, they found a circular orbit (modal value of $e$ = 0.03 and 68\% confidence interval at [0, 0.36]) but did not detect a TTV signal, which might have biased their result for this planet.

The striking eccentricities found for Kepler-278b/c require additional and deeper studies. A high-precision radial velocity follow-up of Kepler-278 is necessary to confirm the eccentricities found in our analysis. From our computed orbital periods, planetary masses, and eccentricities, we estimate a RV semi-amplitude for Kepler-278b of $K$ $\approx$ 14 $\substack{+20.7 \\-4.7}$ m s$^{-1}$ and $K \approx$ 6.6 $\substack{+2.3\\- 5.1}$ m s$^{-1}$ for Kepler-278c. Considering that Kepler-278 is not too faint (V = 11.8 mag), is a slow rotator ($v\sin i$ = 2.5 km s$^{-1}$), and that our estimated RV jitter is 4.3 m s$^{-1}$, the RV signal for Kepler-278b might be detectable with existing precise instruments such as Keck-HIRES. 

If future studies indeed support high-eccentric orbits, it will be necessary to perform a detailed dynamical study of this system in order to constrain its origin and evolution. With the eccentricity values found here, Kepler-278 would join the short list of systems where  both planets present significant eccentricities such as those around Kepler-432 \citep{Quinn2015}. Additionally, they would be among the most eccentric planets around evolved stars. Kepler-278b/c would not follow the trend showing that small planets in \textit{Kepler} multi-planet have low eccentricities \citep{VanEylen2015, VanEylen2019, Mills2019}. In contrast, such values would be in line with the eccentric orbits found for close-in giant planets orbiting evolved stars \citep{Grunblatt2018}, for which orbital decay happens faster than tidal circularization due to tides raised on the evolved host stars \citep{Villaver2009, Villaver2014, Grunblatt2018}. 

Nevertheless, considering the masses, orbital periods and highly eccentric orbits of Kepler-278b/c, it is possible that other mechanisms be at work. The large eccentricities could be the result of planet-planet scattering events \citep[e.g.,][]{Rasio1996, Ford2008}. However, since in this case it would be expected that both the eccentricities and the mutual inclinations be high, the coplanar nature of this system presents a puzzle for this scenario. Moreover, given the semi-major axis and masses of the planets around Kepler-278, a coplanar high eccentricity migration mechanism \citep{Petrovich2015} would not be possible for this system\footnote{One of the configurations in which the coplanar, and high eccentricity migration mechanism could operate is when the two planets present $e$ $\gtrsim$ 0.5 and $M_{p, in}/M_{p, out} (a_{in}/a_{out})^{1/2} \lesssim$ 0.16 \citep{Petrovich2015}, where $M_{p}$ is the planetary mass, $a$ is the semi-major axis, and the indices \textit{in} and \textit{out} refer to the inner and outer planet, respectively. However, for the planets in Kepler-278 this expression is equal to $\sim$1.38.}. Another possible scenario might involve the influence of a stellar binary companion, like the one detected in the adaptive-optics images (see Section \ref{binary}), inducing secular Lidov-Kozai cycles \citep{Lidov1962, Kozai1962}. In this case, the external fourth body should be highly inclined with respect to the orbital plane in order to maintain the mutual inclinations of the planetary orbits \citep{TakedaG2008, Almenara2018b}. Additional scenarios that could produce coplanar highly-eccentric orbits  \citep[e.g., spin-orbit coupling, collisions; see][]{Almenara2018b}, also should be considered. 

Further information that might be key to understand the origin of this system is whether the stellar spin and the orbits of planets are aligned. Although we have derived the rotational velocity, unfortunately, no rotational period has been found for this star from \textit{Kepler} data \citep{McQuillan2013, Mazeh2015}, which could provide this information. Alternatively, the analysis of the rotational splitting of asteroseismic oscillation modes of the available \textit{Kepler} data and those forthcoming from TESS\footnote{The \textit{Transiting Exoplanet Survey Satellite} \citep[TESS;][]{Ricker2015} will observe Kepler-278 between July 18 and August 15 in 2019.} could be used to determine the stellar obliquity, similar to \citet{Huber2013} and \citet{Quinn2015}. This information, coupled with a detailed dynamical study of this system, which is beyond the scope of the current paper, would be key to further constrain the formation, evolution, and stability of the Kepler-278 system.

 \section{Summary and conclusions}
\label{conclusions}
In this paper, we have performed a detailed stellar and planetary characterization study of the remarkable multi-planet systems Kepler-278 and Kepler-391. Our main conclusions are summarized as follows: 

\begin{itemize}

\item[•] Using high-quality spectra collected with Gemini-GRACES, we have refined the stellar parameters (T$_{\mathrm{eff}}$, $\log g$, v$\sin i$, $\log (R'_{HK})$, $M_{\mathrm{\star}}$, $R_{\mathrm{\star}}$, and $\tau_{\star}$), and derived precise chemical abundances of 25 elements (Li, C, N, O,  Na, Mg, Al, Si, S, Ca, Sc, Ti, V, Cr, Mn, Fe, Co, Ni, Cu, Zn, Sr, Y, Zr, Ba, and Ce). Nine of these elements and the carbon isotopic ratios, $^{12}\mathrm{C}/^{13}\mathrm{C}$, were not previously measured. Overall, our stellar parameters agree reasonably well with most of the previous results. However, we find that Kepler-278 is $\sim$15\% less massive than recently reported \citep[][]{Mathur2017, Johnson2017, Fulton2018}.

\item[•] Both the stellar activity index based on the \ion{Ca}{ii} H \& K lines and that obtained from the \ion{Ca}{ii} infrared triplet lines indicate low chromospheric activity for both stars.

\item[•] The stellar parameters of Kepler-278 and Kepler-391 along with their measured carbon $^{12}$C/$^{13}$C isotopic ratio reveal that both stars are just starting their ascent on the RGB. For Kepler-278, this is also confirmed from asteroseismic data.

\item[•] The chemical abundances of light, alpha, Fe-peak, and heavy elements of both stars follow the abundance trends of other evolved thin disk stars in the solar neighborhood. Also, the abundance vs. condensation temperature slopes for both stars are consistent with the average value presented by M16 for subgiants with planets. 

\item[•] The lithium abundance of Kepler-391, A(Li)$_{NLTE}$ = 1.29 $\pm$ 0.09 dex, is slightly below the standard limit of the rare Li-rich giant stars. The evolutionary state of this star, its low $v\sin i$, and lack of other chemical peculiarities would support that the Li content of Kepler-391 is most likely a remnant from the main-sequence phase and not the consequence of a recent planetary engulfment episode or a fresh lithium production phase.

\item[•] The measured stellar metallicities for Kepler-278 ([Fe/H] = 0.22 dex) and Kepler-391 ([Fe/H] = 0.04 dex) are consistent with the tendency that small planets can occur around stars with a wide range of metallicities \citep[e.g.,][]{Petigura2018}. Also, according to the relatively low mineralogical ratio  obtained for both stars ([Mg/Si] $\sim$ 0), none of them would seem to follow the trend of higher [Mg/Si] ratios for hosts with small planets \citep{Adibekyan2015}.

\item[•] Using a photodynamical analysis of the \textit{Kepler} light curves, in combination with our new stellar parameters, we derived improved planetary properties. In particular, for the Kepler-278 system, the increased precision in the transit times obtained from this analysis allowed us to measure for the first time the masses of Kepler-278b ($M_{\mathrm{p}}$ = 56	$\substack{	+37\\	-13}$ $M_{\mathrm{\oplus}}$) and Kepler-278c ($M_{\mathrm{p}}$ = 35	$\substack{+9.9	\\ -21} $ $M_{\mathrm{\oplus}}$). Not only do we confirm the presence of a TTV signal in Kepler-278c, but also detect a previously unreported TTV signal in the inner planet Kepler-278b. For the system Kepler-391, given the lack of detected TTV signals, only upper limits on the masses could be provided.

\item[•] The location of Kepler-278b/c in the mass-radius diagram suggests that their bulk structures might be consistent with significant amount of water content and the presence of H$_{2}$ gaseous envelopes. According to their radii and orbital period, Kepler-278b, Kepler-278c, and Kepler-391c would be classified as warm sub-Neptunes. Kepler-391b is a warm sub-Neptune that resides just inside the rightmost boundary of the hot super-Earth desert \citep{Lundkvist2016}. In case this planet has a bulk structure consistent with a significant amount of water content or $H_{2}$ gaseous envelope, this may be suffering photoevaporation. It is expected that this process increases in a few Myr as its host star continues ascending on the RGB.

\item[•] The photodynamical analysis reveals that the orbits of both planets around Kepler-278 are surprisingly eccentric ($e \sim$ 0.7) and coplanar, which pose a puzzle about its origin. A precise RV follow-up of Kepler-278 is needed to confirm the eccentricity values presented here.

 \end{itemize}

Finally, we note that Kepler-278b/c and Kepler-391b/c are part of the small sample ($\sim$15) of close-in ($a$ $<$ 0.5 AU) transiting planets orbiting RGB stars. Kepler-278 and Kepler-391 are also exceptional  because they are two of the three multi-planet systems detected to date that transit evolved stars. Moreover, the planets orbiting Kepler-278 and Kepler-391 are among the smallest bodies discovered so far around RGB stars.

The fact that \textit{Kepler} observed $\sim$ 16,000 red giants \citep{Yu2018} but only a handful of planets were found highlights the difficulty of finding transiting planets around evolved stars, especially on the RGB. Although this observational bias likely may contribute to the relative paucity of close-in planets detected around giants, it is necessary to examine larger samples of stars in this evolutionary stage in order to constrain other proposed scenarios such as planet engulfment or different planet-formation mechanisms. The TESS mission will observe hundreds of thousands of red giants (with Tmag $<$ 11) from which is expected the detection of 50--100 new planets \citep{Campante2016, Barclay2018} that would be more amenable to spectroscopic and photometric follow-up. A larger sample of planets transiting evolved stars on the RGB, will provide better constraints on the formation and evolution of planets around intermediate and high-mass stars. Moreover, since the expected TESS detection limit in planetary mass is approximately the mass of Neptune \citep{Campante2016}, TESS might give us some clues about the frequency of systems similar to Kepler-278 and Kepler-391.


\begin{acknowledgements}
This study has been partially supported by UNAM-PAPIIT IN-107518. E. J. and R. P. acknowledge the financial support from DGAPA in the forms of Post-Doctoral Fellowships. We thank the anonymous referee for helpful comments and suggestions that improved the paper. This publication has made use of VOSA, developed under the Spanish Virtual Observatory project supported from the Spanish MINECO through grant AyA2017-84089. This research has made use of the NASA Exoplanet Archive, which is operated by the California Institute of Technology, under contract with the National Aeronautics and Space Administration under the Exoplanet Exploration Program. This research has made use of the SIMBAD database, operated at CDS, Strasbourg, France. This paper includes data collected by the \textit{Kepler} mission. Funding for the \textit{Kepler} mission is provided by the NASA Science Mission directorate. Data presented in this paper were obtained from the Mikulski Archive for Space Telescopes (MAST).
\\
\textit{Software}. \texttt{IRAF}, \texttt{OPERA} \citep{Martioli2012}, \texttt{MOOG} \citep{Sneden1973}, $q^{2}$ \citep{Ramirez2014}, \texttt{Teff-LR} \citep{Sousa2010}, \texttt{ARES} \citep{Sousa2015}, \texttt{VOSA} \citep{Bayo2008}, \texttt{iSpec} \citep{Blanco-Cuaresma2014}, \texttt{PARAM} \citep{daSilva2006}, \texttt{INSPECT} \citep{Lind2009}, \texttt{REBOUND} \citep{rein2012}, \texttt{emcee} \citep{goodmanweare2010, Foreman2013}, and \texttt{WHFast} \citep{rein2015}.

\end{acknowledgements}

%
%
\bibliographystyle{aa}
\bibliography{ref.bib} 

\listofobjects

\begin{appendix}

\section{Independent analysis on stellar mass, radius, and age}
\label{appendix-1}
   \begin{figure*}[th!]
   \centering
   \includegraphics[width=.99\textwidth]{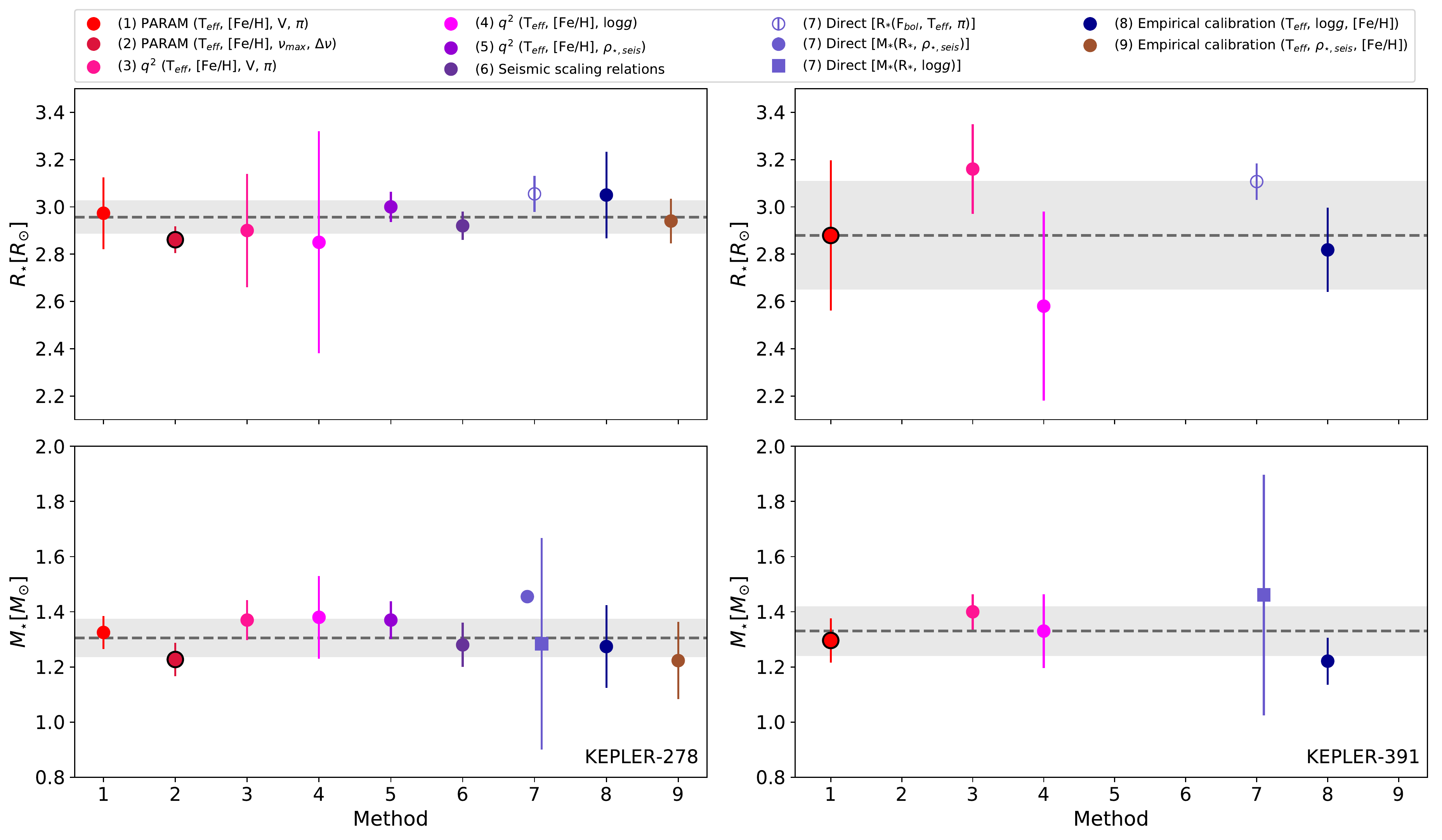}

   \caption{Stellar radii and masses determined by different approaches for Kepler-278 (\textit{left}) and Kepler-391 (\textit{right}). Dashed lines indicate the median values and the shaded areas indicate the standard deviations. The final adopted values for Kepler-278 and Kepler-391 are indicated with black edge-color circles.}
              \label{comp-metodos-fig}%
    \end{figure*}

In order to obtain an independent check on the stellar parameters derived from PARAM 1.3, using both DR2 \textit{Gaia} parallaxes (Method 1) and asteroseismic information (Method 2), we computed masses, radii and ages of both stars with other approaches and stellar models. Fig. \ref{comp-metodos-fig} shows the values obtained for mass and radius with all the methods, whilst Table \ref{comp-metodos-table} summarizes all the estimations including ages and densities. The different methods that we applied are the following:

\textit{$q^{2}$ + YY models}. We determined $M_{\mathrm{\star}}$, $R_{\mathrm{\star}}$, and $\tau_{\star}$ using probability distribution functions with Yonsei-Yale stellar isochrones \citep{Demarque2004}, as described in \citet{Melendez2012} and \citet{Ramirez2014}. This was acomplished via the $q^{2}$ pipeline, using as input T$_{\mathrm{eff}}$, [Fe/H], parallax, and V magnitude (Method 3). With this method, for Kepler-278, $q^{2}$ yields $M_{\mathrm{\star}}$ = 1.370 $\pm$ 0.073 $M_{\mathrm{\odot}}$, $R_{\mathrm{\star}}$ = 2.900 $\pm$ 0.241 $R_{\mathrm{\odot}}$, and $\tau_{\star}$= 4.200 $\pm$ 0.810 Gyr. For Kepler-391, we found $M_{\mathrm{\star}}$ = 1.400 $\pm$ 0.064 $M_{\mathrm{\odot}}$, $R_{\mathrm{\star}}$ = 3.160 $\pm$ 0.189 $R_{\mathrm{\odot}}$, and $\tau_{\star}$= 3.700 $\pm$ 0.655 Gyr. 

As another option, $q^{2}$ allows for the computation of stellar parameters from T$_{\mathrm{eff}}$, [Fe/H], and surface gravity as a luminosity indicator (Method 4). Using this option, $q^{2}$ returned $M_{\mathrm{\star}}$ = 1.380 $\pm$ 0.149 $M_{\mathrm{\odot}}$, $R_{\mathrm{\star}}$ = 2.850 $\pm$ 0.471 $R_{\mathrm{\odot}}$, and $\tau_{\star}$= 3.600 $\pm$ 1.010 Gyr for Kepler-278, whilst for Kepler-391 we found $M_{\mathrm{\star}}$ = 1.330 $\pm$ 0.134 $M_{\mathrm{\odot}}$, $R_{\mathrm{\star}}$ = 2.580 $\pm$ 0.400 $R_{\mathrm{\odot}}$, and $\tau_{\star}$= 4.000 $\pm$ 1.857 Gyr. 

Also, $q^{2}$ includes the possibility of using the stellar density, along with T$_{\mathrm{eff}}$ and [Fe/H] as initial constraints (Method 5), instead of parallax or $\log g$, similar to the method described in \citet{Sozzetti2007} and \citet{Mortier2014}. For Kepler-278, we used the precise asteroseismic stellar density ($\rho_{\star, seis}$) reported by H13, and obtained $M_{\mathrm{\star}}$ = 1.370 $\pm$ 0.086 $M_{\mathrm{\odot}}$, $R_{\mathrm{\star}}$ = 3.000 $\pm$ 0.064 $R_{\mathrm{\odot}}$, and $\tau_{\star}$= 4.200 $\pm$ 1.010 Gyr, which are in good agreement with those determined using the other inputs or models.

   \begin{table*} [t!]
  \small
      \caption[]{Stellar physical parameters obtained by the different techniques for Kepler-278 and Kepler-391.}
         \label{comp-metodos-table}
     \centering
         \begin{tabular}{l l c c c c c}
            \hline\hline																				
\#	&	Method	&	Input parameters	&	M$_{\star}$ [M$_{\odot}$]			&	R$_{\star}$ [R$_{\odot}$]			&	$\tau_{\star}$ [Gyr]			&	$\rho_{\star}$ [g cm$^{-3}$]				\\
\hline	
\multicolumn{7}{c}{Kepler-278} \\	
\hline																					
1	&	PARAM 1.3  (Parsec iscochrones)	&	T$_{\mathrm{eff}}$, [Fe/H], V, $\pi$	&	1.325	$\pm$	0.060	&	2.973	$\pm$	0.152	&	4.466	$\pm$	0.630	&	0.071	$\pm$		0.006	\\
2	&	PARAM 1.3  (Parsec iscochrones)	&	T$_{\mathrm{eff}}$, [Fe/H], $\Delta \nu$, $\nu_{max}$	&	1.227	$\pm$	0.061	&	2.861	$\pm$	0.057	&	5.761	$\pm$	1.019	&	0.074	$\pm$		0.005	\\
3	&	$q^{2}$ (Yonsei-Yale isochrones)	&	T$_{\mathrm{eff}}$, [Fe/H], $\log g$	&	1.370	$\pm$	0.073	&	2.900	$\pm$	0.241	&	4.200	$\pm$	0.810	&	0.079	$\pm$		0.009	\\
4	&	$q^{2}$ (Yonsei-Yale isochrones)	&	T$_{\mathrm{eff}}$, [Fe/H], V, $\pi$	&	1.380	$\pm$	0.150	&	2.850	$\pm$	0.471	&	3.600	$\pm$	1.500	&	0.084	$\pm$		0.019	\\
6	&	Asteroseismic scaling relations \tablefootmark{a}	&	T$_{\mathrm{eff}}$, $\Delta \nu$, $\nu_{max}$	&	1.283	$\pm$	0.077	&	2.922	$\pm$	0.058	&	N/A			&	0.072	$\pm$		0.005	\\
7	&	Direct\tablefootmark{b}	&	R$_{\star}$ (F$_{bol}$, T$_{\mathrm{eff}}$, $\pi$); M$_{\star}$ (R$_{\star}$, $\rho_{\star, seis}$)	&	1.455	$\pm$	0.019	&	3.055	$\pm$	0.076	&	N/A			&	0.072	$\pm$		0.002	\\
7	&	Direct\tablefootmark{b}	&	R$_{\star}$ (F$_{bol}$, T$_{\mathrm{eff}}$, $\pi$); M$_{\star}$ (R$_{\star}$, $\log g$)	&	1.284	$\pm$	0.384	&	3.055	$\pm$	0.076	&	N/A			&	0.063	$\pm$		0.020	\\
8	&	Empirical calibration\tablefootmark{c}	&	T$_{\mathrm{eff}}$, [Fe/H], $\log g$	&	1.274	$\pm$	0.150	&	3.005	$\pm$	0.183	&	N/A			&	0.066	$\pm$		0.011	\\
9	&	Empirical calibration\tablefootmark{d}	&	T$_{\mathrm{eff}}$, [Fe/H], $\rho_{\star, seis}$	&	1.223	$\pm$	0.143	&	2.940	$\pm$	0.095	&	N/A			&	0.068	$\pm$		0.009	\\
	\hline																					
	\multicolumn{7}{c}{Kepler-391} \\																					
	\hline																					
1	&	PARAM 1.3  (Parsec iscochrones)	&	T$_{\mathrm{eff}}$, [Fe/H], V, $\pi$	&	1.296	$\pm$	0.080	&	2.879	$\pm$	0.318	&	4.365	$\pm$	0.899	&	0.077	$\pm$		0.011	\\
2	&	PARAM 1.3  (Parsec iscochrones)	&	T$_{\mathrm{eff}}$, [Fe/H], $\Delta \nu$, $\nu_{max}$	&	N/A			&	N/A			&	N/A			&	N/A				\\
3	&	$q^{2}$ (Yonsei-Yale isochrones)	&	T$_{\mathrm{eff}}$, [Fe/H], $\log g$	&	1.400	$\pm$	0.064	&	3.160	$\pm$	0.189	&	3.700	$\pm$	0.655	&	0.063	$\pm$		0.006	\\
4	&	$q^{2}$ (Yonsei-Yale isochrones)	&	T$_{\mathrm{eff}}$, [Fe/H], V, $\pi$	&	1.330	$\pm$	0.134	&	2.580	$\pm$	0.400	&	4.000	$\pm$	1.857	&	0.109	$\pm$		0.023	\\
5	&	$q^{2}$ (Yonsei-Yale isochrones)	&	T$_{\mathrm{eff}}$, [Fe/H], $\rho_{\star, seis}$	&	N/A			&	N/A			&	N/A			&	N/A				\\
6	&	Asteroseismic scaling relations \tablefootmark{a}	&	T$_{\mathrm{eff}}$, $\Delta \nu$, $\nu_{max}$	&	N/A			&	N/A			&	N/A			&	N/A				\\
7	&	Direct\tablefootmark{b}	&	R$_{\star}$ (F$_{bol}$, T$_{\mathrm{eff}}$, $\pi$); M$_{\star}$ (R$_{\star}$, $\rho_{\star, seis}$)	&	N/A			&	N/A			&	N/A			&	N/A				\\
7	&	Direct\tablefootmark{b}	&	R$_{\star}$ (F$_{bol}$, T$_{\mathrm{eff}}$, $\pi$); M$_{\star}$ (R$_{\star}$, $\log g$)	&	1.461	$\pm$	0.436	&	3.107	$\pm$	0.077	&	N/A			&	0.069	$\pm$		0.022	\\
8	&	Empirical calibration\tablefootmark{c}	&	T$_{\mathrm{eff}}$, [Fe/H], $\log g$	&	1.221	$\pm$	0.085	&	2.818	$\pm$	0.178	&	N/A			&	0.077	$\pm$		0.009	\\
9	&	Empirical calibration\tablefootmark{d}	&	T$_{\mathrm{eff}}$, [Fe/H], $\rho_{\star, seis}$	&	N/A			&	N/A			&	N/A			&	0.000	$\pm$		0.000	\\

           \hline
         \end{tabular} 
 \tablefoot{        
\tablefoottext{a}{Asteroseismic scaling relations from \citet{Kallinger2010}.} \\   
\tablefoottext{b}{Method detailed in \citet{Stassun2017}.} \\
\tablefoottext{c}{Empirical relations of \citet{Torres2010}.}\\
\tablefoottext{d}{Empirical relations of \citet{Enoch2010}.}\\
}  
        \end{table*} 

\textit{Asteroseismic scaling relations}. For Kepler-278, with available asteroseismic information, it was possible to estimate stellar mass and radius from the seismic scaling relations (Method 6) of \citet{Kallinger2010}: 

\begin{equation}
(R_{\star}/R_{\odot}) = (\nu_{max}/\nu_{max, \odot})(\Delta\nu/\Delta\nu_{\odot})^{-2}(T_{eff}/T_{eff, \odot})^{1/2},
\end{equation}
 
 \begin{equation}
(M_{\star}/M_{\odot}) = (\nu_{max}/\nu_{max, \odot})^{3}(\Delta\nu/\Delta\nu_{\odot})^{-4}(T_{eff}/T_{eff, \odot})^{3/2}.
\end{equation}

As before, here we employed our spectroscopic T$_{\mathrm{eff}}$ and asteroseismic parameters of H13. In excellent agreement with the results obtained with \texttt{PARAM}, these equations returned:  $M_{\mathrm{\star}}$ = 1.283 $\pm$ 0.077 $M_{\mathrm{\odot}}$, $R_{\mathrm{\star}}$ = 2.922 $\pm$ 0.058 $R_{\mathrm{\odot}}$.

\textit{Direct mass and radius}. As another test, we derived stellar radii and masses using direct observables (Method 7), following the procedure described in \citet{Stassun2017}. Briefly, the stellar angular radius $\Theta$ is computed from direct observables such as the bolometric flux F$_{bol}$, the effective temperature, and the parallax $\pi$ according to,

\begin{equation}
\Theta = F_{bol}^{1/2} \times \left( \dfrac{2341}{T_{eff}}\right) ^{2} ,
\end{equation}
where F$_{bol}$ is in 10$^{-8}$ erg cm$^{-2}$ s$^{-1}$, T$_{\mathrm{eff}}$ is in K, and $\Theta$ in mas. Then, the linear radius can be obtained via,

\begin{equation}
R_{\star}= 107.47 \dfrac{\Theta}{\pi} ,
\end{equation}
where $\pi$ is in mas and  $R_{\mathrm{\star}}$ in solar radii. We measured $F_{\mathrm{bol}}$ in our \textit{Kepler} stars by fitting  atmosphere models to the observed SEDs using the \texttt{VOSA} interface as is detailed in Section \ref{consistency}. Using the derived $F_{\mathrm{bol}}$, spectroscopic T$_{\mathrm{eff}}$ and DR2 \textit{Gaia} parallaxes in equations (A.4) and (A.5), we obtained $R_{\mathrm{\star}}$ = 3.055 $\pm$ 0.076 $R_{\mathrm{\odot}}$ and $R_{\star}$ = 3.107$\pm$ 0.077  $R_{\mathrm{\odot}}$ for Kepler-278 and Kepler-391, respectively. In both cases the model independent radii are in good agreement with those derived from Yonsei-Yale and PARSEC stellar models.

A direct measure of the stellar mass can be obtained by combining the empirically computed stellar radius with the stellar density obtained from asteroseismology or from the transit light curve analysis. For Kepler-278, we derived $M_{\mathrm{\star}}$ = 1.455 $\pm$ 0.019 $M_{\mathrm{\odot}}$ from the seismic density reported by H13 which agrees with those computed from stellar models. Given that both Kepler-278 and Kepler-391 present low S/N \textit{Kepler} light curves (see Section \ref{planetary-parameters}), we do not obtain stellar densities from these data.

As suggested by \citet{Stassun2017}, another possibility to derive a model-independent stellar mass is by using the direct $R_{\mathrm{\star}}$ in combination with the spectroscopic $\log g$ via the formula:

\begin{equation}
\dfrac{M_{\star}}{M_{\odot}}  = \left(\dfrac{R_{\star}}{R_{\odot}}\right)^{2} 10^{\log g - \log g_{\odot}} ,  
\end{equation}
where $\log g_{\odot}$ = 4.44 dex. For Kepler-278 we derived $M_{\mathrm{\star}}$ = 1.284 $\pm$ 0.384 $M_{\mathrm{\odot}}$ and $M_{\mathrm{\star}}$ = 1.461 $\pm$ 0.436 $M_{\mathrm{\odot}}$ for Kepler-391. 

\textit{Empirical calibration 1}. We computed $M_{\mathrm{\star}}$ and $R_{\mathrm{\star}}$ from the empirical relations of \citet{Torres2010} that link T$_{\mathrm{eff}}$, $\log g$, and [Fe/H] to the stellar mass and radius (Method 8). The relations were calibrated based on the precisely measured masses and radii (3\% or better) of 95 eclipsing binaries. The results based on this calibration are $M_{\mathrm{\star}}$ = 1.274 $\pm$ 0.150 $M_{\mathrm{\odot}}$ and $R_{\mathrm{\star}}$ = 3.005 $\pm$ 0.183 $R_{\mathrm{\odot}}$ for Kepler-278, and of $M_{\mathrm{\star}}$ = 1.221 $\pm$ 0.085 $M_{\mathrm{\odot}}$ and $R_{\mathrm{\star}}$ = 2.818 $\pm$ 0.178 $R_{\mathrm{\odot}}$ for Kepler-391. For both stars, the results are in excellent agreement with those obtained with \texttt{PARAM} and the other methods.

\textit{Empirical calibration 2}. Finally, we tested our results via the empirical calibration of \citet{Enoch2010} which is similar to that of Torres et al., but uses $\rho_{\star}$ instead of $\log g$ as input (Method 9). Also, their equations were calibrated with the same 95 eclipsing binaries from Torres et al. For Kepler-278, we obtain $M_{\mathrm{\star}}$ = 1.223 $\pm$ 0.143 $M_{\mathrm{\odot}}$ and $R_{\mathrm{\star}}$ = 2.940 $\pm$ 0.095 $R_{\mathrm{\odot}}$ from the $\rho_{\star, seis}$ value by H13 as input. These results are in good agreement with those obtained from \texttt{PARAM}.

All values for the stellar radii, masses, ages, and stellar densities that result from the different techniques and models are summarized in Table \ref{comp-metodos-table}. As can be seen from this table and in Fig \ref{comp-metodos-fig}, there is generally good agreement between the masses and radii obtained from \texttt{PARAM} + PARSEC isochrones and those derived from most of the other techniques and stellar models.  

\subsection*{Ages from the [Y/Mg] ratio and lithium abundance}
\label{sec.chemical.ages}
It has been proven that the stellar abundances of Y and Mg can be used to estimate stellar ages of solar-type stars through the [Y/Mg] ratio \citep{Nissen2015, Nissen2016, Tucci2016, Feltzing2017}. Moreover, recently  \citet{Slumstrup2017} found that the [Y/Mg] ratio is also a good indicator of age for evolved stars. Therefore, considering the age-[Y/Mg] relation from \citet{Tucci2016}, we obtain an age of 5.95 $\pm$ 2.57 Gyr for Kepler-278 from [Y/Mg] = $-$0.06, whilst for Kepler-391, with [Y/Mg] = 0.02, we derive an age of 4.14 $\pm$ 2.58 Gyr. 

We also derived the stellar ages employing the lithium-age relation from \citet{Carlos2016}. We estimated that Kepler-278, with a sub-solar lithium abundance, has an age of 5.5 $\pm$ 1.3 Gyr, whilst the Kepler-391's higher lithium abundance implies a sub-solar age of 3.64 $\pm$ 1.2 Gyr. As can be seen from Section  \ref{sec.stellar.parameters} and in the previous section of this Appendix, these independent stellar ages are in good agreement with the values derived from the isochrones analysis.

\section{Additional figures and tables}

\begin{figure*}[th!]
   \centering
 \includegraphics[width=.35\textwidth]{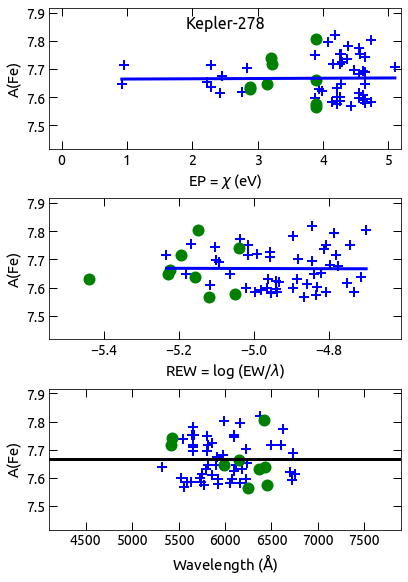}
 \includegraphics[width=.35\textwidth]{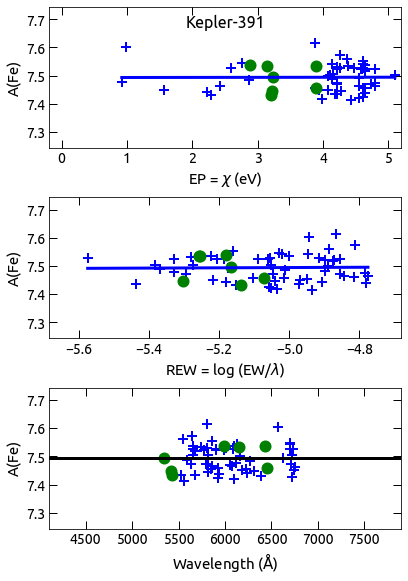}

  \caption{Iron abundance as a function of the spectral lines' excitation potential (upper panels), reduced EWs (middle panels), and wavelength (bottom panels) for Kepler-278 (\textit{left}) and Kepler-391 (\textit{right}). Blue crosses (green circles) correspond to \ion{Fe}{i} (\ion{Fe}{ii}). In the top and middle panels, solid lines are linear fits to the \ion{Fe}{i} data. In the lower panels, the solid lines indicate the average A(Fe) values.}
           \label{spectroscopic.equilibrium}%
  \end{figure*}


   \begin{table*}
  \small
      \caption[]{Elemental abundances of Kepler-278 and Kepler-391.}
         \label{abundances}
     \centering
         \begin{tabular}{l c c c c c c |l c c c c c c}
            \hline\hline
\multicolumn{7}{c|}{Kepler-278} & \multicolumn{7}{c}{Kepler-391}\\	
\hline
Species	&	A(X)	&	[X/H]	&	[X/Fe]	&	$\sigma_{line}$	&	$\sigma_{pars}$	&	$\sigma_{tot}$	&	Species	&	A(X)	&	\text{[X/H]}	&	\text{[X/Fe]}	&	$\sigma_{line}$	&	$\sigma_{pars}$	&	$\sigma_{tot}$	\\
\hline
Li$_{NLTE}$	&	0.87	&	$-$0.18	&	$-$0.40	&	…	&	0.1	&	0.10	&	Li$_{NLTE}$	&	1.29	&	0.24	&	0.20	&	…	&	0.09	&	0.09	\\
C	&	8.54	&	0.11	&	$-$0.11	&	0.03	&	0.09	&	0.10	&	C	&	8.31	&	$-$0.12	&	$-$0.16	&	0.10	&	0.07	&	0.12	\\
N	&	8.17	&	0.34	&	0.12	&	0.09	&	0.13	&	0.16	&	N	&	7.84	&	0.01	&	$-$0.03	&	0.10	&	0.13	&	0.16	\\
O$_{NLTE}$	&	8.84	&	0.15	&	$-$0.07	&	0.04	&	0.11	&	0.12	&	O$_{NLTE}$	&	8.71	&	0.02	&	$-$0.02	&	0.02	&	0.10	&	0.11	\\
$^{12}\mathrm{C}/^{13}\mathrm{C}$	&	$>$40	&		&		&		&		&		&	$^{12}\mathrm{C}/^{13}\mathrm{C}$	&	$>$40	&		&		&		&		&		\\
Na$_{NLTE}$	&	6.63	&	0.39	&	0.17	&	0.01	&	0.05	&	0.05	&	Na$_{NLTE}$	&	6.33	&	0.09	&	0.05	&	0.01	&	0.05	&	0.06	\\
Mg	&	7.84	&	0.24	&	0.02	&	0.02	&	0.02	&	0.03	&	Mg	&	7.64	&	0.04	&	0.00	&	0.05	&	0.03	&	0.06	\\
Al	&	6.83	&	0.38	&	0.16	&	0.01	&	0.04	&	0.04	&	Al	&	6.55	&	0.10	&	0.06	&	0.09	&	0.05	&	0.10	\\
Si	&	7.75	&	0.24	&	0.02	&	0.02	&	0.03	&	0.04	&	Si	&	7.58	&	0.07	&	0.03	&	0.02	&	0.02	&	0.03	\\
S	&	7.21	&	0.09	&	$-$0.13	&	…	&	0.10	&	0.10	&	S	&	7.02	&	$-$0.10	&	$-$0.14	&	…	&	0.09	&	0.09	\\
Ca	&	6.56	&	0.22	&	0.00	&	0.02	&	0.07	&	0.07	&	Ca	&	6.40	&	0.06	&	0.02	&	0.03	&	0.09	&	0.09	\\
\ion{Sc}{II}	&	3.29	&	0.14	&	$-$0.08	&	0.06	&	0.06	&	0.08	&	\ion{Sc}{II}	&	3.12	&	$-$0.03	&	$-$0.07	&	0.02	&	0.04	&	0.05	\\
Ti 	&	5.24	&	0.29	&	0.07	&	0.02	&	0.09	&	0.09	&	Ti 	&	5.07	&	0.12	&	0.08	&	0.02	&	0.10	&	0.10	\\
V	&	4.24	&	0.31	&	0.09	&	0.02	&	0.09	&	0.09	&	V	&	4.00	&	0.07	&	0.03	&	0.05	&	0.11	&	0.12	\\
Cr 	&	5.87	&	0.23	&	0.01	&	0.02	&	0.06	&	0.06	&	Cr 	&	5.73	&	0.09	&	0.05	&	0.04	&	0.07	&	0.08	\\
Mn	&	5.73	&	0.30	&	0.08	&	0.09	&	0.06	&	0.11	&	Mn	&	5.50	&	0.07	&	0.03	&	0.05	&	0.07	&	0.08	\\
Fe	&	7.67	&	0.22	&	0.00	&	0.01	&	0.04	&	0.04	&	Fe	&	7.49	&	0.04	&	0.00	&	0.01	&	0.03	&	0.04	\\
Co	&	5.25	&	0.26	&	0.04	&	0.03	&	0.04	&	0.04	&	Co	&	4.99	&	0.00	&	$-$0.04	&	0.03	&	0.04	&	0.05	\\
Ni	&	6.44	&	0.22	&	0.00	&	0.02	&	0.04	&	0.04	&	Ni	&	6.25	&	0.03	&	$-$0.01	&	0.01	&	0.04	&	0.04	\\
Cu	&	4.60	&	0.41	&	0.19	&	0.15	&	0.04	&	0.15	&	Cu	&	4.40	&	0.21	&	0.17	&	0.11	&	0.03	&	0.12	\\
Zn	&	4.74	&	0.18	&	$-$0.04	&	0.03	&	0.05	&	0.06	&	Zn	&	4.53	&	$-$0.03	&	$-$0.07	&	0.03	&	0.04	&	0.05	\\
Sr	&	2.94	&	0.07	&	$-$0.15	&	0.15	&	0.12	&	0.19	&	Sr	&	2.77	&	$-$0.10	&	$-$0.14	&	0.11	&	0.12	&	0.16	\\
\ion{Y}{II}	&	2.39	&	0.26	&	0.04	&	0.04	&	0.06	&	0.08	&	\ion{Y}{II}	&	2.27	&	0.06	&	0.02	&	0.03	&	0.03	&	0.05	\\
Zr	&	2.89	&	0.31	&	0.09	&	0.03	&	0.06	&	0.07	&	Zr	&	2.69	&	0.11	&	0.07	&	0.10	&	0.04	&	0.11	\\
\ion{Ba}{II}	&	2.37	&	0.19	&	$-$0.03	&	0.03	&	0.06	&	0.07	&	\ion{Ba}{II}	&	2.14	&	$-$0.04	&	$-$0.08	&	0.03	&	0.04	&	0.05	\\
\ion{Ce}{II}	&	1.87	&	0.29	&	0.07	&	0.07	&	0.06	&	0.10	&	\ion{Ce}{II}	&	1.79	&	0.21	&	0.17	&	0.02	&	0.03	&	0.04	\\

           \hline
         \end{tabular} 
           
        \end{table*} 


   \begin{figure*}[th!]
  \centering
      \includegraphics[width=.8\textwidth]{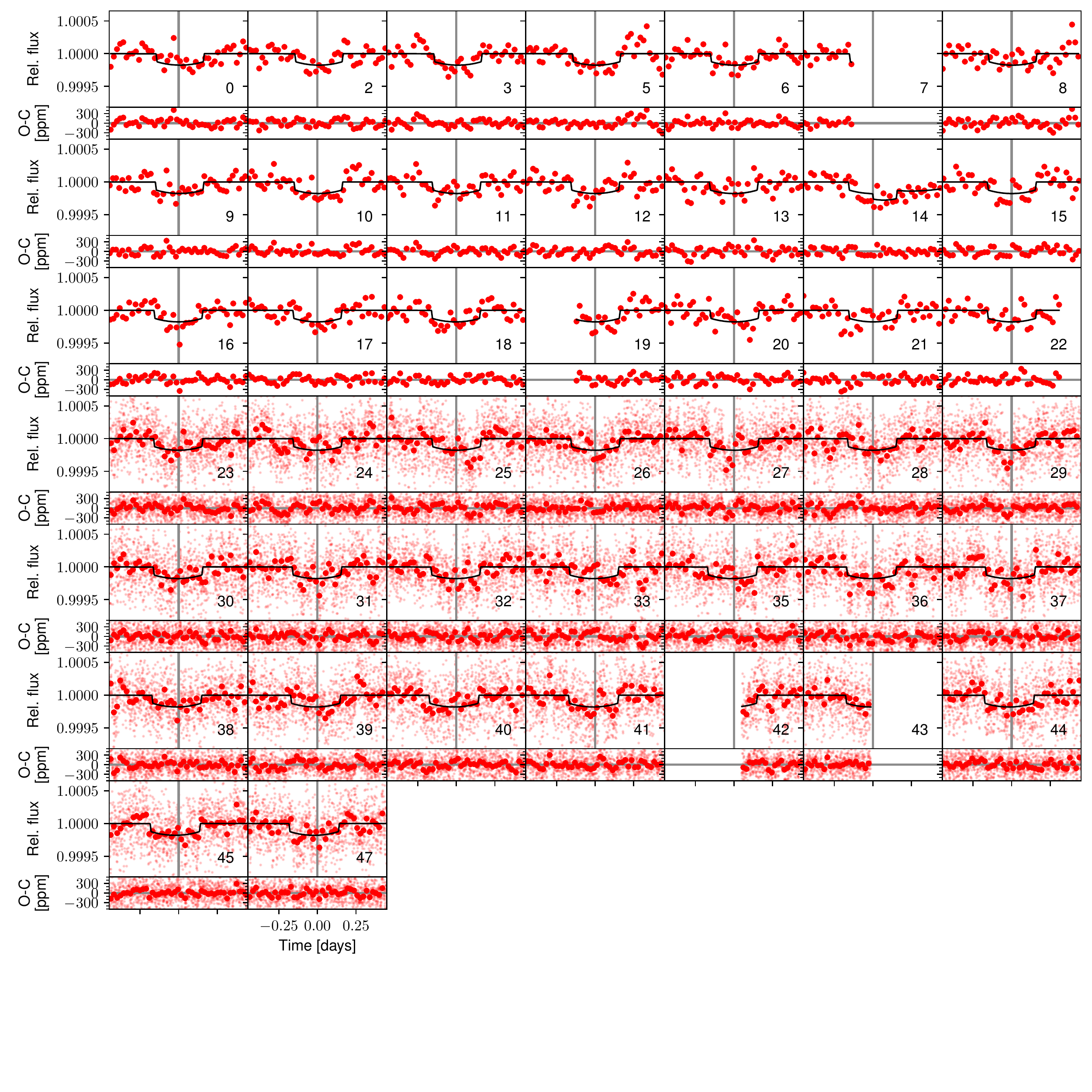}

   \caption{Transits of Kepler-278b observed by \textit{Kepler}. Dots represent the individual short-cadence observations and larger circles are 30-min averaged values. In those panels without short-cadence points, the circles represent the long-cadence data. Each panel is labelled with the transit epoch, and centered relative to a linear ephemeris (indicated by the vertical grey lines). The model distribution is constructed from 1000 random MCMC steps. The black line denotes the median model. In the lower part of each panel the residuals after subtracting the model to the observed data are shown.}
              \label{transit-fit-Kepler278b}%
   \end{figure*}

   \begin{figure*}[th!]
  \centering
    \includegraphics[width=.80\textwidth]{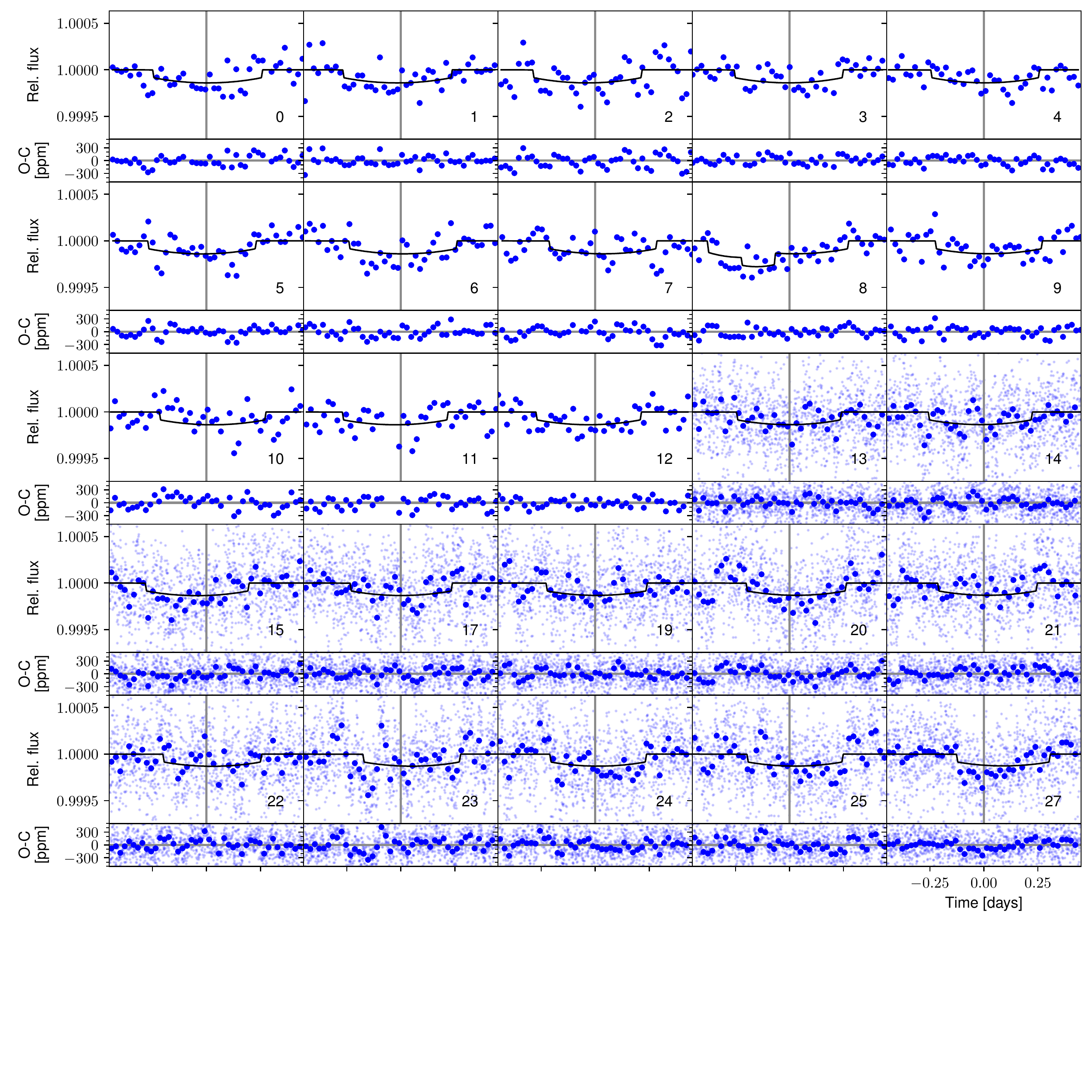}

  \caption{ Idem Fig. \ref{transit-fit-Kepler278b} for Kepler-278c.}
           \label{transit-fit-Kepler278c}%
  \end{figure*}

    \begin{figure*}[th!]
   \centering
     \includegraphics[width=.80\textwidth]{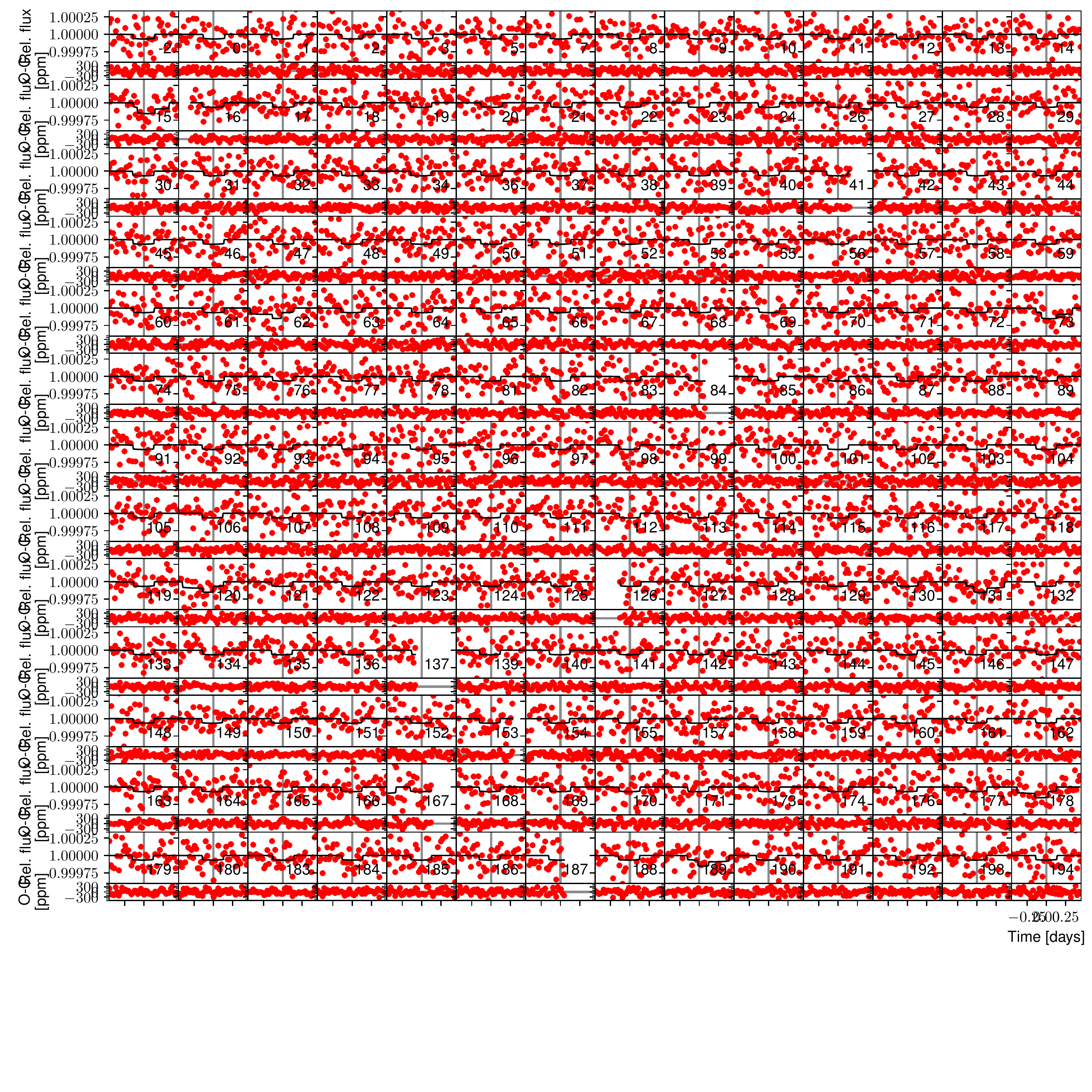}

   \caption{ Idem Fig. \ref{transit-fit-Kepler278b} for Kepler-391b.}
            \label{transit-fit-Kepler391b}%
   \end{figure*}

    \begin{figure*}[th!]
  \centering
 \includegraphics[width=.80\textwidth]{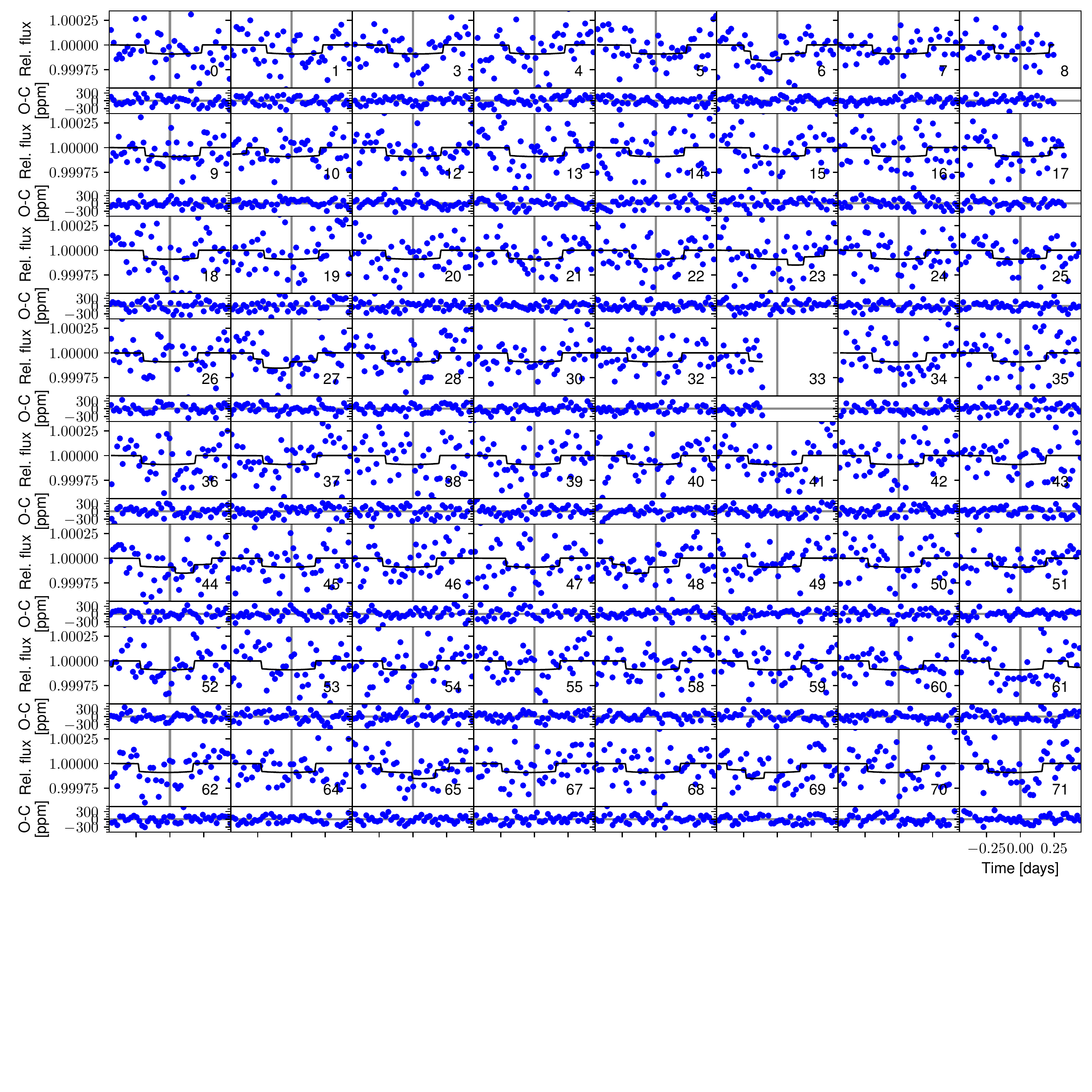}

 \caption{ Idem Fig. \ref{transit-fit-Kepler278b} for Kepler-391c.}
           \label{transit-fit-Kepler391c}%
 \end{figure*}

   \begin{figure*}[th!]
  \centering
      \includegraphics[width=1\textwidth]{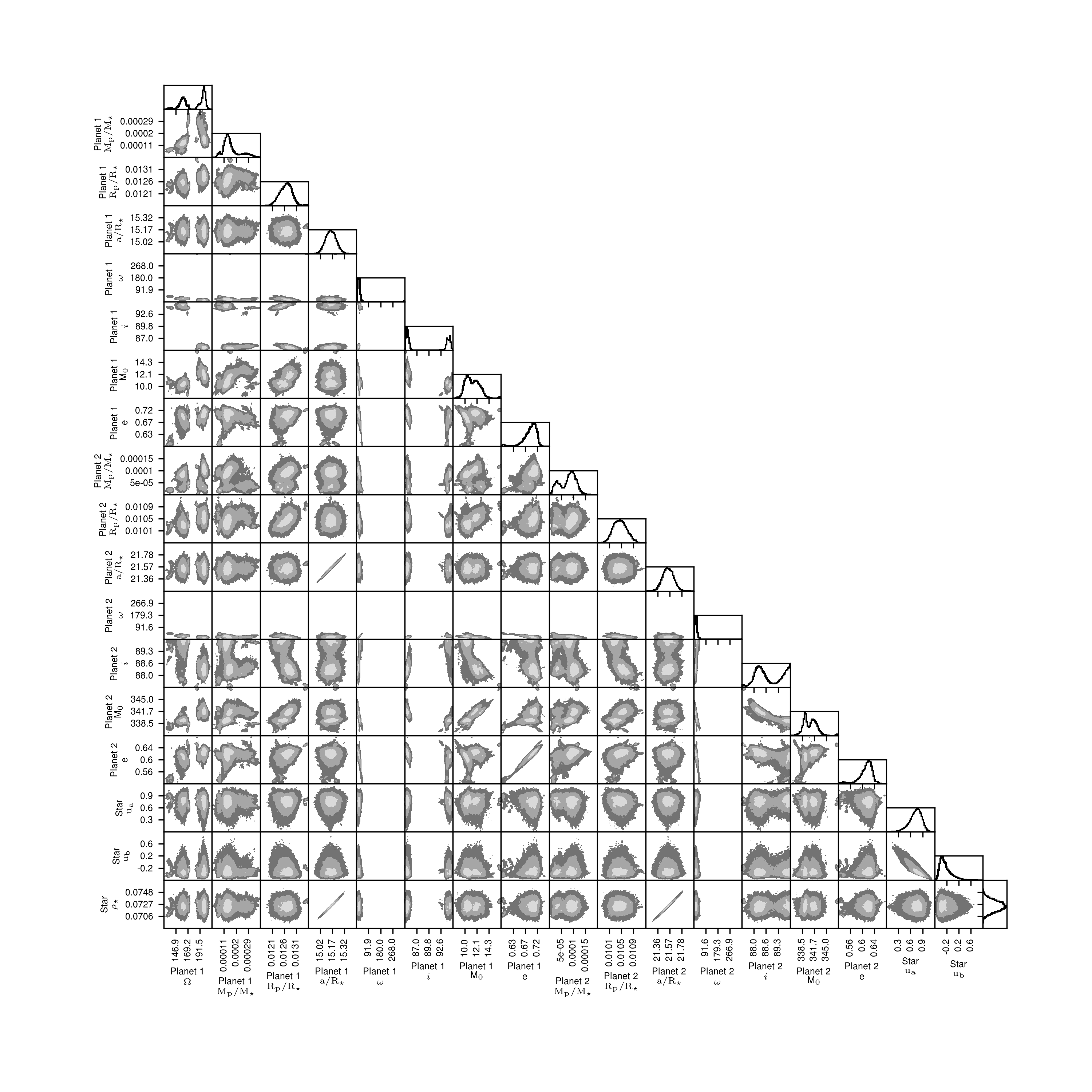}

   \caption{2D projections of the joint posterior samples obtained with the MCMC algorithm for planets around Kepler-278. The 39.3, 86.5, and 98.9 per cent 2D joint confidence regions (in the case of a Gaussian posterior, these regions project onto the one-dimensional (1D) 1, 2, and 3$\sigma$ intervals) are denoted by three different grey levels. The 1D histogram of each parameter is shown at the top of each column, except for the parameter on the last line that is shown at the end of the line. Units are the same as in Table \ref{planetary-parameters-table}.}
              \label{pyramidmodel278}%
   \end{figure*}

   \begin{figure*}[th!]
  \centering
      \includegraphics[width=1\textwidth]{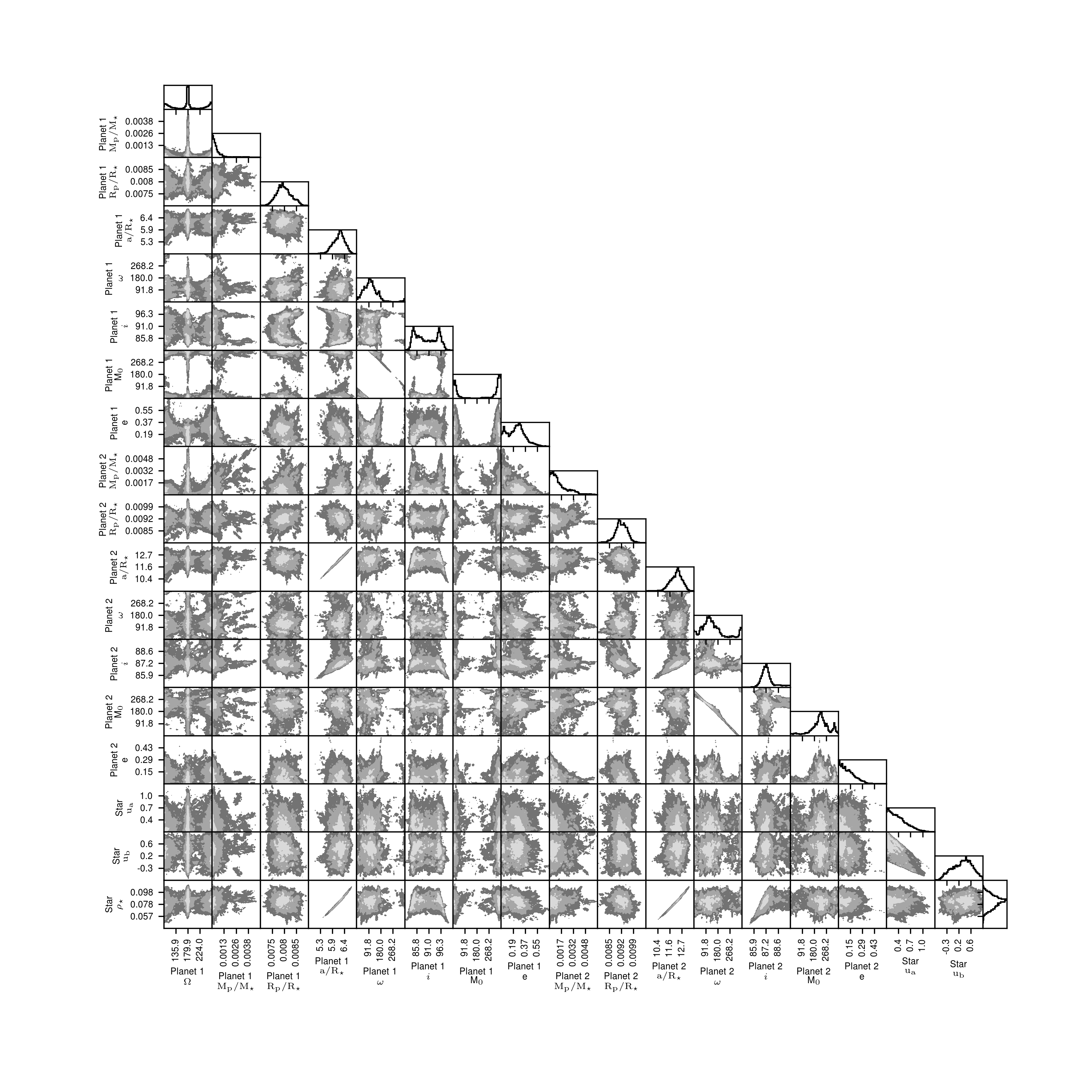}

   \caption{Same as Fig. \ref{pyramidmodel278} but for planets around Kepler-391.}
              \label{pyramidmodel391}%
   \end{figure*}

\end{appendix}

\end{document}